\documentclass[12pt]{article} 
\pdfoutput=1

\newcommand{\refbib}{ref.bib}

\usepackage[margin=0.75in]{geometry}
\usepackage{tikz}
\usetikzlibrary{arrows}
\usepackage{amsmath}
\usepackage{amsthm}
\usepackage{cases}
\usepackage{multirow}
\usepackage{multicol}
\usepackage{booktabs}
\usepackage{setspace}
\usepackage{caption}
\usepackage{tocbibind}
\usepackage[titletoc]{appendix}
\usepackage{listings}
\usepackage{mathtools}
\usepackage{thmtools}
\usepackage{float}
\usepackage{parskip}
\usepackage{amssymb} 
\usepackage[short, nocomma]{optidef}
\usepackage{adjustbox}
\usepackage{stmaryrd}
\usepackage{subcaption} 
\usepackage{accents}
\usepackage{siunitx}
\usepackage{graphicx}
\usepackage{array}

\usepackage{algorithm}
\usepackage[noend]{algpseudocode}

\algrenewcommand\algorithmicdo{\textbf{:}}

\let\oldReturn\Return
\renewcommand{\Return}{\State\oldReturn}

\usepackage[
    backend = biber,
    style = authoryear,
    maxcitenames = 3,
    maxbibnames = 5,
    sorting = nyt,
    dashed = false,
    sortcites = false,
    firstinits=true,
    uniquename=false,
    uniquelist=false,
]{biblatex}

\usepackage[
    bookmarks,
    colorlinks,
    citecolor = black,
    filecolor = black,
    linkcolor = black,
    urlcolor = black
]{hyperref} 

\hypersetup{
}

\newcommand{\articledate}{March 2020}

\numberwithin{equation}{section}
\numberwithin{figure}{section}
\numberwithin{table}{section}

\newcolumntype{C}[1]{>{\centering}m{#1}}

\DeclareSourcemap{
  \maps[datatype=bibtex]{
    \map[overwrite=true]{
      \step[fieldset=urldate, null]
    }
  }
}

\renewbibmacro{in:}{}

\DeclareFieldFormat[article,inbook,incollection,inproceedings,patent,thesis,unpublished]{title}{#1} 

\appto{\biburlsetup}{}

\DeclareFieldFormat{url}{\addspace\url{#1}}

\renewbibmacro*{volume+number+eid}{
  \textbf{\printfield{volume}}
  \setunit*{\addnbthinspace}
  \printfield{number}
  \setunit{\addcomma\space}
  \textbf{\printfield{eid}}
}

\DeclareFieldFormat[article]{number}{\textbf{\mkbibparens{#1}}}

\DeclareFieldFormat[article,inbook,incollection,inproceedings,patent,thesis,unpublished]{pages}{#1}

\AtEveryBibitem{
  \ifentrytype{online}
    {\iffieldundef{title}
      {}
      {}
    {}
  }
  {}
}

\DeclareFieldFormat[online]{note}{#1} 

\declaretheoremstyle[
spaceabove=12pt,
]{theorem}

\newcommand{\note}[1]{}
\newcommand{\todo}[1]{}
\newcommand{\ucite}[1]{(\textbf{??})}
\newcommand{\uparencite}[1]{\ucite{??}}
\newcommand{\utextcite}[1]{\ucite{??}}

\newcommand{\p}[1]{\left( #1 \right)}

\newcommand{\pb}[1]{\left[ #1 \right]}

\newcommand{\pe}[1]{\left\{ #1 \right\}}
\newcommand{\pa}[1]{\left| #1 \right|}
\newcommand{\pp}[1]{\left( #1 \right)_+}

\newcommand{\pn}[1]{\lVert #1 \rVert}

\newcommand{\R}{\mathbb{R}}

\newcommand{\EE}{E}
\newcommand{\PP}{P}

\newcommand{\E}[1]{\EE\pb{#1}}
\newcommand{\Ep}[1]{\EE\p{ #1 } }
\newcommand{\Epb}[1]{\EE\big( #1 \big) }
\newcommand{\Ebb}[1]{\EE\bigg[ #1 \bigg] }

\newcommand{\Ec}[2]{\EE \! \pb{ \left. #1 \right| #2}}

\newcommand{\Pc}[2]{\PP \! \p{ \left. #1 \right| #2}}

\newcommand{\ib}[1]{  \left \llbracket  #1 \right \rrbracket   }

\newcommand{\argmin}{\operatornamewithlimits{argmin}}
\newcommand{\argmax}{\operatornamewithlimits{argmax}}

\begin{document}


\title{ 
Fast 
Lower and Upper Estimates 
for the Price of 
Constrained Multiple Exercise American Options 
by 
Single Pass Lookahead Search 
and 
Nearest-Neighbor Martingale 
} 

\addcontentsline{toc}{section}{Front} 

\author{ 
Nicolas Essis-Breton\footnote{ 
  Address of correspondence to Nicolas Essis-Breton, 
  Department of Mathematics and Statistics at Concordia University, 
  Montreal, Quebec H3G 1M8, Canada, or e-mail: nicolasessisbreton@gmail.com. 
} 
	and 
Patrice Gaillardetz \\ 
Concordia University 
} 

\date{\articledate} 

\maketitle \noindent\hrulefill 

\begin{abstract} 
This article presents fast lower and upper 
estimates for a large class of options: 
the class of constrained multiple exercise American options. 
Typical options in this class are swing options 
with volume and timing constraints, 
and passport options with multiple lookback rights. 
Such options 
are widely traded on energy and financial markets, 
despite the fact 
that there is currently no method to price them. 
These options 
are intractable 
for exact pricing algorithms, 
and there is no approximate pricing algorithm 
that can incorporate all the complex features offered in these options. 
The main contribution 
of this article 
is to propose 
two algorithms 
that fill this gap. 
The first algorithm 
uses the artificial intelligence method 
of lookahead search with three novelties: 
1) we use the lookahead search in a direct scheme, rather than in an iterative scheme: 
neither a value function, nor a policy function are learned; 
2) we exhibit an approximation 
of the lookahead search problem 
in term of a conditional nearest-neighbor basis for the stock path, 
with convergence guarantees provided 
through a Vapnik-Chernovenkis dimension analysis 
and a convergence test based on an energy distance for the filtration. 
3) we solve the lookahead search problem by mixed-integer programming rather than by Q-learning. 
We found that these three novelties 
are keys for fast lower estimate of the considered option class. 
We call the resulting algorithm \textit{Single Pass Lookahead Search}. 
The second algorithm 
uses the dual approach to option pricing 
with three novelties: 
1) we exhibit a martingale basis 
formed by a nearest-neighbor basis for the stock path, 
for which enforcing the martingality condition is practical, 
and with a convergence test provided by an energy distance for the filtration; 
2) through a Vapnik-Chernovenkis dimension analysis, 
we show 
that obtaining a solution of the dual problem 
is much easier 
than obtaining 
a dual martingale 
that generalizes to any out-of-sample test; 
3) we solve the resulting continuous non-linear dual problem 
with a Frank-Wolfe method. 
We found that 
these three novelties 
are key for fast upper estimate of the considered option class. 
We call the resulting algorithm \textit{Nearest-Neighbor Martingale}. 
Several numerical examples illustrate the approaches 
including 
a swing option with four constraints, 
and a passport option with 16 constraints. 
The examples show that 
the proposed approaches 
are versatile, fast, 
and require no adjustment 
when applied to different options. 
The algorithms 
can thus simplify 
the risk management 
of financial derivatives 
when multiple pricing methods are used, 
or enable it 
when no other pricing method is applicable. 

\end{abstract} 

\noindent\hrulefill 

Much progress toward general option pricing method 
was made by 
extending dynamic programming algorithm 
initially designed for 
single exercise option. 
An early example 
of such 
extension 
is the \textit{forest of tree} method (\citeauthor{thompson95} \cite*{thompson95}). 
In this method, 
each tree represents the asset state 
under a particular constraint state. 
A change in the asset state induces 
a move in the asset tree, 
while a change in the constraint state induces 
a jump from one tree to another. 
The method was 
later improved 
by several authors 
(\citeauthor{lavassani01} \cite*{lavassani01}, 
\citeauthor{jaillet04} \cite*{jaillet04}, 
\citeauthor{marshall11} \cite*{marshall11}, 
\citeauthor{pages09} \cite*{pages09}, 
\citeauthor{wilhelm08} \cite*{wilhelm08}, 
\citeauthor{dahlgren05a} \cite*{dahlgren05a}, 
\citeauthor{dahlgren05b} \cite*{dahlgren05b}, 
\citeauthor{carmona08a} \cite*{carmona08a}, 
\citeauthor{carmona08b} \cite*{carmona08b}, 
\citeauthor{osterlee12} \cite*{osterlee12}). 
Another important 
extension of single exercise pricing method 
is 
the extension of the 
least-square Monte Carlo approach (\citeauthor{longstaff01} \cite*{longstaff01}) 
to multiple exercise option (\citeauthor{Meinshausen04} \cite*{Meinshausen04}). 
The extension consists in superposing 
several sets of sample paths, 
each sample path set corresponding 
to a particular state of the constraints. 
In essence, this approach 
could be named \textit{forest of path}. 
The approach also provides 
an upper bound on the option price 
by martingale duality (\citeauthor{rogers02} \cite*{rogers02}, \citeauthor{haugh04} \cite*{haugh04}): 
the optimal martingale is defined 
from the optimal exercise strategy found for the lower bound. 
The approach was later improved by several authors 
(\citeauthor{barrera06} \cite*{barrera06}, 
\citeauthor{bender06} \cite*{bender06}, 
\citeauthor{hambly10} \cite*{hambly10}, 
\citeauthor{bender11a} \cite*{bender11a}, 
\citeauthor{bender11b} \cite*{bender11b}, 
\citeauthor{schoenmakers12} \cite*{schoenmakers12}, 
\citeauthor{selvaprabu13} \cite*{selvaprabu13}, 
and 
\citeauthor{bender15} \cite*{bender15}). 
However, 
the runtime of dynamic programming based algorithm 
degrades quickly with the option complexity, 
in particular the constraint complexity. 
For example, the two previous families of approaches 
demand one tree (or one set of paths) 
for each possible constraint state. 
Further, enumerating the constraint state 
is often tedious, 
and 
these approaches 
do not address this difficulty. 
A fast fully general option pricing method 
that requires no adjustment from one option to another 
has not yet appeared 
in the option pricing literature. 

Our algorithms, Single Pass Lookahead Search (SPLS) 
and 
Nearest-Neighbor Martingale (NNM), 
differ from the previous approaches in several aspects. 
First and foremost, 
they do not use dynamic programming. 
SPLS uses lookahead searches  (\citeauthor{silver17} \cite*{silver17}, \citeauthor{browne12} \cite*{browne12}) 
in a direct scheme 
where no value function and no policy function are learned; 
NNM solves the dual martingale problem directly 
without reusing any information from the lower estimate. 
Second, they are fast and embarrassingly parallel algorithms. 
SPLS finds an optimal exercise strategy 
along a sample path 
with a single pass of lookahead search, 
independently of any other sample path. 
NNM finds an optimal martingale 
with a fast converging Frank-Wolfe scheme (\citeauthor{Frank56} \cite*{Frank56}), 
and each iteration of the scheme 
can be speed-up 
with a parallel computation 
of the dual option value. 
Third, they are fully general 
and applicable with no adjustments 
to a large class of options. 
Both SPLS and NNM 
work 
on a mathematical programming formulation 
of an option pricing problem. 
Such a formulation is intuitive and easy to obtain. 
To achieve these results, 
SPLS constructs 
the decision as step-functions 
on a nearest-neighbor basis for the stock path. 
This basis dispenses 
of building a search tree 
and makes the lookahead search solvable by mixed-integer programming. 
NNM 
constructs 
the 
optimal martingale 
as a conditional nearest-neighbor basis for the stock path. 
This basis 
allows 
the martingality condition 
to be directly enforced in the Frank-Wolfe scheme 
and 
turns each iteration of the scheme 
into the solution of a continuous linear program. 
A conditional nearest-neighbor basis 
is a nearest-neighbor basis with two levels, 
one level for the filtration information, 
and one level for the decision given the filtration information. 

In the literature on 
single exercise American option, 
some recent articles 
use 
the deep learning approach 
to dynamic programming 
(\citeauthor{becker19} \cite*{becker19}), 
and 
solve 
the dual problem directly 
with a Wiener expansion basis 
(\citeauthor{lelong16} \cite*{lelong16}). 
In contradistinction to these articles, 
we use simpler tools 
to provide fast estimates for 
the harder problem 
of constrained multiple exercise American options: 
SPLS uses only lookahead search, 
and NNM uses only a Frank-Wolfe scheme. 
Further, 
our algorithms 
have greater practical implications. 
They are fully general 
and we demonstrate their effectiveness 
on examples 
of unmatched complexity in the literature 
on multiple exercise options. 
First, a swing option 
with two multiple exercise American rights, 
a local and a global volume constraint, 
an exercise limit constraint, 
and a refraction constraint. 
Then, 
a passport option 
on three assets 
with 
constrained trading rights, 
several multiple exercise rights to waive the trading constraints, 
the right to choose between a call payoff or a put payoff, 
a barrier value that cancel the option, 
a multiple exercise reset right, 
and 
an American lookback right. 
All these features 
can be formulated 
in a mathematical program 
for our algorithms 
with five stopping times 
and 16 constraints. 

\section{Constrained Multiple Exercise American Options} \label{cmeao} 

A constrained multiple exercise American option 
is a stochastic control problem of the form 
\begin{maxi!}[0] 
{X,Y}
{ \Ebb{ \sum_{t=0}^{T} e^{-r t} f_t(Y_{0:t}, S_{0:t}) X_t }  \tag{\theparentequation}\label{cmeao_primal}}
{\renewcommand{\theequation}{\theparentequation.\arabic{equation}} \label{}}
{  }
\addConstraint{ g(X, Y, S) }{= 0  \quad \label{cmeao_primal_0}}{}
\end{maxi!} 
where 
$X$ is a multiple exercise stopping time 
adapted to the filtration generated by the stock $S$, 
$Y$ is an adapted control process, 
$f_t$ is the option payoff at time $t$, 
$g$ is a vector constraint that holds almost-surely, 
the expectation is taken under the risk-neutral measure (\citeauthor{BlackScholes73} \cite*{BlackScholes73}, \citeauthor{merton73} \cite*{merton73}, \citeauthor{HarrisonKreps79} \cite*{HarrisonKreps79}), 
$r$ is the risk-free rate, 
and $T$ the maturity. 
The notation $X_{0:t}=(X_{0}, X_{1},  \ldots , X_t)$ denotes the path of a process up to time $t$, 
the option payoff $f_t$ is in $\R$, 
the stock $S_t \in \R^d$ is multi-dimensional, 
$F_t$ is the filtration generated by the stock at time $t$, 
the multiple stopping time $X_t: F_t \to \pe{0,1}$ is a sequence of 0-1 stopping decisions, 
and 
the control process $Y_t: F_t \to \R^a$ is a sequence of vector-valued decisions. 
We refer to the stopping time and the control as the \textit{decisions} 
or the \textit{exercise strategy}. 

A typical option in this class is a vanilla swing option 
that allows a maximum of 10 swings $X$, 
each swing ordering a quantity $Y$ of the underlying $S$ 
up to a global ordering limit of 100. 
This option can be formulated with the following 
stochastic control problem 
\begin{maxi!}[0] 
{X,Y}
{ \Ebb{ \sum_{t=0}^{T} e^{-r t} Y_t S_t X_t }  \tag{\theparentequation}\label{cmeao_vanilla_swing}}
{\renewcommand{\theequation}{\theparentequation.\arabic{equation}} \label{}}
{  }
\addConstraint{ \sum_{t=0}^T X_t }{ \le  10 \quad \label{cmeao_vanilla_swing_0}}{}
\addConstraint{ \sum_{t=0}^T Y_t }{\in [-100, 100] \quad \label{cmeao_vanilla_swing_1}}{}
\addConstraint{ X_t }{\in \pe{0,1}  \quad \label{cmeao_vanilla_swing_2}}{ t=0,1, \ldots ,T}
\addConstraint{ Y_t }{\in \pb{-1,1}  \quad \label{cmeao_vanilla_swing_3}}{ t=0,1, \ldots ,T}
\end{maxi!} 

In \hyperref[meth]{Methods}, we show that the dual problem for this class of options 
can be formulated with 
\begin{mini!}[0] 
{M}
{ \Ebb{ \max_{x,y, g(x,y,S)=0} \sum_{t=0}^{T} e^{-r t} f_t(y_{0:t}, S_{0:t}) x_t - M_t x_t}  \tag{\theparentequation}\label{cmeao_dual}}
{\renewcommand{\theequation}{\theparentequation.\arabic{equation}} \label{}}
{  }
\end{mini!} 
where $M$ is an adapted martingale, 
$x$ is a non-adapted multiple stopping time, 
and 
$y$ is a non-adapted control. 
The proof is similar to the dual derivation in \textcite{rogers02}. 
The key ingredient of our dual formulation 
is that 
the option payoff $f_t$ 
is multiplied by a stopping time, 
so that the stopping time 
can be used as a weight for a penalizing martingale. 
This in turn motivates our choice of primal form \eqref{cmeao_primal}. 

For the previous swing option, 
the dual is given by 
\begin{mini!}[0] 
{M}
{ \Ebb{ \max_{x,y} \sum_{t=0}^{T} e^{-r t} y_t S_t x_t - M_t x_t}  \tag{\theparentequation}\label{cmeao_vanilla_swing_dual}}
{\renewcommand{\theequation}{\theparentequation.\arabic{equation}} \label{}}
{  }
\end{mini!} 
where the maximum is taken over the non-adapted decisions 
satisfying the constraints in \eqref{cmeao_vanilla_swing}. 

Since the stopping time 
in our option class is arbitrary, 
the following primal forms belong to the class 
\begin{align*} 
& \sum_{t=0}^T f_t(Y_{0:t}, S_{0:t}) \prod_i X^i_t , \\
& \sum_{t=0}^T h_t(X_{0:t}, Y_{0:t}, S_{0:t}),
\end{align*} 
where $X_t=(X^1_t, X^2_t,  \ldots )$ 
is a vector-valued multiple stopping time, 
and $h_t$ is a payoff 
that depends on both the stopping time and the control. 
Indeed, for the first form 
take the weighting stopping as $\prod_i X^i_t$. 
For the second form, 
redefine the control as the pair $\widetilde{Y}=(X,Y)$, 
redefine the payoff as $\tilde{h}_t = h_t / \bigvee_i X^i_t$ 
and take the weighting stopping time as 
the logical OR 
$\bigvee_i X^i_t$. 
The dual forms can then be written with 
\begin{align*} 
& \sum_{t=0}^T f_t(Y_{0:t}, S_{0:t}) \prod_i X^i_t - \prod_i X^i_t M_t , \stepcounter{equation}\tag{\theequation}\label{cmeao_form_mul} \\
& \sum_{t=0}^T h_t(X_{0:t}, Y_{0:t}, S_{0:t}) - \bigvee_i X^i_t M_t. \stepcounter{equation}\tag{\theequation}\label{cmeao_form_logical_or}
\end{align*} 
The last dual form is particularly useful 
for option where a weighting stopping time cannot easily be factored from the payoff. 
The last dual form also shows that our option class 
is fully general. 

\section{Single Pass Lookahead Search} \label{spls} 

\subsection{Algorithm} \label{spls_algorithm} 

To present SPLS, we simplify the notation 
and write the residual payoff of a particular exercise strategy 
at time $t$ with 
\begin{align*} 
  J(t, X_{t:T}, S_{t:T}) &= \sum_{s=t}^T e^{-r s} f_s(X_{0:s},S_{0:s}) \stepcounter{equation}\tag{\theequation}\label{spls_residual_payoff} \\
  &\text{s.t.}  \,\, g(X_{t:T}, S) = 0
\end{align*} 
where the exercise strategy $X_{t:T}$ is a vector of stopping times 
and control processes, 
and both the payoff and the constraint are 
conditional on the filtration $F_t$. 
In the payoff $f_t$ and the constraint $g$, 
the past decisions $X_{0:t-1}$ and 
the past stock path $S_{0:t}$ 
are given by the filtration $F_t$, 
the future decision $X_{t:T}$ 
is the proposed exercise strategy, 
and the future stock path $S_{t:T}$ 
is the sample path given as argument. 
We simply write $J(X_{t:T}, S_{t:T})$ when no confusion 
is possible on the anchoring time $t$. 
The key construction of SPLS 
is the lookahead operator $L(X_{t:T}, F_t)$ 
that finds the optimal strategy $X^*_{t:T}$ 
and returns the decision $X^*_t$ at the anchoring time. 
This operator can be written with 
\begin{align*} 
  L(X_{t:T}, F_t) = \p{\argmax_{X_{t:T}} \Ec{J(X_{t:T}, S_{t:T}}{F_t} }_t \stepcounter{equation}\tag{\theequation}\label{spls_lookahead_operator}
\end{align*} 
where $(X)_t=X_t$ is the selection operator for time $t$. 
Given a sample path $S$, 
SPLS extracts an optimal exercise strategy for the sample path 
by repetitive application of the lookahead operator. 
A first decision $X^*_0$ 
is extracted, then a second decision $X^*_1$ 
is extracted conditional on the previous decision $X^*_0$ 
and the observed path $S_{0:1}$, 
and so on. 
This repetitive scheme can be written with 
\begin{align*} 
X^*_0 &= L(X_{0:T}, \sigma(S_{0})), \stepcounter{equation}\tag{\theequation}\label{spls_scheme_0} \\
X^*_1 &= L(X_{1:T}, \sigma(X^*_0, S_{0:1})), \stepcounter{equation}\tag{\theequation}\label{spls_scheme_1} \\
X^*_2 &= L(X_{2:T}, \sigma(X^*_{0:1}, S_{0:2})), \stepcounter{equation}\tag{\theequation}\label{spls_scheme_2} \\
    &  \ldots , \\
X^*_T &= L(X_T, \sigma(X^*_{0:T-1}, S_{0:T})), \stepcounter{equation}\tag{\theequation}\label{spls_scheme_T}
\end{align*} 
where $\sigma(A)$ is the $\sigma$-algebra generated by $A$. 
If we denote by $L_t$ the $t$-th application 
$L(X_{t:T}, \sigma(X^*_{0:t-1}, S_{0:t}))$, 
the option value $V$ can be written with 
\begin{align*} 
V = \Ebb{J(L_{0:T},S_{0:T})}. \stepcounter{equation}\tag{\theequation}\label{spls_v_star}
\end{align*} 

To estimate $V$, 
we use several Monte Carlo projections. 
To present these projections, 
consider the lookahead operator $L(X_{t:T}, F_t)$ at time $t$, 
and let $s$ be anytime between $t$ and $T$. 
First, 
we project 
the decision $X_s$ 
onto the space of step-function of the stock path $S_{0:s}$. 
The step-function is constructed 
with a nearest-neighbor basis for the stock path, 
where the centroids 
are sampled randomly. 
This step-function can be written with 
\begin{align*} 
  \bar{X}_s(S_{0:s}) = \sum_{i=0}^{m_{s-t}} x^{s,i} \ib{S_{t:s} \sim \bar{S}^{s,i}_{t:s} }, \stepcounter{equation}\tag{\theequation}\label{spls_step_fct}
\end{align*} 
where 
$\bar{S}^{s,i}$ is a sample path used for the basis at time $s$, 
$m_{s-t}$ is the number of sample path used at time $s$, 
$\ib{\cdot}$ is the indicator function, 
$a \sim b$ is the event that $a$ belongs 
to the Voronoi cell of $b$, 
and 
$x^{s,i} \in \R$ are the basis weights. 
We write simply $\bar{X}_s$ when no confusion is possible. 
Second, 
we project the lookahead operator 
with a Monte Carlo average of $N$ sample. 
This projected operator can be written with 
\begin{align*} 
  \bar{L}(\bar{X}_{t:T}, F_t) = \p{\argmax_{x^{s,i}} \frac{1}{N} \sum_{n=0}^N J(\bar{X}_{t:T}, S^n_{t:T})}_t \stepcounter{equation}\tag{\theequation}\label{spls_lookahead_operator_bar}
\end{align*} 
where 
$S^n$ is a sample path, 
$\bar{X}_{0:t-1}$ and $S^i_{0:t}$ comes from the filtration $F_t$, 
and 
the maximum runs over the basis weights $x^{s,i}$ 
for time $s=t,t+1, \ldots ,T$ 
and $i=1, 2,  \ldots , m_{s-t}$. 
Note that the projected operator 
is a deterministic program over the nearest-neighbor basis weight 
$x^{s,i}$. 
This deterministic program 
can be solved by any optimization methods, 
and in particular by 
mixed-integer programming. 
The projected optimal exercise strategy can be extracted with 
\begin{align*} 
\bar{X}^*_0 &= \bar{L}(\bar{X}_{0:T}, \sigma(S^n_{0})), \stepcounter{equation}\tag{\theequation}\label{spls_bar_scheme_0} \\
\bar{X}^*_1 &= \bar{L}(\bar{X}_{1:T}, \sigma(\bar{X}^*_0, S^n_{0:1})), \stepcounter{equation}\tag{\theequation}\label{spls_bar_scheme_1} \\
\bar{X}^*_2 &= \bar{L}(\bar{X}_{2:T}, \sigma(\bar{X}^*_{0:1}, S^n_{0:2})), \stepcounter{equation}\tag{\theequation}\label{spls_bar_scheme_2} \\
    &  \ldots , \\
\bar{X}^*_T &= \bar{L}(\bar{X}_T, \sigma(\bar{X}^*_{0:T-1}, S^n_{0:T})). \stepcounter{equation}\tag{\theequation}\label{spls_bar_scheme_T}
\end{align*} 
Third, 
we project the option value 
with a Monte Carlo average of $\widetilde{N}$ sample. 
This projected option value can be written with 
\begin{align*} 
\bar{V} = \frac{1}{\widetilde{N}} \sum_{\widetilde{n}=1}^{\widetilde{N}} J(\bar{L}_{0:T},S^{\widetilde{n}}_{0:T}). \stepcounter{equation}\tag{\theequation}\label{spls_v_star_bar}
\end{align*} 

\subsection{Convergence} \label{spls_convergence} 

The number of sample $\widetilde{N}$ in the option value, 
the number of sample $N$ in the lookahead operator, 
and 
the nearest-neighbor basis size $m=(m_t)_{t=0}^T$ 
are hyperparameters that need to be tuned 
so as to maximize 
the projected option value $\bar{V}$. 
To understand the result of such tuning, 
we analyze the convergence 
of the projection scheme. 
First, we complete a uniform convergence analysis 
with the Vapnik-Chernovenkis dimension as main tool (\citeauthor{vapnik71} \cite*{vapnik71}). 
Then, we complete an energy distance analysis 
(\citeauthor{szekely04} \cite*{szekely04}, \citeauthor{baringhaus04} \cite*{baringhaus04}, \citeauthor{ramdas17} \cite*{ramdas17}) 
The uniform convergence analysis helps to understand 
the interaction between the hyperparameters, 
while the energy convergence analysis 
provides a practical score 
for any choice of hyperparameters. 
As the score demands little computational effort, 
hyperparameters tuning can be done quickly 
by finding the hyperparameters with the best score. 

The uniform convergence of the scheme can be studied 
from two angles. 
A first angle look at convergence of the measure implied by the scheme, 
while a second angle 
look at convergence 
of the scheme to the Bayes-value. 
Convergence to the Bayes-value 
refers to the convergence to the best possible 
option value accessible by the strategy class used in the scheme. 
Convergence in measure provides many insights for parameters tuning, 
but does not quantify the rate of convergence to the option value. 
The benefits and drawbacks of the other mode of convergence are reversed. 
The first mode of convergences shows that SPLS 
is a consistent estimator for the option value, 
in that the scheme converges in probability to the option value. 
The second mode of convergence shows 
that SPLS is universally Bayes-value consistent, 
in that, 
for any stochastic dynamic of the underlying, 
any payoffs, and any constraints, 
there exists a value of the hyperparameters 
that approximate arbitrarily well 
the best possible option value accessible by the projected strategy class. 
If we assume that the optimal exercise strategy 
can be approximated arbitrarily well by a projected strategy, 
this last result implies 
that SPLS converges uniformly to the option value. 

Completing an energy distance analysis has its roots 
in the stability theory of stochastic optimal control. 
Within this theory, 
a projection scheme is judged good 
if the scheme preserves 
the optimum of the control problem. 
Such a stability property 
can be guaranteed 
when the projected stochastic space 
is close to the true stochastic space 
in some variant of the Wasserstein distance 
(\citeauthor{heitsch06} \cite*{heitsch06}, \citeauthor{pichler12} \cite*{pichler12}). 
Such a convergence certificate 
has two important advantages. 
First, 
when tuning hyperparameters, 
the option price should be computed 
only for hyperparameters 
that can be certified. 
Second, such a certificate 
allows to use low dimensional projection 
while maintaining accuracy. 
When the certificate is fast to compute, 
these advantages accelerate 
both hyperparameters tuning 
and the projection scheme computation. 
Unfortunately, 
Wasserstein-based distances are not fast to compute, 
especially in high dimension \parencite{ramdas17}. 
To correct this shortcoming, 
we use the energy distance instead. 
We show that SPLS converges in measure 
when the energy distance is small, 
and we propose a practical 
hyperparameters tuning algorithm based on the energy distance. 

\subsubsection{Uniform Convergence in Measure} \label{spls_uniform_convergence_in_measure} 

Convergence in measure can be decomposed into three key drivers. 
First, the projected strategy space 
needs to be dense in the space of strategy. 
This driver is related to the nearest-neighbor basis size $m$ 
and the unknown size $\alpha$ 
of the smallest nearest-neighbor basis needed to replicate accurately 
the optimal strategy $X^*$. 
Second, 
once a dense projected strategy space is found, 
the projected lookahead operator needs 
to be accurate. 
This driver is related 
to the interaction between the number of sample in the lookahead operator $N$, 
and the nearest-neighbor basis size $m$. 
Third, 
once an accurate projected lookahead operator is found, 
the projected option value 
needs to be accurate. 
This driver is related 
to the number of sample $\widetilde{N}$ 
in the option value $\bar{V}$. 
With these three drivers, we can show 
that the projected option value $\bar{V}$ 
is a consistent estimator 
for the option value $V$. 

 \medskip \noindent \text{\bf Projected Strategy} \medskip 

Consider the lookahead operator $L(X_{t:T}, F_t)$ at time $t$, 
and let $s$ be anytime between $t$ and $T$. 
For the projected strategy convergence, 
let $\alpha_{s-t}$ be 
the smallest number of Voronoi cell needed 
to tesselate perfectly 
the set $\{(S_{0:s}, X^*_s)\}$. 
For example, 
for a vanilla American option 
let $B_i$ 
be any Voronoi cell of a perfect tesselation, 
then $X^*_s(B_i) \in \{\{0\},\{1\}\}$, 
and $X^*_s(B_i) \in \{\{0,1\}\}$ is not possible. 
In other words, 
a perfect tesselation 
separates the optimal decision 
into cell with unambiguous decision. 
Cells with unambiguous decision 
are crucial 
for the projected strategy space 
to contain the optimal strategy. 
To quantify 
this crucial requirement, 
we look at how big the nearest-neighbor basis size $m_{s-t}$ 
needs to be in order 
to approximate 
uniformly well 
any tesselation of size $\alpha_{s-t}$. 

To write our convergence rate, 
let $(\bar{B}_i)_{i=1}^{m_{s-t}}$ 
be the tesselation generated by the nearest-neighbor basis at time $s$. 
Also, 
for a given tesselation of size $\alpha_{s-t}$, 
let 
\begin{align*} 
  A \in 2^{\prod_{i=1}^{\alpha_{s-t}} B_i } \setminus \{\{B_{1}\},\{B_{2}\}, \ldots ,\{B_{\alpha_{s-t}}\}, \{\cup_{i=1}^{\alpha_{s-t}} B_i\}\},
\end{align*} 
be the power set 
of all permutation of the cells in the tesselation, 
excluding the permutation that contains a single cell of the tesselation, 
and excluding the union of the cells. 
In \hyperref[meth]{Methods}, we show that the uniform convergence rate of the projected strategy 
can then be written with 
\begin{align*} 
P\bigg( \sup_A \frac{1}{m_{s-t}} \sum_{i=1}^{m_{s-t}} \ib{\bar{B}_i \in A}  > \epsilon \bigg)
 \le
2 {m_{s-t}}^{3 d (s-t) \alpha^2_{s-t}} e^{- 2 \epsilon^2 m_{s-t}},
\stepcounter{equation}\tag{\theequation}\label{spls_projected_strategy_rate}
\end{align*} 
where $\epsilon>0$. 
This convergence rate 
says that the empirical probability 
that a projected Voronoi cell $\bar{B}_i$ 
overlaps several cells of any tesselation 
of size $\alpha_{s-t}$ 
decreases exponentially fast 
with $m_{s-t}$. 
The key ingredients 
for the proof 
are 
the shatter coefficient 
of the family of set $A$ (\citeauthor{vapnik71} \cite*{vapnik71}), 
and Hoeffending inequality. 
In regards to hyperparameters tuning, 
this convergence rate 
says 
that 
the projected strategy space is dense 
as soon as $m_{s-t}$ is higher than 
$3 d (s-t) \alpha^2_{s-t}$. 
In particular, a good 
projected strategy space is accessible 
well before $m_{s-t}$ is taken as infinity. 

 \medskip \noindent \text{\bf Projected Lookahead Operator} \medskip 

Assuming that the projected strategy space is dense, 
the projected lookahead operator 
will be close to the true lookahead opertor 
if 
the distribution of the projected residual payoff $J(\bar{X}_{t:T},S^n)$ is uniformly accurate. 
This distribution 
is defined by the joint distribution of the projected strategy 
and the projected stock path. 
Since the projected strategy is a function 
of several nearest-neighbor basis, 
and 
since the stock path takes value in a real plane, 
this joint distribution is characterized 
by set of the form 
$A \in \p{\prod_{s=t}^{T} \{\bar{B}^{s,i}\}_{i=1}^{m_{s-t}}, A'}$ 
where $\bar{B}^{s,i}$ is the Voronoi cell 
generated at time $s$ by the centroid $\bar{S}^{s,i}_{t:s}$, 
and $A'$ is an interval in $\R^{d(T-t)}$. 
To write the convergence rate, 
let 
$\nu(A) = P(S_{t:T} \in A)$ 
be the probability 
that a sample path $S_{t:T}$ 
is in both components of $A$, 
and let 
$\nu_N(A) = \frac{1}{N} \sum_{n=1}^N \ib{S^n_{t:T} \in A}$ 
be the empirical probability. 
A direct application 
of the Vapnik-Chernovenkis inequality (\citeauthor{vapnik71} \cite*{vapnik71}, \citeauthor{devroye96} \cite*{devroye96}) 
gives 
the following rate of convergence 
\begin{align*} 
P\bigg(
  \sup_A |
  \nu_N(A)
    -
    \nu(A)
  |> \epsilon
\bigg)
   \le
    8 N^{ 3 d^2 (T-t)^2 \sum_{s=t}^T m^2_{s-t} }
    e^{ - \epsilon^2 N / 32},
\stepcounter{equation}\tag{\theequation}\label{spls_projected_lookahead_operator_rate}
\end{align*} 
with $\epsilon>0$. 
See \hyperref[meth]{Methods} for more details. 
This rate of convergence 
says that 
the square of the nearest-neighbor basis size 
affects the residual payoff distribution. 
For hyperparameters tuning, 
this implies 
that the basis sizes 
should be chosen as small as possible, 
so that 
an accurate distribution is accessible 
with a moderate number of sample $N$. 
Indeed, 
recall that to compute 
the projected option value 
several samples of $J(\bar{L}_{0:T},S_{0:T})$ 
are needed: 
the larger $N$, 
the heavier the computation of each projected lookahead operator $\bar{L}_t$. 
A good way to satisfy this requirement, 
is to use a monotonically increasing basis size, 
for example $m=(1,\alpha_1,\alpha_2, \ldots ,\alpha_T)$, 
rather than a constant basis size such as $m=(1,\alpha_T,\alpha_T, \ldots ,\alpha_T)$. 
Of course, since 
the optimal nearest-neighbor basis size $\alpha$ 
is not known, 
hyperparameters tuning 
should be performed 
by enforcing a monotone constraint on the basis size. 

 \medskip \noindent \text{\bf Projected Option Value} \medskip 

Since the projected option value 
is simply a Monte Carlo expectation, 
the uniform convergence of the projected option value 
is given by the uniform law of large numbers 
of \textcite{vapnik81}. 
Since the goal of this section is to understand 
the hyperparameters impact, 
we make the simplifying assumption 
that the projected lookahead operator 
is a uniform Lipschitz function 
of the filtration. 
This assumption will be removed in the next section. 
Assume further that the residual payoff $J(X,S)$ 
is a Lipschitz function of the decision $X$. 
The convergence rate can then be written with 
\begin{align*} 
P\bigg( \sup_{\bar{L}} | \bar{V} - \Epb{ J(\bar{L}, S) } |  > \epsilon \bigg)
 \le
8 \p{\frac{C K}{\epsilon}}^{ (a+d) T^2} e^{- \epsilon^2 \widetilde{N} / (128 B^2)},
\stepcounter{equation}\tag{\theequation}\label{spls_projected_option_value_rate}
\end{align*} 
where 
$C$ is the Lipschitz constant of 
the projected lookahead operator, 
$B$ is an upper bound on the residual payoff, 
$K$ is the Lipschitz constant of the residual payoff, 
and 
$a$ is the dimension of a decision, 
so that $X \in \R^{a (T+1)}$. 
See \hyperref[meth]{Methods} for more details. 
For a particular option and a particular stochastic dynamic of the stock, 
we can assume that the Lipschitz constant $C$ 
is universal. 
The rate of convergence is then uniform 
with respect to the projected lookahead operator $\bar{L}$, 
and 
the hyperparameter $\widetilde{N}$ 
is independent, 
in term of convergence impact, 
of the number of sample in the lookahead operator $N$ and 
the nearest-neighbor basis size $m$. 
For hyperparameters tuning, 
this implies that 
once a value of $\widetilde{N}$ is found 
such that the projected option value converges, 
the same value of $\widetilde{N}$ 
can be used to tune the two other hyperparameters $N$ and $m$. 

With the three previous convergence rates, 
we can prove 
that 
the projected option value $\bar{V}$ 
converges in probability to the option value $V$ 
when the hyperparameters are taken as very large. 
This result can be written with 
\begin{align*} 
  \lim P(|\bar{V} - V|>\epsilon) = 0,
  \stepcounter{equation}\tag{\theequation}\label{spls_consistency}
\end{align*} 
where the limit 
is taken with $\widetilde{N}$, $N$ and $m$ 
going to infinity, 
with $m_{\max}^2 < N $, 
and 
$m_{\max}=\pn{m}_\infty$ 
the largest nearest-neighbor size. 
See \hyperref[meth]{Methods} for more details. 
In the proof, we make 
the assumptions 
that the convergence rates 
\eqref{spls_projected_strategy_rate} 
and 
\eqref{spls_projected_lookahead_operator_rate} 
control 
the denseness of the projected strategy space 
and 
the accuracy of the projected lookahead operator. 
The next section presents a stronger result 
that implies consistency 
without these assumptions. 

\subsubsection{Uniform Convergence in Bayes-Value} \label{spls_uniform_convergence_in_bayes-value} 

Convergence in Bayes-value 
is obtained 
by assuming that the residual payoff is Lipschitz in the strategy. 
The first step towards this result 
is to obtain the rate of convergence 
of the average residual payoff 
to the expected residual payoff. 
This rate of convergence can be written with 
\begin{align*} 
& P(\sup_{\bar{L}} | \frac{1}{\widetilde{N}} \sum_{\widetilde{n}=1}^{\widetilde{N}} J(\bar{L}_{0:T}, S^{\widetilde{n}}) - \Epb{J(\bar{L}_{0:T},S)} | > \epsilon)  \\
& \le  8
   \exp\p{
    8 T
    (c d T m_{\max}^3)^{T}
    \p{\frac{K X_{\max} S_{\max}}{\epsilon}}^{c (a+d) T^2 m_{\max}^3}
    e^{ - \epsilon^2 N / (512 B^2 K^2)}
    - \epsilon^2 \widetilde{N} / (128 B^2)
  }
,
\stepcounter{equation}\tag{\theequation}\label{spls_bayes_value_rate}
\end{align*} 
where the supremum is taken over the class 
of strategy implied by the projected lookahead operator, 
$B$ is an upper bound for the residual payoff, 
$K$ is the Lipschitz constant of the residual payoff, 
$c$ is a universal constant, 
$X_{\max}$ 
is an upper bound for the decision, 
and 
$S_{\max}$ 
is an upper bound for the stock. 
The key ingredients of the proof 
are the uniform law of large numbers, 
and a careful estimates of the covering number 
for the residual payoff. 
The Lipschitz assumption 
allows to express this covering number 
in terms of a covering number 
for the projected lookahead operator. 
The resulting covering number 
can then be estimated with another application 
of the uniform law of large numbers. 
Further, this rate of convergence can be used 
to prove consistency of SPLS. 
See \hyperref[meth]{Methods} for more details. 

With the previous rates of convergence, 
we can show the universal convergence 
of the projected option value 
to the Bayes-value. 
This result can be written with 
\begin{align*} 
  \lim \bar{V} = V,
    \stepcounter{equation}\tag{\theequation}\label{spls_bayes_value}
\end{align*} 
where 
the lookahead operator in the option value $V$ 
is restricted to 
the class of strategy implied by 
the projected lookahead operator, 
the limit is taken with $\widetilde{N}$, $N$ and $m$ 
going to infinity, 
and $m_{\max}^3 < o(N)$. 
By assuming that the class of projected strategy 
is dense in the space of strategy, 
this result implies 
that SPLS converges uniformly 
to the option value. 

\subsubsection{Energy Convergence} \label{spls_energy_convergence} 

We start by reviewing the definition of energy distance 
and by applying this distance to a Voronoi tesselation. 
Then, we present an energy-based hyperparameters tuning 
algorithm for SPLS. 

Let $X$ and $Y$ be two random variables. 
The energy distance $ED(X,Y)$ 
between $X$ and $Y$ is defined as 
\begin{align*} 
 ED(X,Y) = 2\Ep{\pn{X-Y}} - \Ep{\pn{X-X'}} - \Ep{\pn{Y-Y'}},
\end{align*} 
where $X'$ and $Y'$ have the same distribution than $X$ and $Y$, 
all variables are independent, 
and the norm can be defined freely. 
We use the $\ell_1$-norm. 
Given a sample $(X_i)_{i=1}^n$ 
and $(Y_j)_{j=1}^m$, the energy distance can be estimated with 
\begin{align*} 
  \widehat{ED}(\{X_i\},\{Y_i\}) = \frac{2}{nm} \sum_{i=1}^n \sum_{j=1}^m \pn{X_i - Y_j}
  - \frac{1}{n^2} \sum_{i=1}^n \sum_{j=1}^n \pn{X_i - X_j}
  - \frac{1}{m^2} \sum_{i=1}^m \sum_{j=1}^m \pn{Y_i - Y_j}
  .
\end{align*} 
The energy distance has the property 
that $X$ and $Y$ have the same distribution 
if and only if $ED(X,Y)=0$. 

For the projected lookahead operator, 
the convergence rate \eqref{spls_projected_lookahead_operator_rate} 
shows that 
the distribution that is crucial to guarantee the accuracy 
of the projected lookahead operator 
is the joint distribution 
of the strategy and the stock path. 
Since this joint distribution 
is the distribution of a stochastic process, 
the distribution can be characterized 
by its conditional distribution. 
As the projected lookahead operator 
is a Monte Carlo average on the stock path, 
any discrepancy in the conditional distribution 
originates from the projected strategy. 
In particular, 
if the number of sample path $N$ 
used in the projected lookahead operator 
is not adequate 
with respect to to the nearest-neighbor basis size $m$, 
the conditional distribution 
will concentrate 
in the wrong cell of 
the Voronoi tesselation induced 
by the projected strategy. 
One way to measure the accuracy of the projected lookahead operator 
is hence 
to verify that the conditional distribution 
concentrates in the appropriate Voronoi cells. 

To measure the accuracy of the conditional distribution concentration, 
consider the projected lookahead operator $L_t$ at time $t$, 
a sample path $S$, 
and any time $s$ between time $t$ and $T-1$. 
Denote the conditional cell by $\pi S_s=(i_s, i_{s+1})$. 
The conditional cell is 
the index of the Voronoi cell 
at time $s$ and $s+1$ for the stock path $S$. 
The index $i_s$ 
denotes one of the Voronoi cell $\{\bar{B}^{s,i}\}_{i=1}^{m_{s-t}}$, 
and similarly for $i_{s+1}$. 
For an entire sample $S^{(N)}_{t:T} = (S^n_{t:T})_{n=1}^N$, 
denote by $\pi S^{(N)}_{t:T}$, the resulting sample of conditional cells. 

For a fix nearest-neighbor size $m$, 
a large sample $N'$ 
of conditional cells 
can be considered 
as the reference conditional distribution. 
The adequacy of the number of sample paths $N$ 
in the projected lookahead operator 
can then be measured 
by the energy distance 
between 
the reference sample 
and 
the conditional cells sample induced by the projected lookahead operator. 
This energy distance can be written with 
\begin{align*} 
  \widehat{ED}(\bar{L}_t) = \widehat{ED}(\pi {S^\prime_{t:T}}^{(N^\prime)}, \pi S^{(N)}_{t:T}).
\end{align*} 
By extension, the total energy of the projected lookahead operator $\bar{L}$ 
can be defined as the expected average energy. 
This total energy can be written with 
\begin{align*} 
  \widehat{ED}(\bar{L}) = \frac{1}{\widetilde{N} T} \sum_{n=1}^{\widetilde{N}} \sum_{t=0}^T \widehat{ED}(\bar{L}_t).
  \stepcounter{equation}\tag{\theequation}\label{spls_total_energy}
\end{align*} 
For process with independent and identically distributed increments, the total energy 
can be estimated with the energy of the lookahead at time zero, giving 
\begin{align*} 
  \widehat{ED}(\bar{L}) = \frac{1}{\widetilde{N}} \sum_{n=1}^{\widetilde{N}} \widehat{ED}(\bar{L}_0).
\end{align*} 
Whenever applicable, the second form for the total energy 
is preferable as this form is faster to compute. 

By construction, 
a total energy of zero 
is a certificate that the projected lookahead operator is accurate. 
A total energy of zero is hence an equivalent 
condition to the convergence rate \eqref{spls_projected_lookahead_operator_rate} 
and can be used as a substitute 
to prove the consistency of SPLS \eqref{spls_consistency}. 

Using the total energy, hyperparameters tuning can be done as follows. 
First, fix a sample size $\widetilde{N}$ 
and an energy threshold $\delta$. 
The sample size $\widetilde{N}$ can be small 
as the energy certificate guarantees the convergence. 
The energy threshold $\delta$ is the maximum total energy 
that we are willing to tolerate 
in order 
to consider the projected lookahead operator accurate. 
In numerical experiments, 
we found that there is a clear shift 
in the total energy once 
the lookahead is no more accurate, 
and 
it is not necessary to develop a formal hypothesis test to define this threshold. 
Such a threshold can be found 
by computing the projected option value \eqref{spls_v_star_bar} on a small set of hyperparameters. 
The threshold is then given by the hyperparameters 
with the maximum projected option value. 
Second, the projected option value 
is computed conditionally on the energy threshold 
on a grid of the form 
\begin{align*} 
  (N=1,2, \ldots ,N_{\max}) \times (m=1,2, \ldots ,m_{\max}),
\end{align*} 
The notation $m=x$ means that the nearest-neighbor size 
is increasing with $m_0=1$ and $m_T=x$. 
At each grid point, 
the total energy \eqref{spls_total_energy} is computed 
and the projected option value 
is computed only if the total energy is below the threshold $\delta$. 
Third, 
the maximum projected option value 
found through the grid search 
is taken as a lower estimate for the option price. 

\section{Nearest Neighbor Martingale} \label{nnm} 

\subsection{Algorithm} \label{nnm_algorithm} 

To present NNM, assume that any martingale is defined 
as a random walk with increments that are conditionally zero-mean. 
Such a martingale $M$ can be written with 
\begin{align*} 
  M_0 &= 0, \\
  M_t &= M_{t-1} + I_t, \,\,\,\,\, t=1,2, \ldots ,T, \\
  \Ep{I_t|F_{t-1}} &= 0, \,\,\,\,\, t=1,2, \ldots ,T,
\end{align*} 
where $I_t$ is the random walk increment at time $t$. 
The martingality condition 
$\Ep{M_t|F_{t-1}} = M_{t-1}$ 
is enforced by the increment being conditionally zero-mean. 
Denote by $D(M, S)$ 
the dual payoff 
for a particular martingale $M$ 
and a sample path $S$. 
This dual payoff can be written with 
\begin{align*} 
  D(M, S) = \max_{x,y, g(x,y,S)=0} \sum_{t=0}^{T} e^{-r t} f_t(y_{0:t}, S_{0:t}) x_t - M_t x_t.
\end{align*} 
Given the probability measure $P$ of the sample path, 
the Rogers operator 
gives the martingale that minimizes 
the expectation of the dual payoff. 
This operator can be written with 
\begin{align*} 
  R(P) = \argmin_{M} \Ebb{D(M,S)}.
\end{align*} 
The option value is then given by the expectation 
of the dual payoff with the martingale 
given by the Rogers operator 
\begin{align*} 
  V = \Ebb{D(R(P),S)}.
\end{align*} 
See \hyperref[meth]{Methods} for the equivalence of the dual and primal problem. 

To estimate $V$, 
we use several Monte Carlo projections 
and a relaxation of the Rogers operator. 
First, 
we project 
the martingale increment $I_t$ 
onto the space 
of conditional nearest-neighbor basis for the stock paths. 
The centroid for such a basis 
is a pair $(\bar{S}^{t,i}_{0:t-1}, \bar{S}^{t,i,j}_{t})$ 
where the path $\bar{S}^{t,i}_{0:t-1}$ is sampled randomly, 
and the next stock price $\bar{S}^{t,i,j}_{t}$ 
is sampled conditional on the path $\bar{S}^{t,i}_{0:t-1}$. 
The resulting step-function can be written with 
\begin{align*} 
\bar{I}_t(S_{0:t}) = \sum_{i=1}^{p_t} \sum_{j=1}^{q_t} x^{t,i,j} \ib{S_{0:t-1} \sim \bar{S}^{t,i}_{0:t-1}, S_t \sim \bar{S}^{t,i,j}_{t}},
\end{align*} 
where 
$p_t$ is the number of stock path used, 
$q_t$ is the number of conditional stock price used, 
and 
$x^{t,i,j} \in \R$ are the basis weights. 
We write simply $\bar{I}_t$ when no confusion is possible. 
Second, we relax the Rogers operator 
to a Frank-Wolfe iteration. 
To describe the iteration, 
let $\bar{Z}$ be any projected martingale. 
The relaxed Rogers operator 
produces an improved projected martingale 
$\bar{M}$ 
by solving the following linear program 
\begin{align*} 
\bar{R}(\bar{Z}) &= \argmin_{x^{t,i,j}} \frac{1}{N} \sum_{n=1}^N D(\bar{M}, S^n) \stepcounter{equation}\tag{\theequation}\label{nnm_rogers_relaxed_obj} \\
\text{s.t.} &
  \sum_{n=1}^N \sum_{j=1}^{q_t}
    x^{t,i,j} \ib{S^n_{0:t-1} \sim \bar{S}^{t,i}_{0:t-1}, S^n_{t} \sim \bar{S}^{t,i,j}_{t}} = 0
      \,\,\,\,\, t=1,2, \ldots ,T, \,\, i=1,2, \ldots ,p_t \stepcounter{equation}\tag{\theequation}\label{nnm_rogers_relaxed_mc} \\
& \bar{M}_0 = 0 \stepcounter{equation}\tag{\theequation}\label{nnm_rogers_relaxed_zero} \\
& x^{t,i,j} - z^{t,i,j} \in [-\eta, \eta] \,\,\,\,\, t=1,2, \ldots ,T, \,\, i=1,2, \ldots ,p_t, \,\, j=1,2, \ldots ,q_t \stepcounter{equation}\tag{\theequation}\label{nnm_rogers_relaxed_lr}
\end{align*} 
where $\eta>0$ is the learning rate, 
$x^{t,i,j}$ are the basis weights for the increment of the martingale $\bar{M}$, 
and 
$z^{t,i,j}$ are the basis weights for the increment of the martingale $\bar{Z}$. 
The constraints \eqref{nnm_rogers_relaxed_mc} and \eqref{nnm_rogers_relaxed_zero} 
ensure that the improved martingale satisfies 
the martingality condition, 
while the constraint \eqref{nnm_rogers_relaxed_lr} 
ensures that the improved martingale 
is a small change to the given martingale. 
The relaxed operator is iterated $m$ 
times 
to obtain the projected optimal martingale $\bar{R}^m(\bar{Z})$. 
Third, we project the dual option value with another sample 
$\{\widetilde{S}^{\widetilde{n}}\}_{n=1}^{\widetilde{N}}$. 
This projected dual value can be written with 
\begin{align*} 
  \bar{V} = \frac{1}{\widetilde{N}} \sum_{\widetilde{n}=1}^{\widetilde{N}} D(\bar{R}^m(\bar{Z}), \widetilde{S}^{\widetilde{n}}).
  \stepcounter{equation}\tag{\theequation}\label{nnm_projected_dual_value}
\end{align*} 
As the next section will show, 
the computational effort required in the relaxed Rogers operator 
to obtain a good projected dual value 
is often prohibitive. 
For this reason, the projected dual value is further relaxed 
by using the same sample 
$\{S^n\}_{n=1}^{N}$ 
that is used for the relaxed operator. 
This relaxed dual value can be written with 
\begin{align*} 
  \bar{\bar{V}} = \frac{1}{N} \sum_{n=1}^{N} D(\bar{R}^m(\bar{Z}), S^n).
  \stepcounter{equation}\tag{\theequation}\label{nnm_relaxed_dual_value}
\end{align*} 
As the next section will show, 
estimating the dual option value with the relaxed dual value 
is not an heuristic. 
The relaxed dual value converges faster to the dual option value 
then the projected dual value. 
However, the optimal martingale associated to the relaxed dual value 
does not generalize well to out-of-sample test. 
In other words, 
the relaxed dual value 
approximates the distribution of the sample path 
well enough 
to obtain a good point estimate, 
but not well enough 
to obtain a good estimate of a stochastic process. 

\subsection{Convergence} \label{nnm_convergence} 

For hyperparameters tuning insights, 
we first look at the uniform convergence in measure of NNM, 
which shows that the relaxed dual value is a consistent estimator of the dual value. 
Second, we look at the uniform convergence in Bayes-value, 
which shows that NNM is universally Bayes-value consistent. 
Third, we complete an energy convergence analysis 
and provide an energy-based algorithm 
for hyperparameters tuning. 
\subsubsection{Uniform Convergence in Measure} \label{nnm_uniform_convergence_in_measure} 

The convergence in measure of NNM is driven by three major factors. 
First, the projected martingale space needs to dense, 
second, the relaxed Rogers operator needs to be accurate, 
and 
third, 
the projected dual value needs to be accurate. 
The relaxed operator accuracy 
depends on whether 
the operator is used 
for the projected dual value (a stochastic process estimate) 
or 
the relaxed dual value (a point estimate). 

 \medskip \noindent \text{\bf Projected Martingale} \medskip 

Denote by 
$(p,q)=((p_t, q_t), t=1, 2,  \ldots , T)$ 
the nearest-neighbor basis size. 
Without loss of generality, 
we can assume that the optimal martingale 
is a nearest-neighbor martingale of size $(\alpha, \beta)$. 
For example, 
for a vanilla American put 
in a binomial world (\citeauthor{crr79} \cite*{crr79}) 
$(\alpha, \beta) = ((2^{t-1},2), t=1, 2,  \ldots , T)$, 
and 
a good projected martingale can be obtained with 
$(p,q) = ( (\mathcal{O}(t), 2), t=1, 2,  \ldots , T)$, 
see \hyperref[meth]{Methods} for more details. 
The projected martingale space 
can be considered dense in the space of martingale 
as soon as the projected space 
approximates uniformly well any 
martingale of size $(\alpha, \beta)$. 

We consider separately the convergence of the conditioning part 
and the current part of the nearest-neighbor basis. 
For the conditioning part, let $(\bar{B}_i)_{i=1}^{p_t}$ 
be the tesselation generated by the conditioning part of the nearest-neighbor basis at time $t$. 
For any conditioning tesselation of size $\alpha_t$ 
let 
\begin{align*} 
  A \in 2^{\prod_{i=1}^{\alpha_t } B_i }
    \setminus
    \{\{B_{1}\},\{B_{2}\}, \ldots ,\{B_{\alpha_t}\}, \{\cup_{i=1}^{\alpha_t } B_i\}\},
\end{align*} 
be the power set 
of all permutation of the cell $B_i$ in the tesselation, 
excluding the permutation that contains a single cell of the tesselation, 
and excluding the union of the cells. 
In \hyperref[meth]{Methods}, we show that the uniform convergence rate of the conditioning part 
can be written with 
\begin{align*} 
P\bigg( \sup_A \frac{1}{p_t} \sum_{i=1}^{p_t } \ib{\bar{B}_i \in A}  > \epsilon \bigg)
 \le
2 p_t^{3 d t \alpha_t^2  } e^{- 2 \epsilon^2 p_t },
\stepcounter{equation}\tag{\theequation}\label{nnm_conditioning_part_rate}
\end{align*} 
where $\epsilon>0$. 

For the current part, 
let $(\bar{B}_i)_{i=1}^{q_t}$ 
be the tesselation generated by the current part of the nearest-neighbor basis at time $t$. 
For any current part tesselation of size $\beta_t$ 
let 
\begin{align*} 
  A \in 2^{\prod_{i=1}^{\beta_t } B_i } \setminus \{\{B_{1}\},\{B_{2}\}, \ldots ,\{B_{\beta_t}\}, \{\cup_{i=1}^{\beta_t } B_i\}\},
\end{align*} 
be the power set 
of all permutation of the cell $B_i$ in the tesselation, 
excluding the permutation that contains a single cell of the tesselation, 
and excluding the union of the cells. 
In \hyperref[meth]{Methods}, we show that the uniform convergence rate of the current part 
can be written with 
\begin{align*} 
P\bigg( \sup_A \frac{1}{q_t} \sum_{i=1}^{q_t } \ib{\bar{B}_i \in A}  > \epsilon \bigg)
 \le
2 q_t^{3 d \beta_t^2} e^{- 2 \epsilon^2 q_t },
\stepcounter{equation}\tag{\theequation}\label{nnm_next_part_rate}
\end{align*} 
where $\epsilon>0$. 

These convergence rates 
say that the empirical probability 
that a conditional Voronoi cell $\bar{B}_i$ 
overlaps several cells of any conditional tesselation 
of size $(\alpha_t,\beta_t)$ 
decreases exponentially fast 
with $p_t$ and $q_t$. 
In regard to hyperparameters tuning, 
this convergence rate 
says 
that 
the projected martingale space is dense 
as soon as $p_t$ is higher than 
$3 d t \alpha_t^2$, 
and as soon 
as $\beta_t$ is higher than 
$3 d \beta_t^2$. 

 \medskip \noindent \text{\bf Relaxed Rogers Operator} \medskip 

Assuming that the projected martingale space is dense, 
the convergence of NNM 
lies in the accuracy of the relaxed Rogers operator. 
When the relaxed Rogers operator is used 
to estimate the projected dual value, 
the goal is to find a projected martingale 
that generalizes well to any out-of-sample 
computation of the dual value. 
Within a probabilistic setting, 
this goal is equivalent 
to estimate accurately 
the distribution of pair of the form $(\bar{M}, M)$, 
where $\bar{M}$ is a projected martingale, 
and $M$ is the optimal martingale. 
When the relaxed Rogers operator is used 
to estimate the relaxed dual value, 
the goal is to estimate the dual option value, 
and this goal is equivalent of estimating 
the distribution of pair of the form $(\bar{M}, D)$ 
where 
$\bar{M}$ is a projected martingale, 
and $D$ is the dual payoff value. 
The first usage is a stochastic process estimate, 
while the second usage is a point estimate. 

To write the convergence rates, 
let $B = \prod_{t=0}^T \{ \bar{B}^t_i \}_{i=0}^{p_t q_t}$ 
be the tesselation generated by the projected martingale. 
The optimal martingale $M$ is a vector in $\R^T$, 
and the dual payoff $D$ is a real scalar. 

When the relaxed Rogers operator is used for a stochastic process estimate, 
denote by $A$ set of the form $B \times [a,b]$, 
with $[a,b]$ an interval in $\R^T$, 
and let $\nu$ be the probability that a sample stock path and 
the optimal martingale are in $A$, $\nu(A) = P((S,M) \in A)$. 
The corresponding empirical probability 
can be written with $\nu_N(A) = \frac{1}{N} \sum_{i=1}^N \ib{S \in B, M \in [a,b]}$. 
A direct application 
of the Vapnik-Chernovenkis inequality (\citeauthor{vapnik71} \cite*{vapnik71}, \citeauthor{devroye96} \cite*{devroye96}) 
gives 
the following rate of convergence 
\begin{align*} 
P\bigg(
  \sup_A |
  \nu_N(A)
    -
    \nu(A)
  |> \epsilon
\bigg)
   \le
    8 N^{ 3 d T \sum_{t=1}^T t p_t^2 q_t^2 }
    e^{ - \epsilon^2 N / 32},
\stepcounter{equation}\tag{\theequation}\label{nnm_relaxed_rogers_operator_rate_1}
\end{align*} 
with $\epsilon>0$. 
See \hyperref[meth]{Methods} for more details. 

When the relaxed Rogers operator is used for a point estimate, 
denote by $A$ set of the form $B \times [a,b]$, 
with $[a,b]$ an interval in $\R$, 
and let $\nu$ be the probability that a sample stock path and 
the dual payoff are in $A$, $\nu(A) = P((S,D) \in A)$. 
The corresponding empirical probability 
can be written with $\nu_n(A) = \frac{1}{N} \sum_{i=1}^N \ib{S \in B, D \in [a,b]}$. 
A direct application 
of the Vapnik-Chernovenkis inequality (\citeauthor{vapnik71} \cite*{vapnik71}, \citeauthor{devroye96} \cite*{devroye96}) 
gives 
the following rate of convergence 
\begin{align*} 
P\bigg(
  \sup_A |
  \nu_n(A)
    -
    \nu(A)
  |> \epsilon
\bigg)
   \le
    8 N^{ 3 d \sum_{t=1}^T t p_t^2 q_t^2 }
    e^{ - \epsilon^2 N / 32},
\stepcounter{equation}\tag{\theequation}\label{nnm_relaxed_rogers_operator_rate_2}
\end{align*} 
with $\epsilon>0$. 
See \hyperref[meth]{Methods} for more details. 

The first observation that comes from these convergence rates 
is that to guarantee the accuracy of a stochastic process estimate 
the sample size $N$ needs to be $T$ times bigger than 
for a point estimate. 
As each iteration of the relaxed Rogers operator 
demands to compute the dual payoff for each sample stock path, 
and demands to solve a large linear program, 
this factor often makes a point estimate 
much faster to obtain. 
Second, these convergence rates 
show that the product of the nearest-neighbor basis size 
slowdown the convergence of the relaxed operator. 
The basis size should hence be chosen as small as possible. 
In particular, 
the conditioning size $p_t$ 
and 
the current size $q_t$ 
can be taken as monotonically increasing. 

 \medskip \noindent \text{\bf Projected Dual Value} \medskip 

Once the relaxed Rogers operator is accurate, 
NNM accuracy lies 
in an accurate estimation 
of the projected dual value. 
Since the goal of this section is to understand 
the hyperparameters impact, 
we use a probabilistic perspective 
where the relaxed Rogers operator 
is viewed as a martingale, 
and we look at how well 
the dual value associated to any martingale can be estimated. 
With this assumption, 
analyzing 
the convergence of the projected dual value 
\eqref{nnm_projected_dual_value} 
or the relaxed dual value 
\eqref{nnm_relaxed_dual_value} 
is equivalent. 
We use the notation of the projected dual value. 

Let $\bar{M}$ be a projected martingale, 
the convergence rate of the projected dual value 
to the dual value follows 
by assuming that the dual payoff is a Lipschitz function of the martingale, 
and by using the uniform law of large numbers. 
This rate can be written with 
\begin{align*} 
  & P(\sup_{\bar{M}} |\frac{1}{\widetilde{N}} \sum_{\widetilde{n}=1}^{\widetilde{N}} D(\bar{M}, S^{\widetilde{n}})  - \Epb{D(\bar{M},S)}| > \epsilon) \\
  & \le
  8
  (c d T p_{\max}^3 q_{\max}^3)^T \p{\frac{U K}{\epsilon}}^{c d T^2 p_{\max}^3 q_{\max}^3 }
  e^{- \epsilon \widetilde{N} / (128 B^2)},
  \stepcounter{equation}\tag{\theequation}\label{nnm_relaxed_dual_value_rate}
\end{align*} 
where 
$(p_{\max}, q_{\max})=(\pn{p}_\infty, \pn{q}_\infty)$ is the largest nearest-neighbor size, 
$U$ is an upper bound on the projected martingale, 
$K$ is the Lipschitz constant of the dual payoff, 
$B$ is an upper bound on the dual payoff, 
and 
$c$ is a universal constant. 
See \hyperref[meth]{Methods} for more details. 
For hyperparameters tuning, 
this convergence rate 
says that the nearest-neighbor basis size $(p,q)$ 
has a direct impact on the convergence of the projected dual value. 
An appropriate sample size $\widetilde{N}$ should hence be chosen 
by considering 
the biggest nearest-neighbor basis 
that will be used in hyperparameters tuning. 
Once such a sample size $\widetilde{N}$ is found, 
the projected dual value will be accurate 
for all the nearest-neighbor basis size considered. 

With the previous convergence rates, 
we can prove 
that the projected dual value $\bar{V}$ 
and the relaxed dual value $\bar{\bar{V}}$ 
converge in probability to the option value $V$ 
when the hyperparameters are taken as very large. 
These results can be written with 
\begin{align*} 
  \lim P(|\bar{V} - V|>\epsilon) = 0, \stepcounter{equation}\tag{\theequation}\label{nnm_consistency_1}\\
  \lim P(|\bar{\bar{V}} - V|>\epsilon) = 0, \stepcounter{equation}\tag{\theequation}\label{nnm_consistency_2}
\end{align*} 
where the limit 
is taken with $\widetilde{N}$, $N$ and $(p,q)$ 
going to infinity. 
The increasing rate for 
the nearest-neighbor basis size 
is 
$p_{\max}^3 q_{\max}^3 < o(N)$ 
for the projected dual value, 
and 
$p_{\max}^2 q_{\max}^2 < o(N)$ 
for the relaxed dual value. 
This difference in increasing rate 
further shows 
that 
the computational effort 
behind 
the projected dual value 
and 
the relaxed dual value 
are substantially different. 
See \hyperref[meth]{Methods} for more details. 

\subsubsection{Uniform Convergence in Bayes-Value} \label{nnm_uniform_convergence_in_bayes-value} 

For the projected dual value, 
convergence in Bayes-value 
is obtained by assuming that the dual payoff is Lipschitz in the martingale, 
and by considering a more general relaxed Rogers operator. 
To define this operator, 
denote by $S^{(N)}=(S^n)_{n=1}^N$ a random sample of size $N$ of the stock path. 
The general relaxed Rogers operator $\bar{R}(S^{(N)})$ makes no assumption 
on the optimization method used and can be written with 
\begin{align*} 
\bar{R}(S^{(N)}) = \argmin_{\bar{M}} \frac{1}{N} \sum_{n=1}^N D(\bar{M}, S^n),
\end{align*} 
where the minimization is subject to the same constraint than in \eqref{nnm_rogers_relaxed_obj}. 
The rate of convergence to the Bayes-value can then be written with 
\begin{align*} 
& P(\sup_{\bar{R}(S^{(N)})} | \frac{1}{\widetilde{N}} \sum_{n=1}^{\widetilde{N}} D(\bar{R}(S^{(N)}), S^n) - \Epb{D(\bar{R}(S^{(N)}), S)} | > \epsilon)  \\
& \le  8
    (c d T p_{\max}^3 q_{\max}^3 N^3)^T \p{\frac{U K}{\epsilon}}^{c d T^2 p_{\max}^3 q_{\max}^3 N^3}
    e^{- \epsilon^2 \widetilde{N} / (128 B^2)}
  \stepcounter{equation}\tag{\theequation}\label{nnm_projected_dual_value_bayes_rate}
,
\end{align*} 
where the supremum is taken over the class of martingale 
implied by the general relaxed Rogers operator, 
$U$ is an upper bound on the projected martingale, 
$K$ is the Lipschitz constant of the dual payoff, 
$B$ is an upper bound on the dual payoff, 
and 
$c$ is a universal constant. 

For the relaxed dual value, there is no out-of-sample test 
and the convergence in Bayes-value can be analyzed 
directly with a probabilistic view of the relaxed Rogers operator. 
The convergence rate \eqref{nnm_relaxed_dual_value_rate} 
can hence be taken as the Bayes-value rate 
of the relaxed dual value. 

With the previous rates of convergence, 
we can show the universal convergence 
of the projected dual value 
and the relaxed dual value 
to the Bayes-value. 
These results can be written with 
\begin{align*} 
  \lim \bar{V} = V, \stepcounter{equation}\tag{\theequation}\label{nnm_bayes_value_1} \\
  \lim \bar{\bar{V}} = V, \stepcounter{equation}\tag{\theequation}\label{nnm_bayes_value_2}
\end{align*} 
with 
the Rogers operator 
restricted to 
the class of martingale implied by 
the relaxed Rogers operator. 
For the projected dual value $\bar{V}$, 
the limit is taken with 
$\widetilde{N}$, $N$, and $(p,q)$ 
going to infinity, 
with an increasing rate of $p_{\max}^3 q_{\max}^3 N^3 < o(\widetilde{N})$. 
For the relaxed dual value $\bar{\bar{V}}$, 
the limit is taken with 
$N$ and $(p,q)$ 
going to infinity, 
with an increasing rate of $p_{\max}^3 q_{\max}^3 < o(N)$. 
The difference in increasing rate 
between the projected dual value 
and the relaxed dual value 
supports 
the difference in computational effort 
between the two methods. 
By assuming that the class of projected martingale 
is dense in the space of strategy, 
these Bayes-value consistency results imply 
that NNM converges uniformly 
to the dual option value. 

\subsubsection{Energy Convergence} \label{nnm_energy_convergence} 

As the projected dual value needs a stochastic process estimate, 
an energy score for this method is difficult to design. 
Indeed, the measure that needs to be estimated 
is given in \eqref{nnm_relaxed_rogers_operator_rate_1} 
and is for set of the form $(\bar{M}, M)$, 
where $\bar{M}$ is a projected martingale, 
and $M$ is the optimal martingale. 
The projected martingale is 
a step-function on a nearest-neighbor basis of the stock, 
and the quality of its distribution can be measured 
by looking at the energy of the tesselation. 
For the optimal martingale however 
there is no simple basis of comparison. 
Especially so, that the optimal martingale is unknown. 
This shortcoming does not apply 
to the relaxed dual value 
as from \eqref{nnm_relaxed_rogers_operator_rate_2} 
the distribution that needs to be estimated is for set of the form 
$(\bar{M})$. 

To define an energy score for the relaxed dual value, 
we adapt the energy score of SPLS 
(Section \ref{spls_energy_convergence}). 
To this end, 
fix a time $t$, 
and 
consider the conditional tesselation $\cup_{i=1}^{p_t} \cup_{j=1}^{q_t} B_{i,j}$ 
induced by the projected martingale. 
A Voronoi cell in this tesselation is of the form 
$B_{i,j} = U_i \times W_j$ 
where $U_i$ is the conditioning part, 
and $W_j$ is the current part. 
As the tesselation already has a conditional nature, 
the conditional cell at time $t$ for a stock path $S$ 
that falls in the cell $B_{i,j}$ 
can be defined with $\pi (S_{0:t-1},S_t) = (i, j)$, 
where $i$ is the index of the conditioning part cell, 
and $j$ is the index of the current part cell. 
For a sample of stock paths 
$S^{(N)}=(S^n)_{n=1}^N$, 
denote by $\pi S^{(N)}$ 
the corresponding sample of conditional cells. 
The energy of the relaxed dual value can then be written with 
\begin{align*} 
  \widehat{ED}(\bar{R}(\bar{Z})) = \widehat{ED}(\pi S^{(N)}, \pi {S^\prime}^{(N^\prime)}),
\end{align*} 
where ${S^\prime}^{(N^\prime)}$ is another independent sample with $N'$ 
much bigger then $N$. 
As the energy distance detects any discrepancy 
in distribution, 
an energy of zero 
is an equivalent condition 
to \eqref{nnm_relaxed_rogers_operator_rate_2} 
and can be used as substitute 
to prove the consistency 
of the relaxed dual value \eqref{nnm_consistency_2}. 

In numerical experiments, 
we found that 
when the nearest-neighbor size $(p,q$) 
is too high compared to the sample size $N$, 
the relaxed dual value does not converge. 
The Frank-Wolfe iteration in the relaxed Rogers operator is able to 
continually decrease the relaxed dual value. 
In contrast, 
when the nearest-neighbor size is appropriate 
for the sample size, 
the Frank-Wolfe iteration converges. 
Defining an appropriate energy threshold 
can hence be done by detecting 
where the Frank-Wolfe iteration diverges. 
A simple way to detect this divergence 
is to compare the Frank-Wolfe iteration 
to the lower bound found by SPLS. 
Alternatively, 
instead of defining an energy threshold, 
the SPLS lower bound can be used 
to define a barrier 
past which the Frank-Wolfe iteration is considered divergent. 
For example, with an SPLS value of $v$, 
the Frank-Wolfe iteration is considered divergent 
whenever the relaxed dual value falls below $v$. 
As SPLS is a lower estimate of the option price, 
the lower bound barrier may be setted slight higher 
then the actual SPLS value. 
For example, 
the lower bound barrier can be setted to $\alpha v$, 
with $\alpha>1$. 
We call the factor $\alpha$ 
the lower bound repulsion factor. 

Using the filtration energy, hyperparameters tuning can be done as follows. 
First, fix a sample size $N$ 
and an energy threshold $\delta$. 
The sample size $N$ can be small 
as the energy certificate guarantees the convergence. 
The energy threshold $\delta$ is the maximum energy 
that we are willing to tolerate 
to consider the relaxed dual value accurate. 
This threshold can be selected 
by detecting where the relaxed dual value \eqref{nnm_relaxed_dual_value} diverges. 
Second, the filtration energy is computed 
on a grid of the form 
\begin{align*} 
  (p=1,2, \ldots ,p_{\max}) \times (q=1,2, \ldots ,q_{\max}),
\end{align*} 
The notation $p=x$ means that the conditioning size 
is increasing with $p_0=1$ and $p_T=x$, 
and similarly for the notation $q=x$. 
Third, 
the minimum relaxed dual value 
found through the grid search 
is taken as an upper estimate for the option price. 

\section{An American Option} \label{aa} 

This section and the following 
applies SPLS and NNM 
to options of increasing complexity. 
For each option, 
we present 
the price of the option under different pricing parameters. 
The first two examples can be valued 
by exact pricing algorithm, 
and we compare our algorithms 
to option prices that are available in the literature. 
See \hyperref[meth]{Methods} for a description 
of our implementation of SPLS and NNM. 

Consider a single exercise American put option 
on a stock $S$ with a strike price of $K$. 
The option can be exercised 50 times per year, 
up to the maturity $T$. 
The stochastic program for this option 
can be written with 
\begin{maxi!}[0] 
{X}
{ \Ebb{ \sum_{t=0}^T e^{-r t} \pp{K - S_t} X_t }  \tag{\theparentequation}\label{aao_aao}}
{\renewcommand{\theequation}{\theparentequation.\arabic{equation}} \label{}}
{  }
\addConstraint{ \sum_{t=0}^T X_t }{ \le  1 \quad \label{aao_aao_cst}}{}
\end{maxi!} 
where $X$ is the exercise decision, 
and $r$ is the risk-free rate. 
The risk-neutral dynamic for the stock price $S$ 
is a geometric Brownian motion 
and can be described by the following stochastic differential equation 
\begin{align*} 
  dS = r S dt + \sigma S dW, \stepcounter{equation}\tag{\theequation}\label{aao_sde}
\end{align*} 
where $\sigma$ is the stock volatility, 
and $W$ is a standard Brownian motion. 
Our benchmark for this example 
are the finite difference prices 
computed in \textcite{longstaff01}. 
These prices are computed 
by an implicit finite difference scheme, 
with 40,000 time steps per year 
and 1,000 steps for the stock price. 

Table \ref{aao_option_price} presents the option price 
under different pricing parameters. 
Both SPLS and NNM uses a $1,000$ simulations. 
A first observation from this table 
is that the price estimates are in average within $5\%$ 
of the option price. 
This range of precision is adequate 
for an estimation algorithm, 
especially if this precision is maintained 
with much more complex options. 
A second observation from this table 
is that there is some volatility in the estimates, 
and that there is no simple rule on the pricing parameters 
that allow to predict this volatility. 
This situation is made on purpose. 
Table \ref{aao_option_price} uses a very coarse hyperparameters grid, 
and uses a low number of simulation to keep the estimates fast. 
The table hence shows the performance 
that comes out-of-box with the algorithms. 
Later, we will see that with a finer hyperparameters tuning 
and more computing time, 
the volatility in the estimates disappears. 
A third observation from this table 
is that the energy level of the estimates is low. 
This low energy level indicates 
that the projection scheme behind the algorithms 
agrees in distribution with the distribution of the pricing problem. 
Even if a low number of simulations was used, 
this agreement in distribution gives 
confidence 
that the estimates are the best possible 
within the chosen projected space dimension. 

\sisetup{zero-decimal-to-integer} 

\begin{table}[ht] 

\singlespacing 
\tabcolsep=0.11cm 
\centering 

\begin{tabular}{*{6}{c}} 

	\toprule 

	$S_{0}$ & $\sigma$ & $T$ & Finite Difference & SPLS & NNM \\ 

	\midrule 

	36 & 
	0.20 & 
	1 & 
	4.478 & 
	4.251 [0.08] (-0.05) & 
	4.337 [0.07] (-0.03) 
	\\ 

	 & 
	 & 
	2 & 
	4.840 & 
	4.591 [0.00] (-0.05) & 
	4.877 [0.05] (0.01) 
	\\ 

	 & 
	0.40 & 
	1 & 
	7.101 & 
	6.901 [0.00] (-0.03) & 
	7.090 [0.11] (-0.00) 
	\\ 

	 & 
	 & 
	2 & 
	8.508 & 
	8.462 [0.00] (-0.01) & 
	8.814 [0.06] (0.04) 
	\\ 

	\hline 

	38 & 
	0.20 & 
	1 & 
	3.250 & 
	3.232 [0.00] (-0.01) & 
	3.329 [0.05] (0.02) 
	\\ 

	 & 
	 & 
	2 & 
	3.745 & 
	3.608 [0.00] (-0.04) & 
	3.861 [0.05] (0.03) 
	\\ 

	 & 
	0.40 & 
	1 & 
	6.148 & 
	5.989 [0.09] (-0.03) & 
	6.164 [0.15] (0.00) 
	\\ 

	 & 
	 & 
	2 & 
	7.670 & 
	7.573 [0.00] (-0.01) & 
	7.729 [0.05] (0.01) 
	\\ 

	\hline 

	40 & 
	0.20 & 
	1 & 
	2.314 & 
	2.325 [0.00] (0.00) & 
	2.393 [0.06] (0.03) 
	\\ 

	 & 
	 & 
	2 & 
	2.885 & 
	2.753 [0.00] (-0.05) & 
	3.085 [0.05] (0.07) 
	\\ 

	 & 
	0.40 & 
	1 & 
	5.312 & 
	5.308 [0.09] (-0.00) & 
	5.419 [0.13] (0.02) 
	\\ 

	 & 
	 & 
	2 & 
	6.920 & 
	6.946 [0.00] (0.00) & 
	7.392 [0.05] (0.07) 
	\\ 

	\hline 

	42 & 
	0.20 & 
	1 & 
	1.617 & 
	1.655 [0.08] (0.02) & 
	1.705 [0.07] (0.05) 
	\\ 

	 & 
	 & 
	2 & 
	2.212 & 
	2.055 [0.00] (-0.07) & 
	2.131 [0.05] (-0.04) 
	\\ 

	 & 
	0.40 & 
	1 & 
	4.582 & 
	4.493 [0.08] (-0.02) & 
	4.633 [0.07] (0.01) 
	\\ 

	 & 
	 & 
	2 & 
	6.248 & 
	5.943 [0.00] (-0.05) & 
	6.066 [0.08] (-0.03) 
	\\ 

	\hline 

	44 & 
	0.20 & 
	1 & 
	1.110 & 
	1.058 [0.09] (-0.05) & 
	1.113 [0.07] (0.00) 
	\\ 

	 & 
	 & 
	2 & 
	1.690 & 
	1.612 [0.00] (-0.05) & 
	1.919 [0.05] (0.14) 
	\\ 

	 & 
	0.40 & 
	1 & 
	3.948 & 
	3.894 [0.08] (-0.01) & 
	4.016 [0.08] (0.02) 
	\\ 

	 & 
	 & 
	2 & 
	5.647 & 
	5.422 [0.00] (-0.04) & 
	5.606 [0.07] (-0.01) 
	\\ 

	\bottomrule 

\end{tabular} 

\captionsetup{font={footnotesize}} 
\caption{ 
	\label{aao_option_price} 
	Comparison of SPLS and NNM 
	with a finite difference method 
	for a single exercise American put option. 
	The option can be exercised 50 times per year, 
	the strike price is 40, 
	and the risk-free rate is 0.06. 
	The initial stock price $S_{0}$, 
	the volatility $\sigma$, 
	and the maturity $T$ are as indicated. 
	The finite difference numbers 
	are from \textcite{longstaff01}. 
	The filtration energy is reported in bracket, 
	and the relative difference to the finite difference price in parenthesis. 
	SPLS uses 
	$\widetilde{N}=1,000$ simulations, 
	an energy tolerance of $0.1$, 
	an energy validation sample of size $N'=1,000$, 
	and 
	an hyperparameters grid size of 25 with $(N_{\max}, m_{\max})=(200,200)$. 
	NNM uses 
	$N=1,000$ simulations, 
	an energy tolerance of $0.4$, 
	an energy validation sample of size $N'=10,000$, 
	a lower bound barrier of $1.02$, 
	and 
	an hyperparameters grid size of 30 with $(p_{\max},q_{\max})=(10,500)$. 
} 

\end{table} 

\subsection{SPLS Analysis} \label{aao_spls_analysis} 

\begin{figure}[ht] 

  \includegraphics[width=\textwidth, height=3.5in]{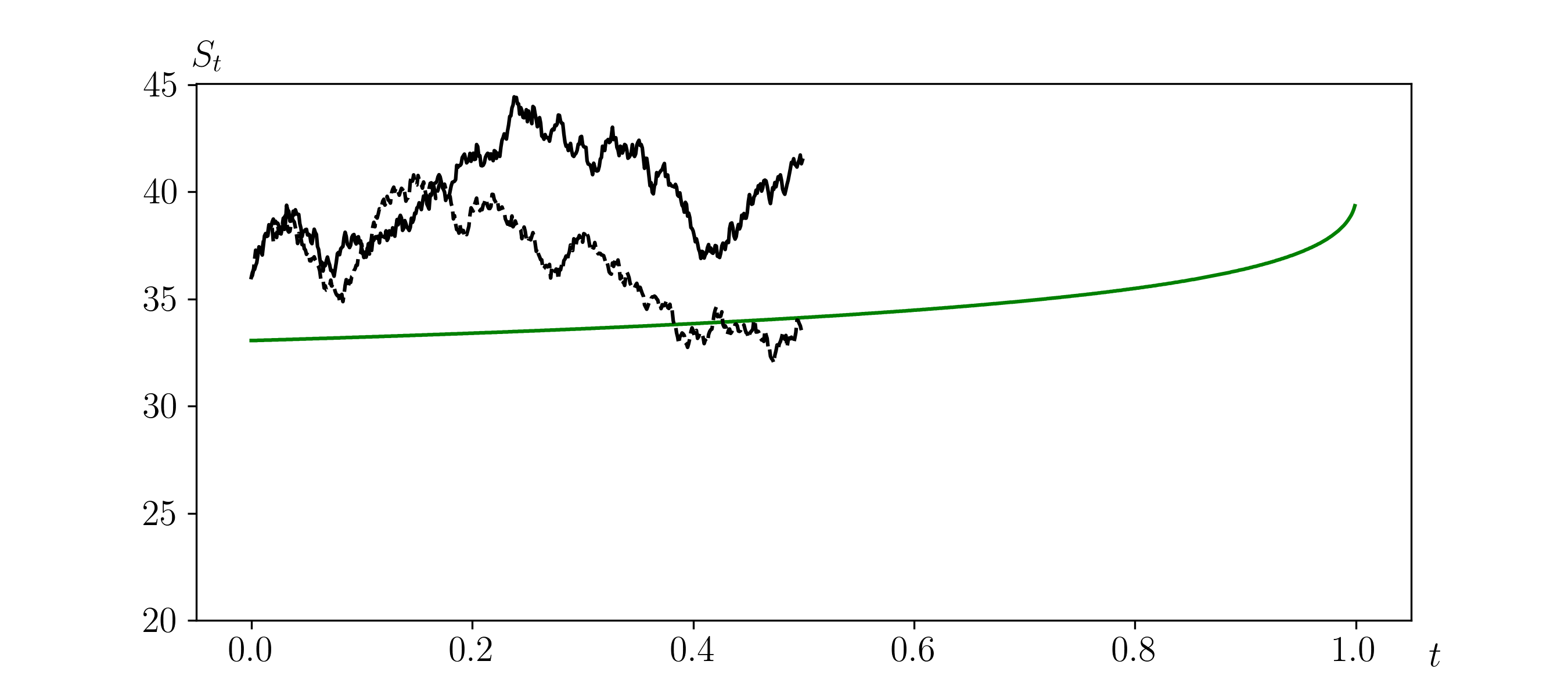} 

  \includegraphics[width=\textwidth, height=3.5in]{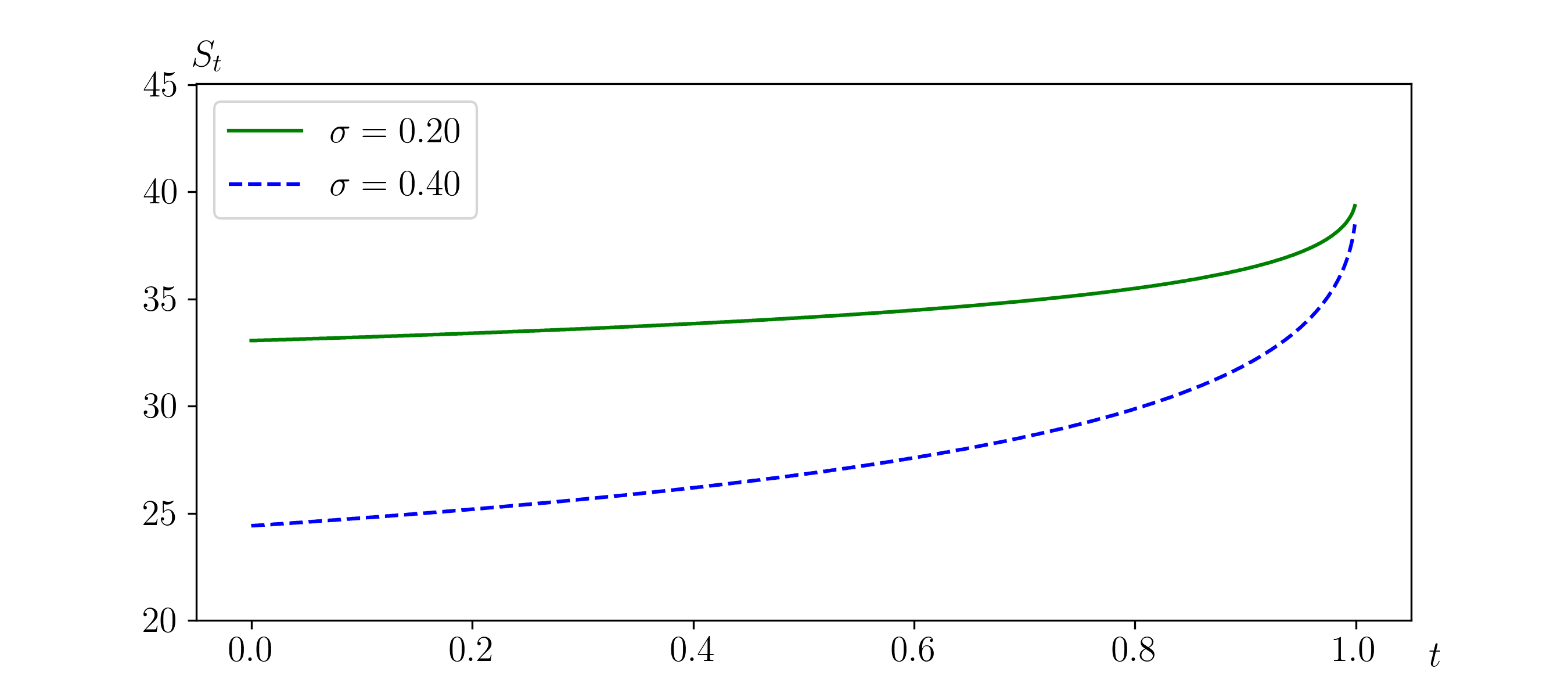} 

  \captionsetup{font={footnotesize}} 
  \caption{ 
    \label{aao_exercise_boundary_example} 
    Exercise boundary 
    for an American put 
    under different volatility. 
    The exercise boundary 
    is a function of the time to maturity, 
    and the last observed value of the stock. 
    The x-axis is the time period $t$, 
    and the y-axis is the stock price $S_t$ 
    at time $t$. 
    In the top figure, 
    two sample stock paths are shown: 
    one that crosses the boundary, 
    and for which the optimal strategy is to exercise the option 
    at the crossing moment; 
    a second one that never hits 
    the boundary, 
    and for which the optimal strategy is to never exercise the option. 
    The option can be exercised 50 times per year, 
    the strike price is 40, 
    the risk-free rate is 0.06, 
    and the maturity is one year. 
    The volatility $\sigma$ is as indicated. 
    The exercise boundary 
    is found 
    with an implicit finite difference scheme 
    with 1,000 time steps per year 
    and 10,000 steps for the stock price. 
  } 

\end{figure}

For an American put, 
an exercise strategy is optimal 
if the strategy captures the exercise boundary. 
Indeed, 
the optimal exercise strategy 
depends only 
on which side of the exercise boundary 
the last observed value of the stock fall 
(\citeauthor{carr92} \cite*{carr92}). 
See Figure \ref{aao_exercise_boundary_example} 
for an example 
of exercise boundary 
under different volatility. 
Looking at how SPLS approximates 
the exercise boundary 
should hence 
provide many insights on the algorithm, 
and will be the starting point of our analysis. 
Then, 
we look 
at 
the convergence of SPLS 
with respect to the projected strategy quality 
and the filtration energy. 
Finally, 
we look at the volatility of the SPLS estimates. 

\subsubsection{Exercise Boundary} \label{aao_exercise_boundary} 

Consider the projected lookahead operator 
at time $t$ 
\eqref{spls_lookahead_operator_bar}. 
This operator $\bar{L}(\bar{X}_{t:T},F_t)$ 
is a function of the strategy class and the filtration, 
and is hence a random variable. 
For this example, the operator is a random variable in $\pe{0,1}$. 
One way to understand the operator 
is to look at the distribution of the operator 
and to contrast this distribution with the exercise boundary. 
Indeed, 
even if the operator uses the filtration, 
the operator will be accurate 
only if it maps accurately 
the filtration $F_t$ 
to the last observed stock value $S_t$, 
so that the distribution of the operator captures the exercise boundary. 

\begin{figure}[ht] 

  \includegraphics[width=\textwidth]{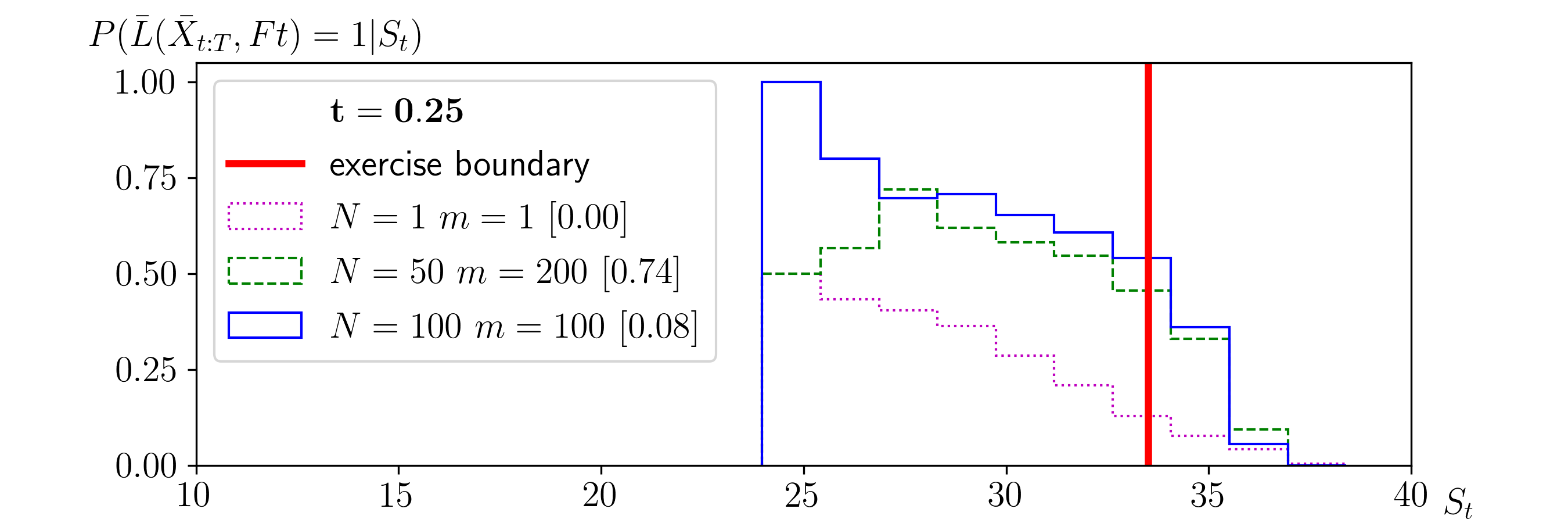} 

  \includegraphics[width=\textwidth]{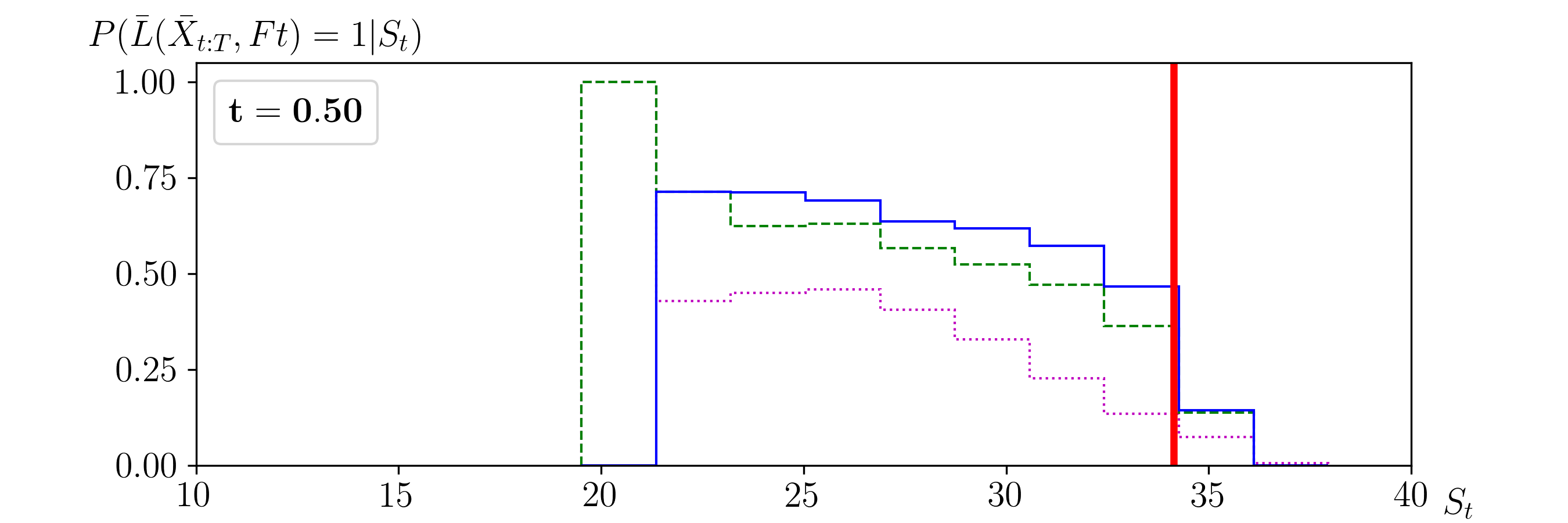} 

  \includegraphics[width=\textwidth]{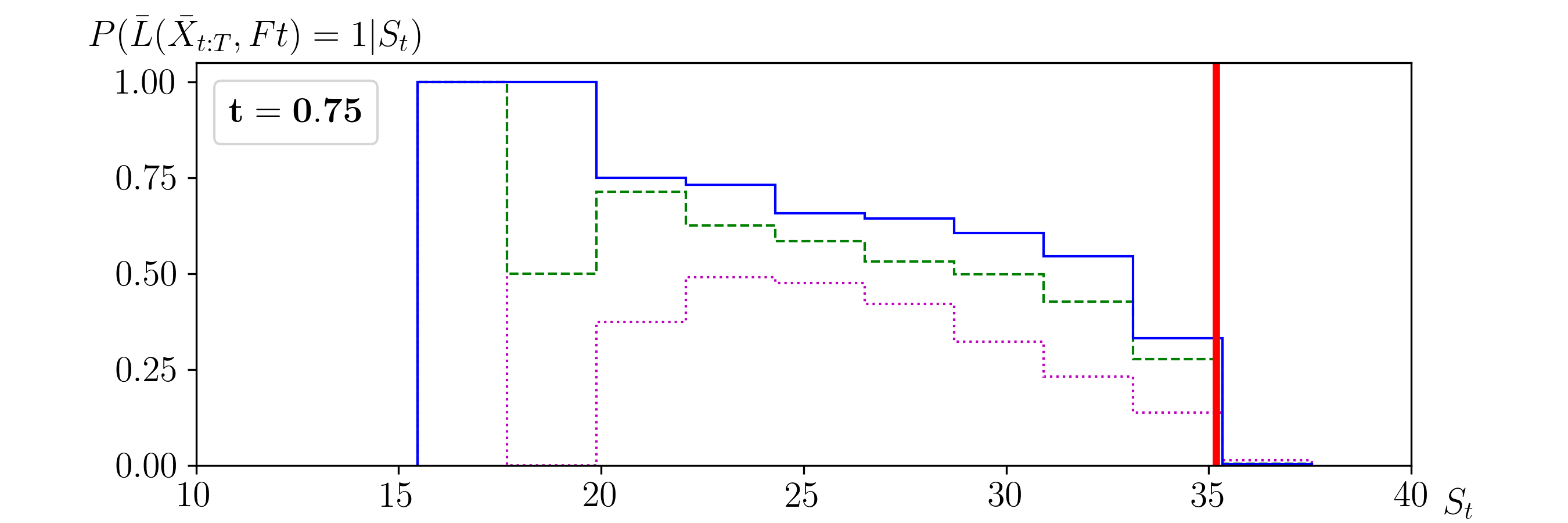} 

  \captionsetup{font={footnotesize}} 
  \caption{ 
    \label{aao_distribution_exercise_boundary} 
    Distribution 
    of the projected lookahead operator 
    at different time points. 
    The time $t$, 
    the number of samples $N$ in the projected operator, 
    and the nearest-neighbor basis size $m$ 
    are as indicated. 
    The energy of the hyperparameters 
    is indicated in square bracket, 
    the exercise boundary is indicated 
    with a vertical line. 
    The x-axis is the stock price $S_t$, 
    and the y-axis is the empirical conditional probability 
    that the projected operator 
    exercises the option. 
    The empirical probability 
    is estimated with $100,000$ simulations, 
    and 
    with an initial stock price $S_{0}$ 
    drawn uniformly at random from the interval $[35,45]$. 
    The option can be exercised 50 times per year, 
    the strike price is $40$, 
    the risk-free rate is $0.06$, 
    the volatility is $0.20$, 
    and the maturity is one year. 
  } 

\end{figure}

Figure \ref{aao_distribution_exercise_boundary} 
shows the distribution of the projected lookahead operator 
at different time period. 
A first observation from this figure 
is that 
when the filtration energy is low, 
the projected operator accuracy increases, 
and the option is exercised with high probability when 
the stock crosses the exercise boundary. 
A second observation is that 
for some hyperparameters, 
the projected operator 
sometimes exercises the option before 
the stock crosses the exercise boundary. 
While with some others hyperparameters, 
such a too early exercise is not observed, 
but many exercise opportunities are missed. 
These distributions 
are two extremes, 
and the convergence theory 
of the previous section 
guarantees 
the existence of intermediate distributions. 
A third observation is that 
hyperparameters with similar filtration energy 
may perform differently. 
This is because 
the filtration energy 
certifies 
that 
the projected operator distribution 
has converged, 
but 
certifies nothing for the projected strategy convergence. 
This situation is related to 
the difference in convergence impact 
between the projected operator convergence \eqref{spls_projected_lookahead_operator_rate} 
and 
the projected strategy convergence \eqref{spls_projected_strategy_rate}. 
The next section analyses 
more the difference between these two convergence factors. 

  \clearpage 

\subsubsection{Projected Strategy} \label{aao_projected_strategy} 

\begin{figure}[ht] 

  \includegraphics[width=\textwidth]{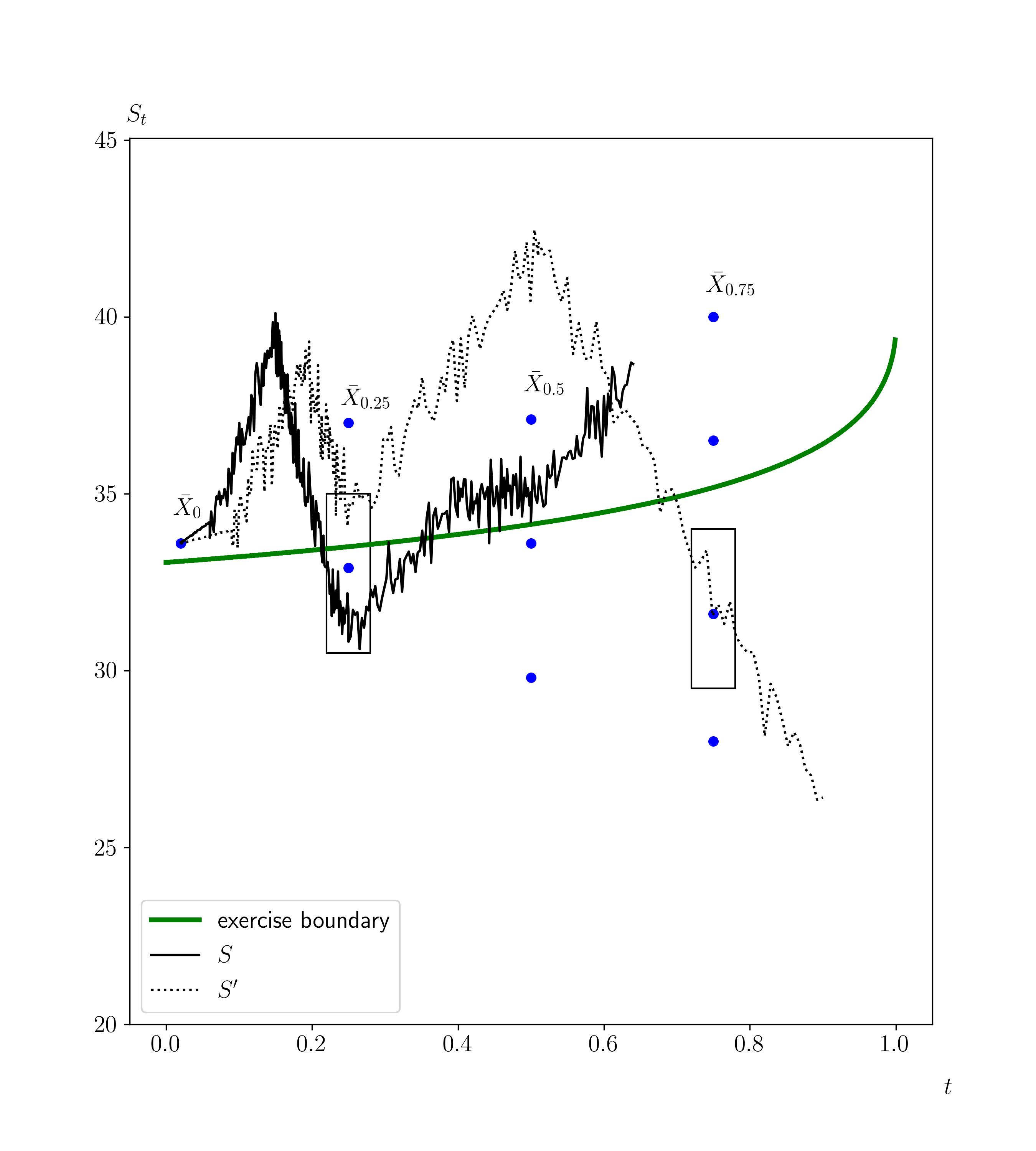} 

  \captionsetup{font={footnotesize}} 
  \caption{ 
    \label{aao_nnb_granularity} 
    A projected strategy with ambiguous cell. 
    For the sample path $S$, 
    the optimal strategy is 
    to exercise the option at time $0.25$, 
    while for the sample path $S'$, 
    the optimal strategy 
    is to exercise the option at time $0.75$. 
    However, both 
    sample paths 
    fall in the same Voronoi cell at time $0.25$, 
    and exercising the optimal decision 
    for both sample paths 
    is infeasible 
    for the projected lookahead operator. 
    The x-axis is the time period $t$, 
    and the y-axis is the stock price $S_t$ 
    at time $t$. 
    For the projected strategy $\bar{X}$, 
    the Voronoi centroids 
    are indicated by dots, 
    and the Voronoi cells are indicated by rectangles. 
    The exercise boundary is as indicated. 
  } 

\end{figure}

The quality of a projected strategy depends 
on the granularity of the underlying nearest-neighbor basis. 
The basis should be fine enough 
to differentiate on which side of the 
exercise boundary the last observed stock price falls. 
As SPLS enforces the exercises constraints 
almost surely, we need in fact a stronger granularity: 
the basis should be fine enough 
to differentiate on which side of the exercise boundary 
every previous stock prices fall. 
Indeed, for a lookahead at time $t=0.1$, 
consider three projected decisions 
$(\bar{X}^2_{0.25}, \bar{X}^1_{0.5}, \bar{X}^3_{0.75})$, 
where $\bar{X}^1_{0.25}$ 
means that the Voronoi cell 
underlying the decision is the cell generated by the first centroid 
$\bar{S}^{0.25,1}_{0.1:0.25}$, 
see \eqref{spls_step_fct} for the notation. 
Consider also two sample paths 
$(S^*_{0.25}, S_{0.5}, S_{0.75})$ 
and 
$(S^\prime_{0.25}, S^\prime_{0.5}, {S^\prime}^*_{0.75})$, 
where $S^*_{0.5}$ means that the sample path hits the exercise boundary at time $t=0.5$. 
Now, note that 
if the two sample paths fall in the same Voronoi cell, 
the following two constraints cannot be enforced 
if the projected decisions are optimal 
\begin{align*} 
  \bar{X}^2_{0.25}(S^*_{0.25}) + \bar{X}^1_{0.5}(S_{0.5}) + \bar{X}^3_{0.75}(S_{0.75})  \le  1, \\
  \bar{X}^2_{0.25}(S^\prime_{0.25}) + \bar{X}^1_{0.5}(S^\prime_{0.5}) + \bar{X}^3_{0.75}({S^\prime}^*_{0.75})  \le  1.
\end{align*} 
By construction, the decisions $\bar{X}^1_{0.25}$ and $\bar{X}^1_{0.75}$ 
must be the same for every path that fall in the underlying Voronoi cell, 
and the two constraints will be of the form $(2 \le 1)$, 
if the projected decisions are optimal. 
Figure \ref{aao_nnb_granularity} illustrates this situation. 

The projected strategy convergence rate 
\eqref{spls_projected_strategy_rate} 
guarantees 
that the previous ambiguous situation 
is rare in probability 
when the nearest-neighbor basis size $m$ is large. 
Two questions then arise: 
how big a basis is needed to avoid ambiguity? 
and 
how much ambiguity can be tolerated in the projected lookahead operator? 
These questions 
are answered 
by the Bayes-value convergence rate 
\eqref{spls_bayes_value_rate}. 
Here, we provide a different insight on these questions, 
by defining an ambiguity metric specific to single exercise American option. 

For the option of this example, 
a specific measure of ambiguity 
is the number of ambiguous constraints. 
For a projected lookahead at time $t$, 
and a nearest-neighbor basis size $m$, 
this measure can be written with 
\begin{align*} 
  \widehat{AM}_t = \frac{1}{{N_a}^2} \sum_{i=1}^{N_a} \sum_{j=1}^{N_a}
    \ib{
      c(\bar{X}, S^i) \text{ and } c(\bar{X}, S^j)
      \text{ are infeasible if } \bar{X} \text{ is optimal}
    },
\end{align*} 
where $c(\bar{X},S^i)$ 
is the exercise constraint 
$\bar{X}_0(S^i) + \bar{X}_1(S^i)+ \ldots +\bar{X}_T(S^i)  \le 1$, 
and $S^i$ is a path in a sample of size $N_a$. 
The event in the metric 
can be checked 
by taking every projected decision as optimal. 
For example, the decision $\bar{X}_1(S^i)$ 
is taken as exercising the option 
only if the sample path $S^i$ hits the exercise boundary at time $1$. 
Intuitively, the ambiguity metric will be low 
if the nearest-neighbor basis 
as many Voronoi cells around the exercise boundary, 
so that the decision $\bar{X}_t$ 
can differentiate 
between a sample path for which 
the option has already been exercised, 
and a sample path for which the option is exercisable for the first time at time $t$. 
A total ambiguity metric can be defined 
as the average ambiguity 
\begin{align*} 
  \widehat{AM} = \frac{1}{T} \sum_{t=0}^T \widehat{AM}_t.
\end{align*} 
The total ambiguity is a proxy score 
for the convergence of the projected strategy. 
This score can be used as a substitute 
to the projected strategy convergence rate \eqref{spls_projected_strategy_rate} 
to prove the convergence in probability of SPLS 
\eqref{spls_consistency}. 

\begin{figure}[ht] 

  \includegraphics[width=0.8\textwidth]{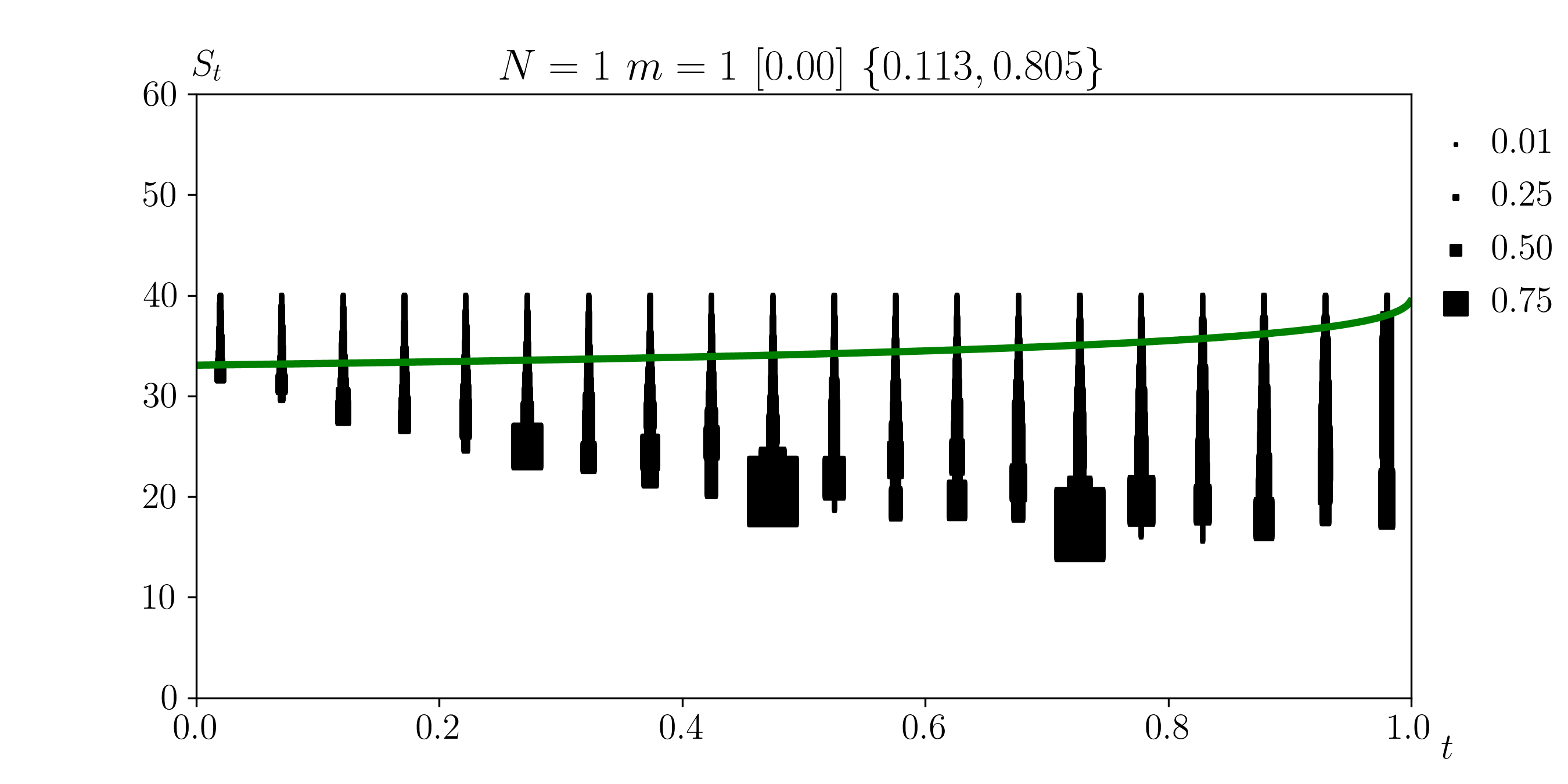} 
  \includegraphics[width=0.8\textwidth]{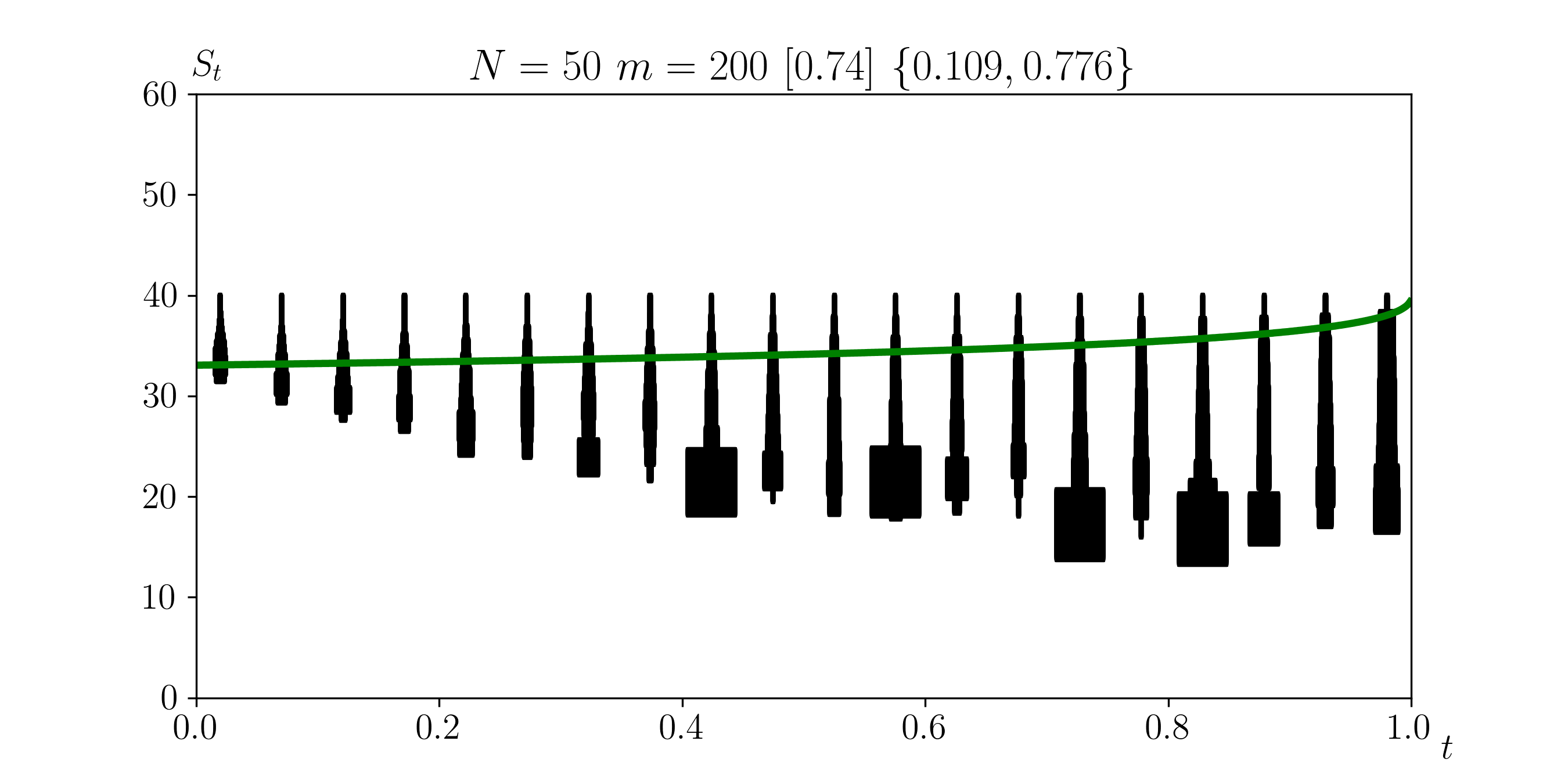} 
  \includegraphics[width=0.8\textwidth]{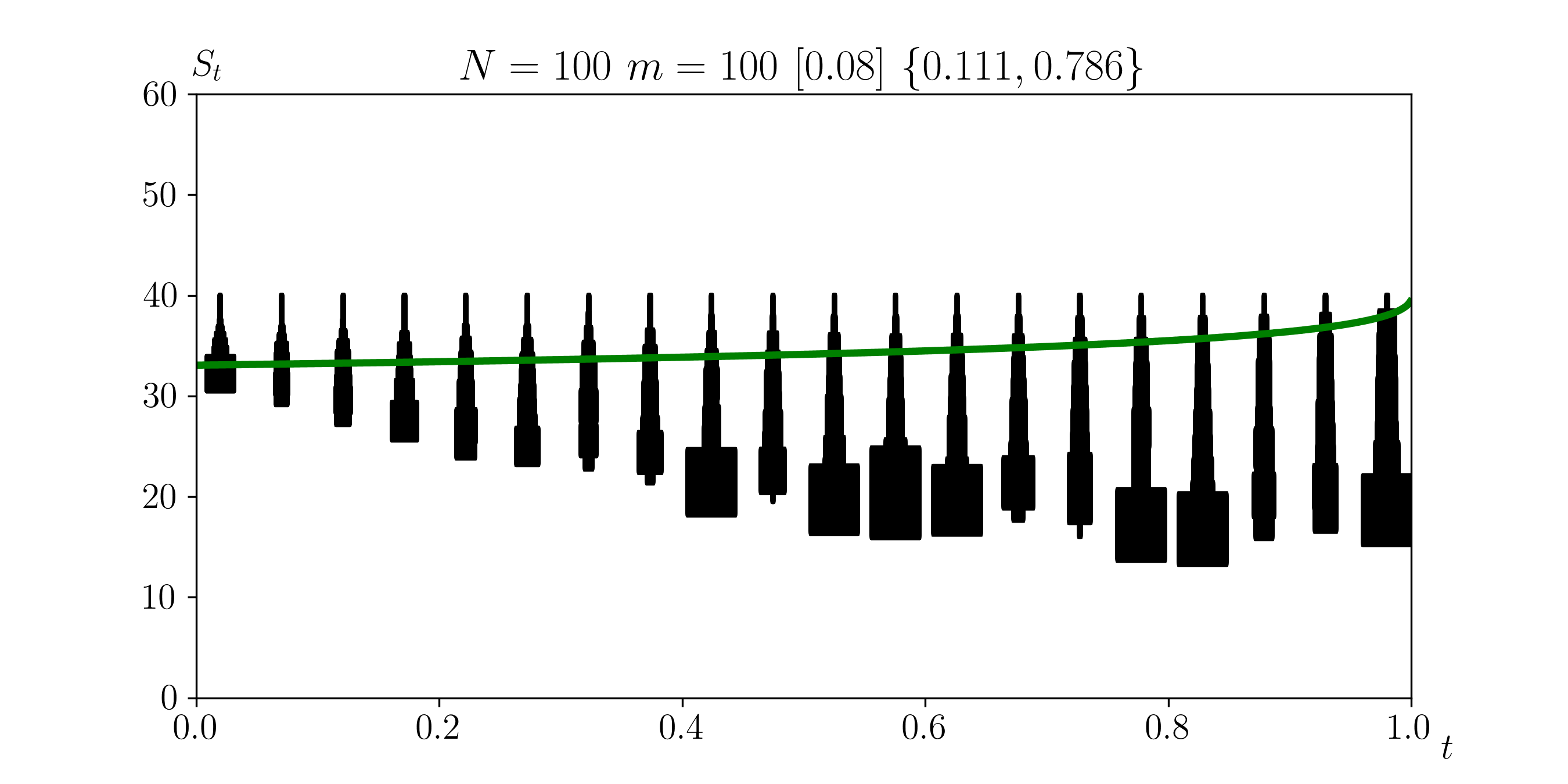} 

  \captionsetup{font={footnotesize}} 
  \caption{ 
    \label{aao_ambiguity} 
    Distribution 
    of the projected lookahead operator over time. 
    The empirical conditional probability 
    of exercise $P(\bar{L}(\bar{X}_{t:T},F_t)|S_t)$ 
    is represented in the figure with a linewidth gradient. 
    The exercise boundary is the bold curve. 
    The sample size $N$ in the projected lookahead operator, 
    and 
    the nearest-neighbor basis size $m$ 
    are as indicated. 
    The filtration energy is indicated in square bracket. 
    The ambiguity metric is indicated in braces. 
    The first number in the brace 
    is the ambiguity metric computed 
    with a sample of $1,000$ simulations, 
    while the second number is computed 
    with $10,000$ simulations. 
    The simulations are done 
    with an initial stock price $S_{0}$ 
    draw uniformly at random from the interval $[35,45]$. 
    The x-axis is the time period $t$, 
    and the y-axis is the stock price $S_t$ 
    at time $t$. 
  } 

\end{figure}

Figure \ref{aao_ambiguity} 
shows 
the exercise distribution 
and the ambiguity metric 
for different hyperparameters. 
For a projected lookahead at time $t$, 
the exercise distribution 
is an histogram 
of the empirical conditional probability 
$P(\bar{L}(\bar{X}_t,F_t)|S_t)$, 
see Figure \ref{aao_exercise_boundary} for reference. 
For Figure \ref{aao_ambiguity}, 
the histogram probabilities are represented 
with the point size: 
point with larger size 
have a higher conditional probability of exercise. 
This convention can be written with 
\begin{align*} 
  \text{point size at time } t &\propto P(\bar{L}_t(\bar{X}_t,F_t)=1|S_t),
\end{align*} 
where the conditional exercise probability 
is estimated with a sample 
of projected lookahead operator. 

A first observation from Figure \ref{aao_ambiguity} 
is that 
when the stock is closed to be in-the-money, 
the projected operator may exercises too early. 
However, 
the conditional exercise probability of a too early exercise is small, 
and 
the probability that 
the stock is close to be in-the-money without crossing the exercise boundary 
is also small. 
This two factors 
allows the SPLS prices in Table \ref{aao_option_price} 
to be good estimates of the option value. 
A second observation is that 
the ambiguity metric 
depends on the sample size $N'$ used to compute the metric. 
For the sample size used in the projected lookahead operator, 
which is of the order of $100$, 
the ambiguity metric of order $1,000$ 
suggests that 
very few sample paths lead to ambiguous constraints. 
A third observation is 
that the ambiguity metric 
differentiates 
hyperparameters quality. 
This can be seen 
by the hyperparameter $(N,m)=(100,100)$ 
always exercising when the option is in-the-money, 
while the hyperparameter $(1,1)$ 
exercises less. 
When two hyperparameters 
have a low ambiguity metric 
such as 
$(100,100)$ 
and $(50,200)$, 
the best hyperparameter 
have a lower filtration energy. 
These observations can be summarized as follows. 
The nearest-neighbor size $m$ 
defines a discretization of the strategy space, 
while the sample size $N$ 
defines a distribution on this discretization. 
A finer discretization is better 
only if 
the distribution quality is maintained 
with a higher sample size. 

  \clearpage 

\subsubsection{Estimates Volatility} 

\begin{figure}[ht] 

  \includegraphics[width=\textwidth]{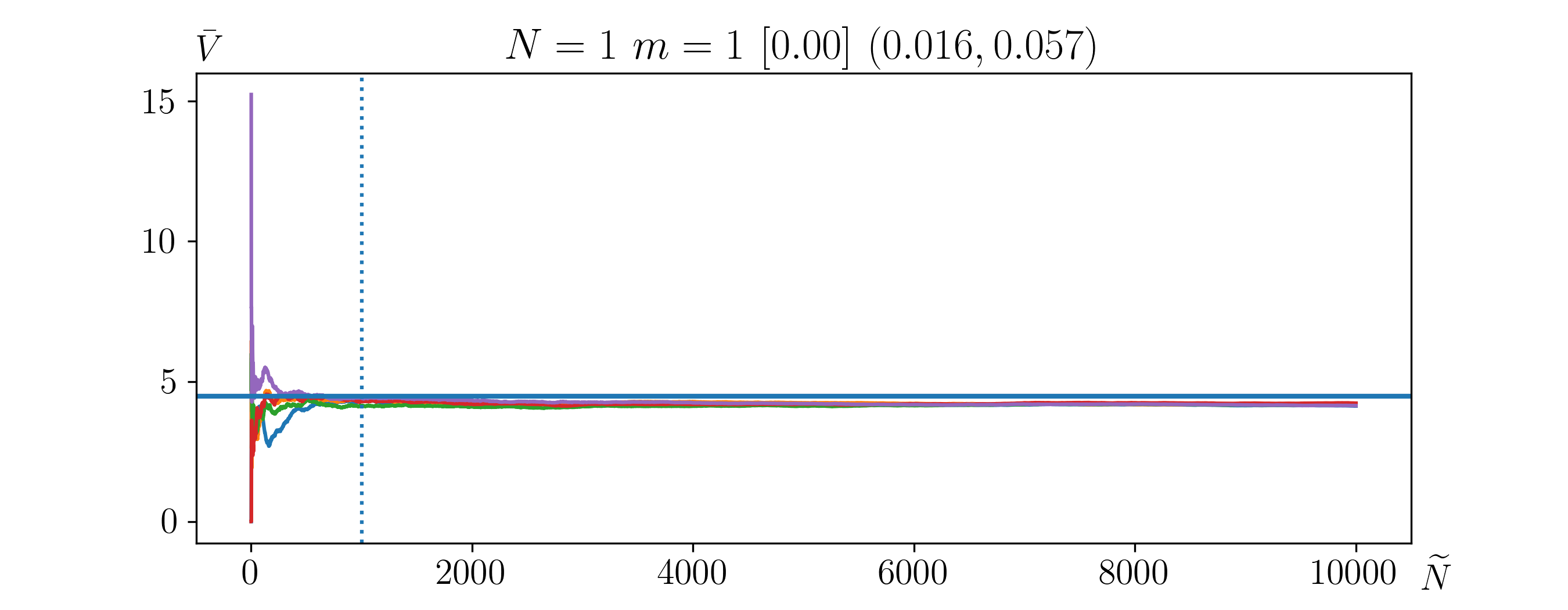} 

  \includegraphics[width=\textwidth]{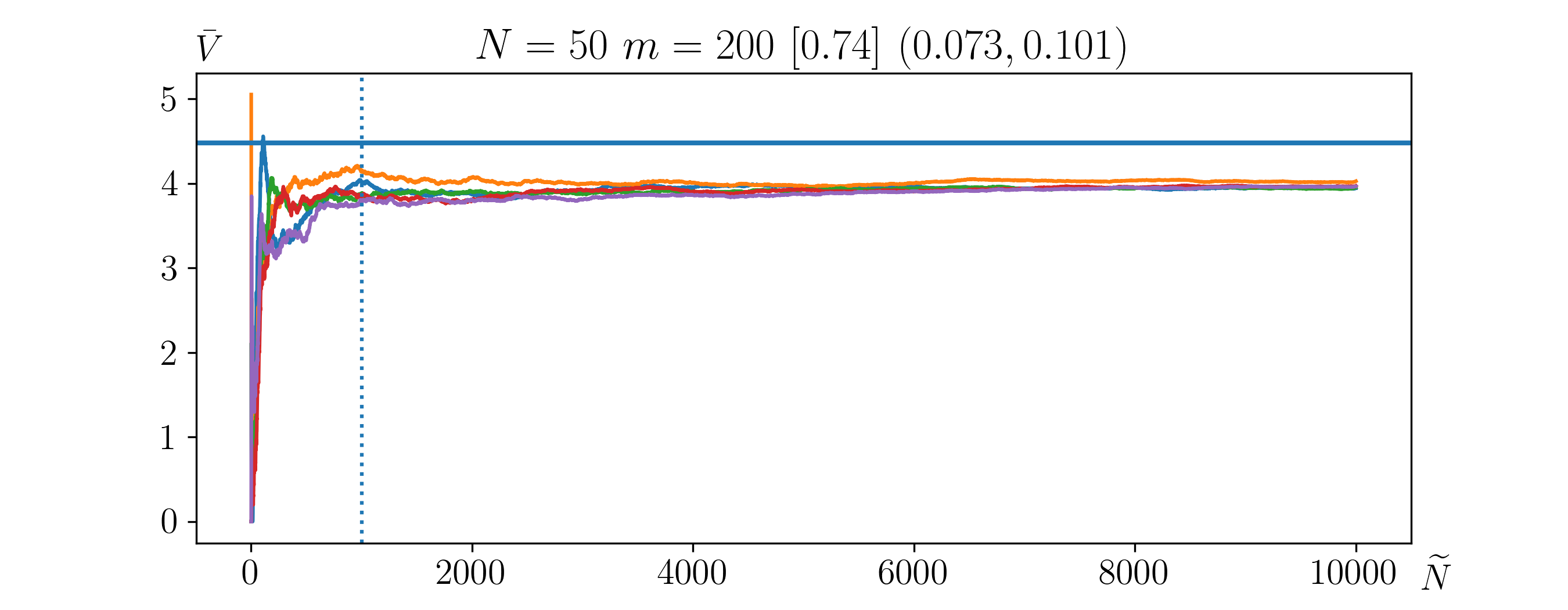} 

  \includegraphics[width=\textwidth]{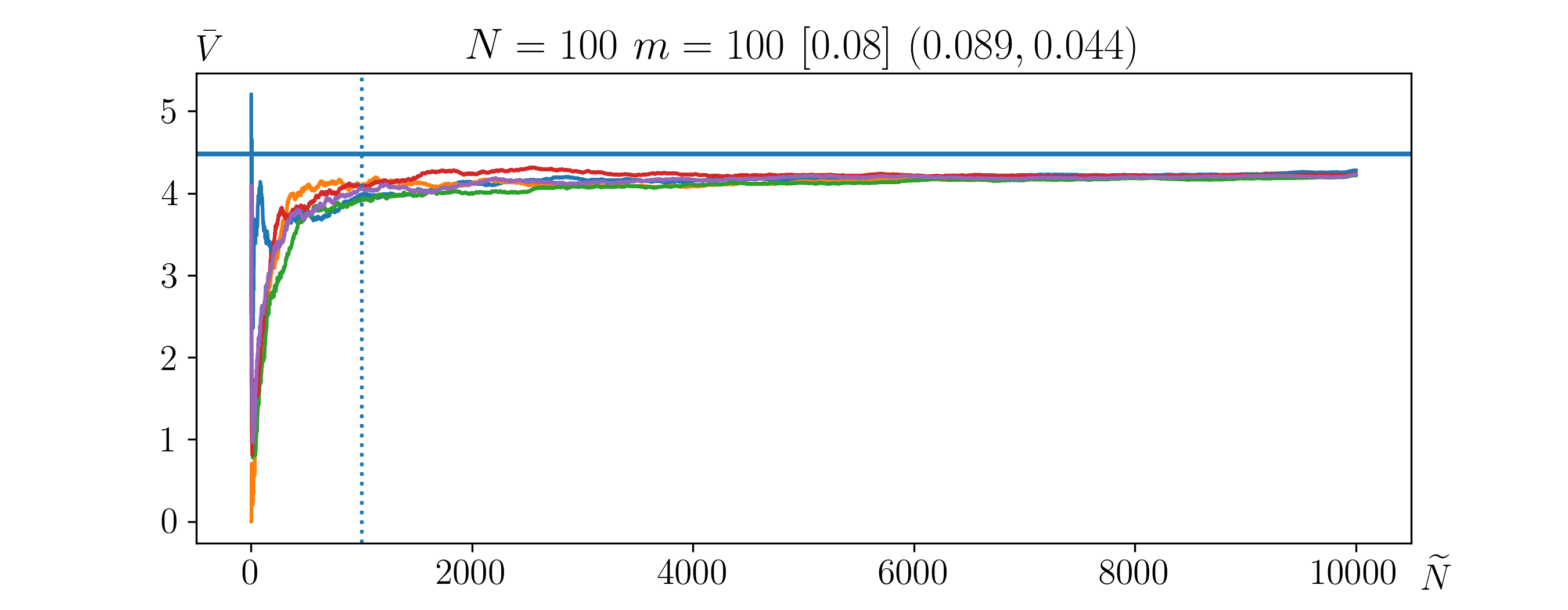} 

  \captionsetup{font={footnotesize}} 
  \caption{ 
    \label{aao_estimates_volatility_spls} 
    Convergence of the projected option value 
    for different random seed. 
    The number of sample $\widetilde{N}$ in the projected option value, 
    the sample size $N$ in the projected lookahead operator, 
    and the nearest-neighbor basis size $m$ 
    are as indicated. 
    The energy of the hyperparameters 
    is indicated in square bracket, 
    and 
    the finite difference price is the horizontal line. 
    The best relative difference 
    between the project option value 
    and the finite difference price 
    is indicated in parenthesis. 
    The first number in parenthesis is 
    the best difference after 
    a $1,000$ sample, 
    and the second number is after $10,000$. 
    The vertical line indicates the $1,000$ sample level. 
    The option can be exercised 50 times per year, 
    the strike price is $40$, 
    the risk-free rate is $0.06$, 
    the volatility is $0.20$, 
    the maturity is one year, 
    and the initial stock price is $36$. 
  } 

\end{figure}

With the projected lookahead operator being a random variable, 
SPLS uses a random strategy to approximate the optimal strategy: 
at each time period, 
the exercise decision is drawn at random from a certain probability distribution. 
As no value function and no policy function are learned 
in the course of sampling, 
the sampling distribution is fixed 
and should converge quickly to an estimate of the option value. 
Figure \ref{aao_estimates_volatility_spls} 
confirms this intuition. 
When started from any random seed, 
any given set of hyperparameters converges quickly to the same value. 
This observation 
is supported 
by the projected option value convergence rate 
\eqref{spls_projected_option_value_rate} 
and 
the Bayes-value convergence rate 
\eqref{spls_bayes_value_rate}. 

By using a random strategy that converges quickly, 
one approach in using SPLS 
is to use several small samples, 
instead of a single large sample. 
This a bootstrap procedure: 
use a small sample size $\widetilde{N}=1,000$, 
instead of a large sample size $\widetilde{N}=10,000$, 
and estimate the price with the highest estimate obtained. 
Figure \ref{aao_estimates_volatility_spls} 
and 
Table \ref{aao_option_price} 
show that this approach works, 
especially if different hyperparameters 
are used in each small sample. 
This approach is massively parallel 
and can provide fast and accurate estimates. 

\begin{figure}[ht] 

  \includegraphics[width=\textwidth]{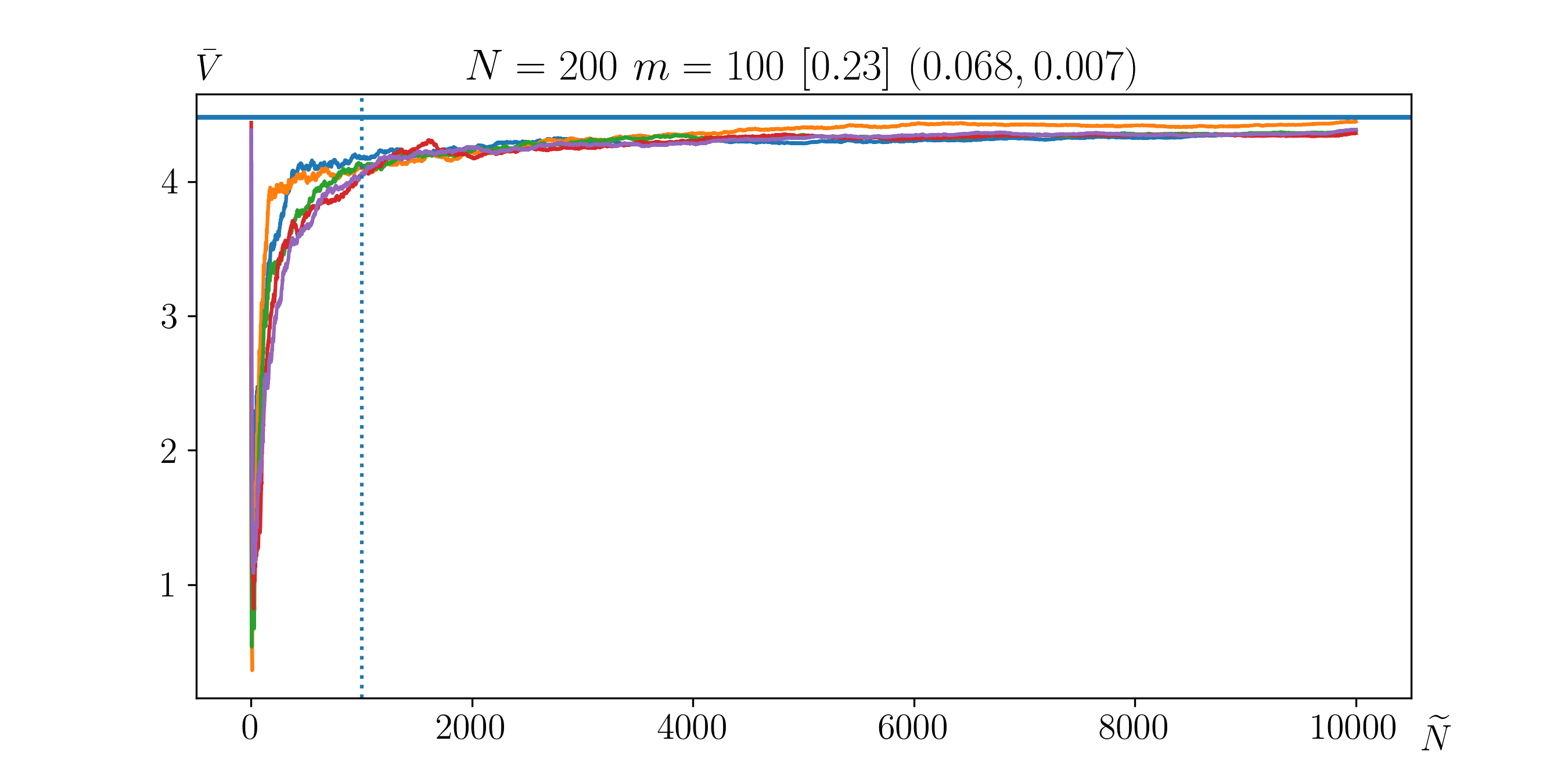} 
  \includegraphics[width=\textwidth]{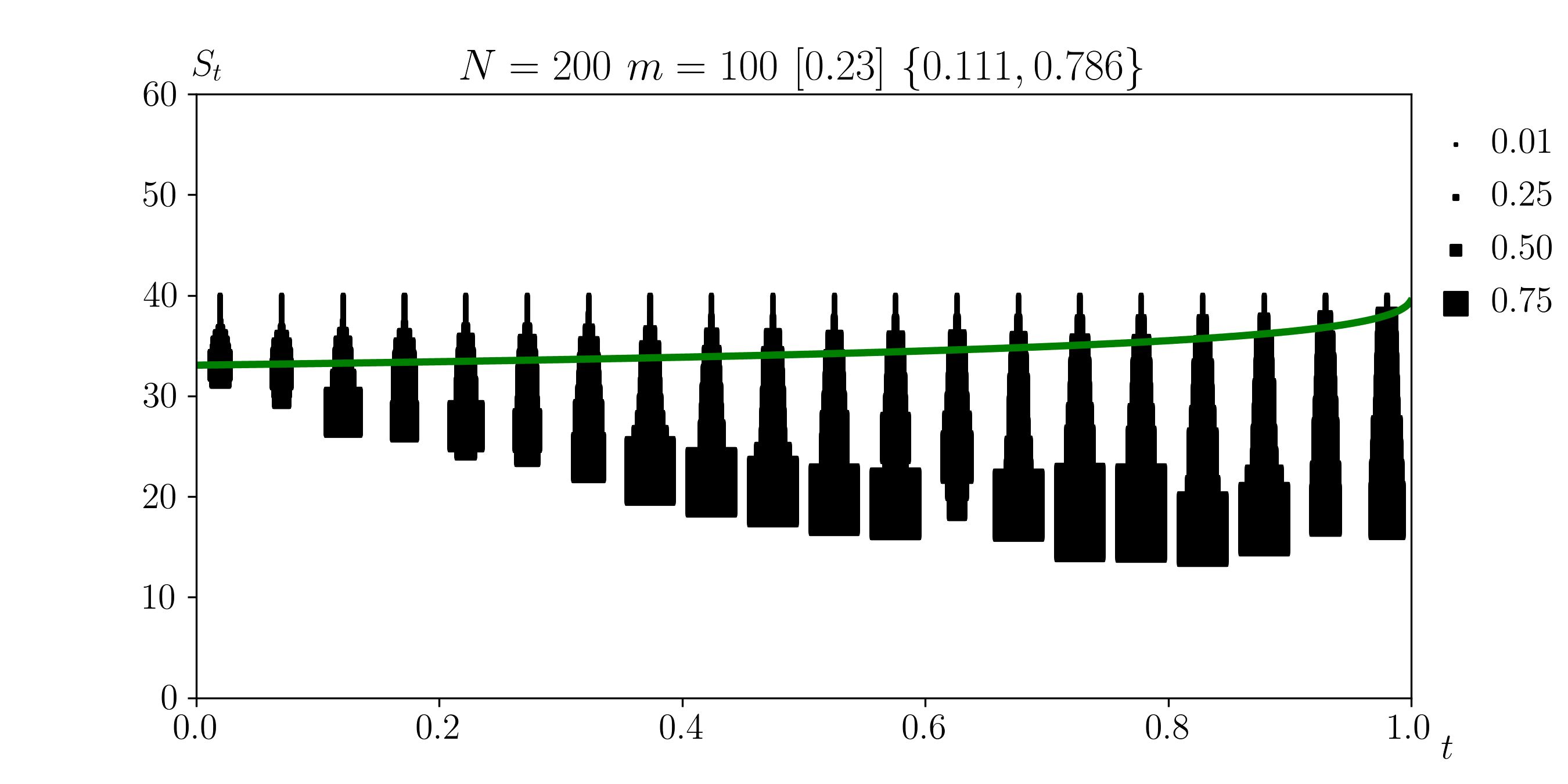} 

  \captionsetup{font={footnotesize}} 
  \caption{ 
    \label{aao_estimates_volatility_spls_200_100} 
    \textbf{Top}. 
    Convergence of the projected option value 
    for different random seeds 
    with a high accuracy projection scheme. 
    The number of samples $\widetilde{N}$ in the projected option value, 
    the sample size $N$ in the projected lookahead operator, 
    and the nearest-neighbor basis size $m$ 
    are as indicated. 
    The energy of the hyperparameters 
    is indicated in square bracket, 
    and 
    the finite difference price is the horizontal line. 
    The best relative difference 
    between the project option value 
    and the finite difference price 
    is indicated in parenthesis. 
    The first number in parenthesis is 
    the best difference after 
    a $1,000$ sample, 
    and the second number is after $10,000$. 
    The vertical line indicates the $1,000$ sample level. 
    The energy validation sample is of size $N'=2,000$, 
    The option can be exercised 50 times per year, 
    the strike price is $40$, 
    the risk-free rate is $0.06$, 
    the volatility is $0.20$, 
    the maturity is one year, 
    and the initial stock price is $36$. 
    \textbf{Bottom}. 
    Distribution 
    of the projected lookahead operator over time 
    for the hyperparameters set in the top figure. 
    The empirical conditional probability 
    of exercised $P(\bar{L}(\bar{X}_{t:T},F_t)|S_t)$ 
    is represented in the figure with a linewidth gradient. 
    The exercise boundary is the bold curve. 
    The sample size $N$ in the projected lookahead operator, 
    and 
    the nearest-neighbor basis size $m$ 
    are as indicated. 
    The filtration energy is indicated in square bracket. 
    The ambiguity metric is indicated in braces. 
    The first number in the brace 
    is the ambiguity metric computed 
    with a sample of $1,000$ simulations, 
    while the second number is computed 
    with $10,000$ simulations. 
    The simulations are done 
    with an initial stock price $S_{0}$ 
    drawn uniformly at random from the interval $[35,45]$. 
    The x-axis is the time period $t$, 
    and the y-axis is the stock price $S_t$ 
    at time $t$. 
  } 

\end{figure}

The convergence theory of the previous section 
guarantees that there exists hyperparameters 
for SPLS 
that can approximate the price of any option 
to any degree of accuracy. 
Such hyperparameters can be found 
with a fine grained hyperparameters tuning. 
For instance, 
Table \ref{aao_option_price} could be obtained 
with a higher precision. 
To show that this is possible, 
Figure \ref{aao_estimates_volatility_spls_200_100} 
shows the convergence of a high accuracy hyperparameter set 
for one of the prices in Table \ref{aao_option_price} for which 
the estimation error was 5\%. 
The distribution of the projected lookahead operator 
is also displayed. 

  \clearpage 

\subsection{NNM Analysis} \label{aao_nnm_analysis} 

For any option, 
a dual martingale is optimal 
if 
the martingale 
is close 
to the martingale part 
of the value function 
(\citeauthor{rogers07} \cite*{rogers07}, \citeauthor{rogers02} \cite*{rogers02}, \citeauthor{haugh04} \cite*{haugh04}, \citeauthor{andersen04} \cite*{andersen04}). 
They are many senses 
in which two stochastic processes can be close. 
For an American put, 
we can look at the behavior 
of a dual martingale 
across 
the exercise boundary 
by quantizing the martingale. 
In such a perspective, 
two martingales are close 
if the distribution of their behavior 
across the exercise boundary is similar. 
This sense of closeness provides 
many insights 
on NNM 
and will be the starting point of our analysis. 
Then, 
we look at the volatility of the NNM estimates. 

\subsubsection{NNM Martingale} \label{aao_nnm_martingale} 

\begin{figure}[ht] 

  \includegraphics[width=\textwidth]{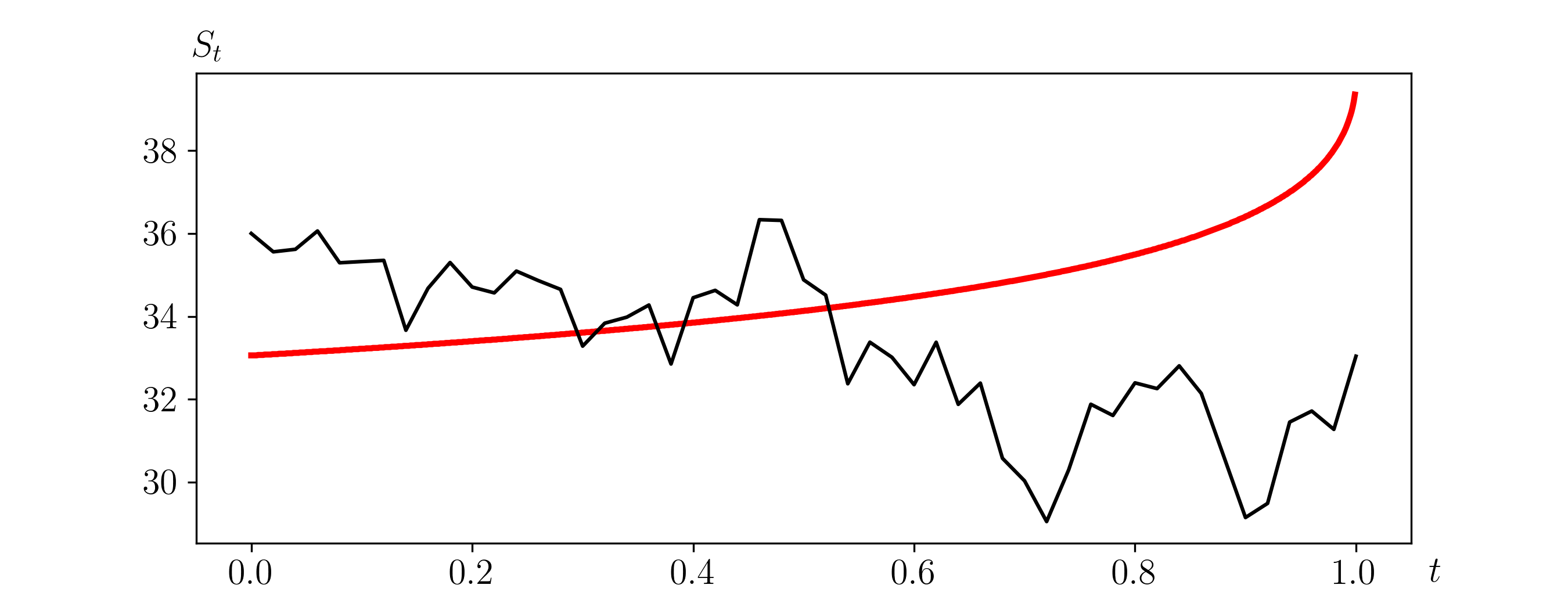} 
  \includegraphics[width=\textwidth]{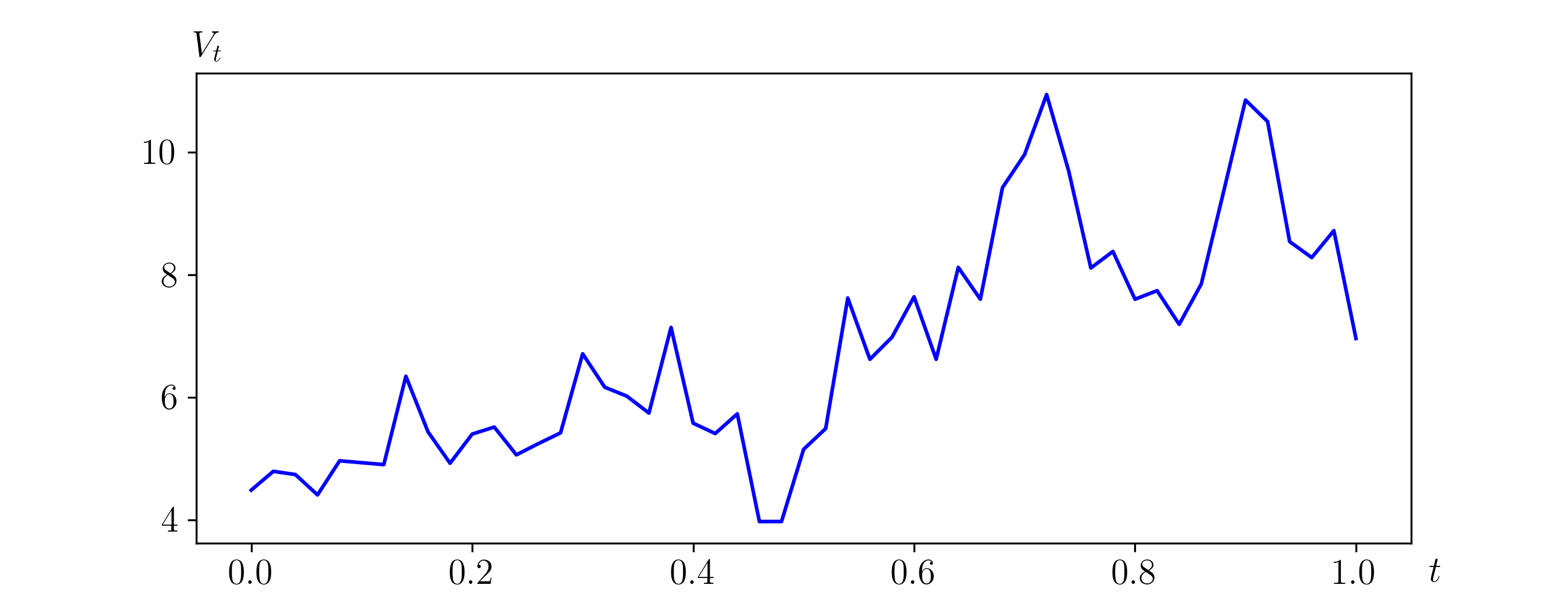} 
  \includegraphics[width=\textwidth]{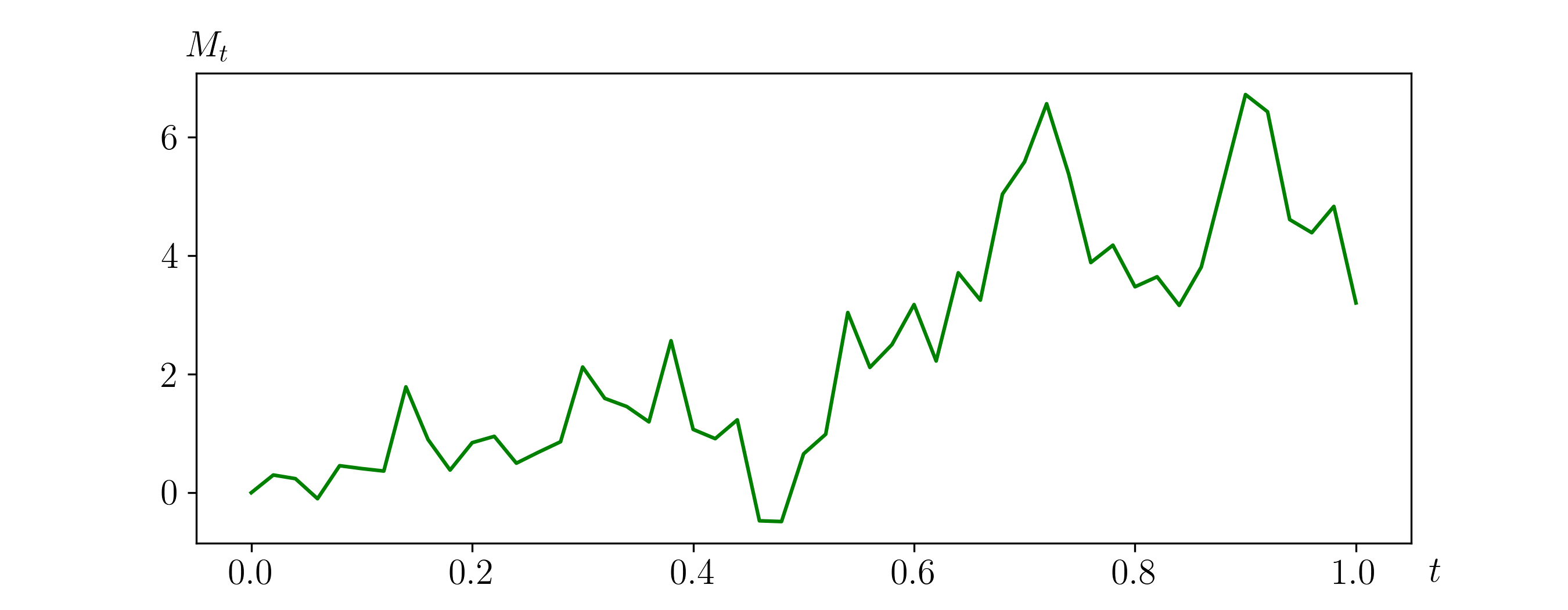} 

  \captionsetup{font={footnotesize}} 
  \caption{ 
    \label{aao_martingale_part_example} 
    A sample stock path, 
    the corresponding value function path, 
    and 
    the corresponding martingale part increment path. 
    In the top graph, 
    the exercise boundary 
    is the bold line. 
    The x-axis is the time period $t$, 
    and the y-axis is either 
    the stock price $S_t$, 
    the value function $V_t$, 
    or the martingale part $M_t$ of the value function. 
    The option can be exercised 50 times per year, 
    the strike price is 40, 
    the risk-free rate is 0.06, 
    the maturity is one year, 
    and the volatility is $0.20$. 
    The exercise boundary 
    is found 
    with an implicit finite difference scheme 
    with 1,000 time steps per year 
    and 10,000 steps for the stock price. 
    The martingale part is found 
    by nested simulation at each time period. 
    Each nested simulation uses $10,000$ 
    sample of the next period stock price. 
    See \hyperref[meth]{Methods} for more details. 
  } 

\end{figure}

An NNM martingale is the martingale produced 
by the projected Rogers operator for some choice of hyperparameters. 
We will compare the behavior of several NNM martingales 
to the optimal martingale across the exercise boundary. 
For an American put, the optimal martingale can be obtained as follows. 
With an implicit finite difference method, 
obtain the value function. 
Then, 
obtain the martingale part of the value function 
by nested simulation along a sample stock path. 
The resulting martingale is optimal. 
See \hyperref[meth]{Methods} for more details. 
Figure \ref{aao_martingale_part_example} 
shows 
an example of sample stock path, 
the corresponding value function path, 
and the corresponding martingale part path. 

\begin{figure}[ht] 

	\begin{center} 
		\textbf{Optimal Martingale \{1.617\}} 
	\end{center} 

	\includegraphics[width=0.98\textwidth]{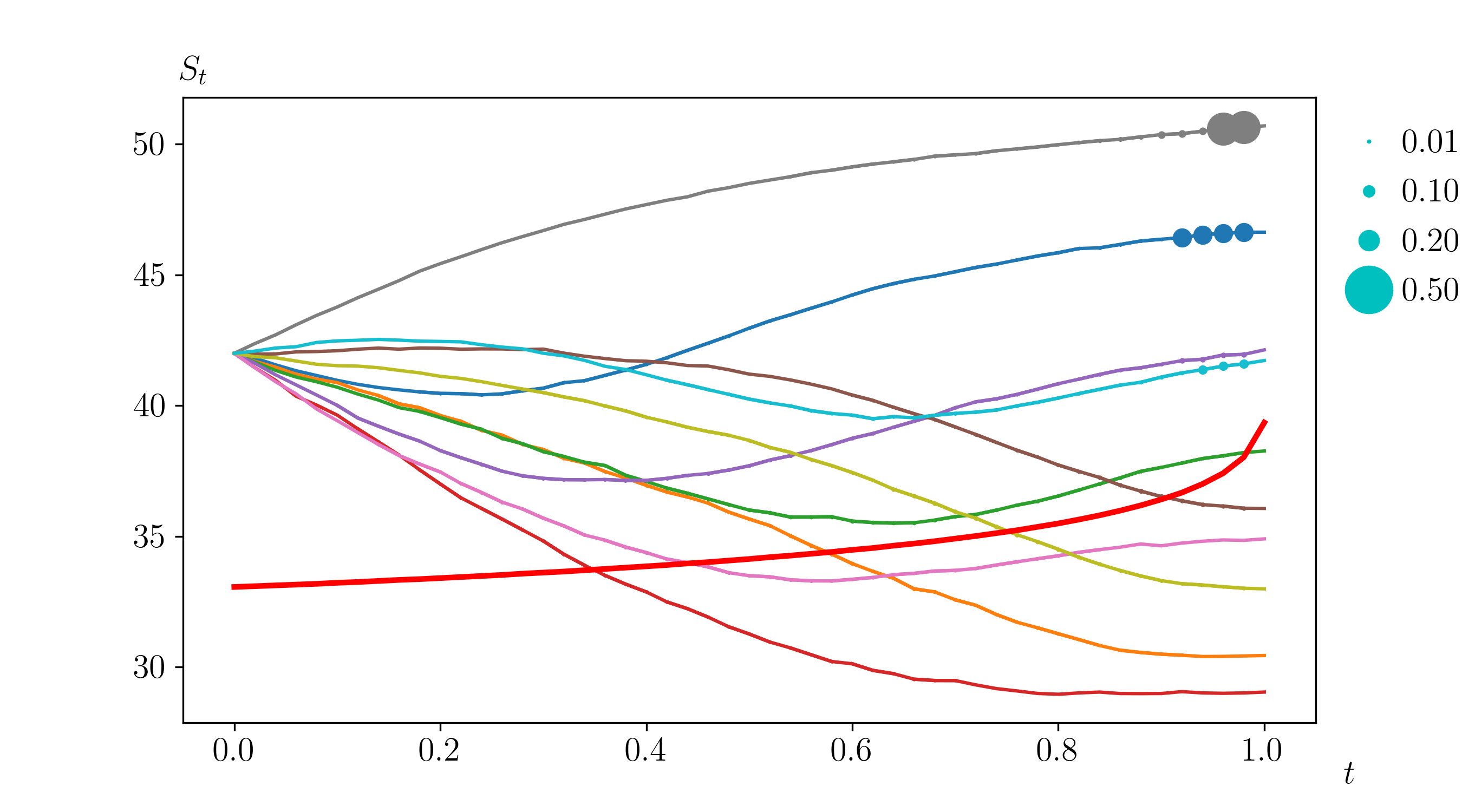} 
    \includegraphics[width=0.98\textwidth]{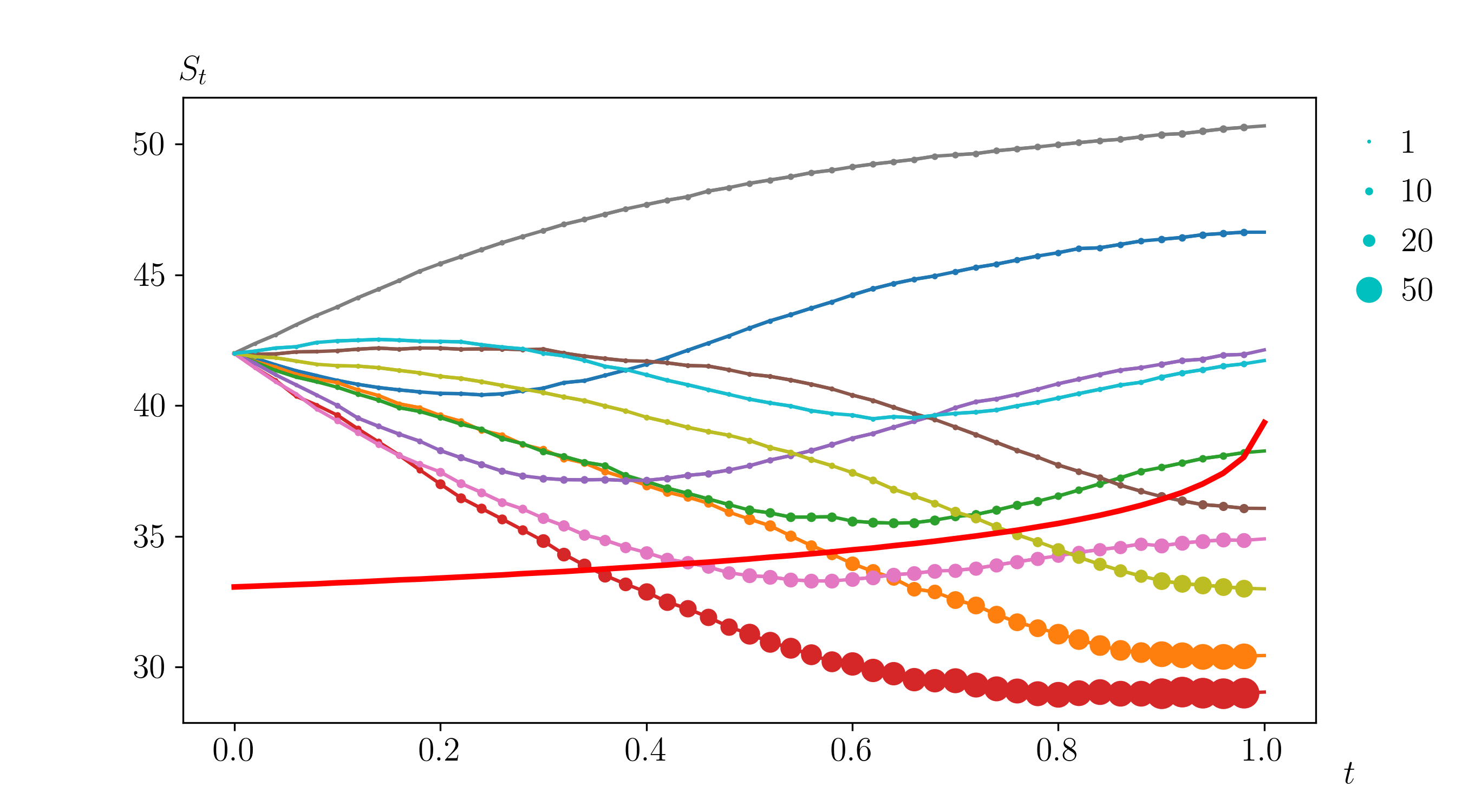} 

  \captionsetup{font={footnotesize}} 
  \caption{ 
    \label{aao_optimal_martingale_part} 
    The dual exercise distribution 
    and the $\ell_1$-average 
    of the optimal martingale 
    over a quantization of the optimal martingale. 
    The sample stock paths shown 
    are the implicit stock path centroids 
    implied 
    by the Voronoi tesselation 
    of the optimal martingale. 
    The Voronoi tesselation 
    is obtained with a sample of $10,000$ 
    sample martingale path 
    by finding 10 centroids with Lloyd's method. 
    In the figure title, 
    the price of the option is in brace. 
    The exercise boundary 
    is the bold line. 
    The x-axis is the time period $t$, 
    and the y-axis is the stock price $S_t$. 
    In the top graph, 
    the distribution of the dual exercise is represented 
    with a linewidth gradient. 
    In the bottom graph, 
    the $\ell_1$-average of the optimal martingale 
    is represented with a linewidth gradient. 
    The option can be exercised 50 times per year, 
    the strike price is 40, 
    the risk-free rate is 0.06, 
    the maturity is one year, 
    the volatility is $0.20$, 
    and the initial stock price is $42$. 
    The exercise boundary 
    is found 
    with an implicit finite difference scheme 
    with 1,000 time steps per year 
    and 10,000 steps for the stock price. 
    The martingale part is found 
    by nested simulation at each time period. 
    Each nested simulation uses $10,000$ 
    sample of the next period stock price. 
    See \hyperref[meth]{Methods} for more details. 
  } 

\end{figure}

To study the distribution of a martingale 
across the exercise boundary, 
we quantize a sample of martingale path 
with a small number of centroids. 
As each martingale path 
is generated by a stock path, 
the martingale centroids 
define an implicit tesselation 
on the sample stock path. 
This implicit tesselation can be used 
to map any martingale metric 
to the regions of the exercise boundary. 
We call such a mapped martingale metric a \textit{dual metric}. 
See \hyperref[meth]{Methods} for more details. 
Figure \ref{aao_optimal_martingale_part} 
shows 
the distribution 
of three dual metrics: 
the typical stock trajectory, 
the dual exercise decision 
and 
the $\ell_1$-average of the optimal martingale. 
The typical stock trajectory 
within a cell of the implicit tesselation, 
is the average of the stock path that falls within the cell. 
In the top figure, 
the linewidth gradient is the distribution 
of the dual exercise over time. 
To define the dual exercise, 
recall that to each stock path is associated 
the dual payoff 
\begin{align*} 
  \max_x \sum_{t=0}^T e^{-r t} \pp{K - S^i_t} x_t   - x_t M^i_t, \stepcounter{equation}\tag{\theequation}\label{aao_dual_exercise_time}
\end{align*} 
where $S^i$ is a sample path, 
and $M^i$ is the corresponding martingale path. 
Let $t^i$ be the optimal exercise time in \eqref{aao_dual_exercise_time}, 
and 
let 
$S^i_{t^i}$ be the stock price at the dual exercise time $t^i$. 
The distribution of the dual exercise time 
is an histogram of a sample $\{(S^i_{t^i}, t^i)\}$ of such dual exercise. 
The figure presents this distribution 
along the Voronoi cell. 
In the bottom figure, 
the linewidth gradient 
is the $\ell_1$-average value of the optimal martingale 
over time. 
To define this $\ell_1$-average, 
consider a particular Voronoi cell, 
and 
let $\{M^i\}$ 
be the set of martingale path that falls within the cell. 
The $\ell_1$-average at time $t$ can be written with 
\begin{align*} 
  \frac{1}{n} \sum_{i=1}^n \pa{M^i_t}, \stepcounter{equation}\tag{\theequation}\label{aao_dual_abs_average}
\end{align*} 
where $n$ is the number of martingale sample path 
that falls within the cell. 

Figure \ref{aao_optimal_martingale_part} 
shows 
that the optimal martingale 
forces the dual exercise 
to be in the out-of-the-money region. 
To achieve this, 
the martingale takes large positive value 
in the in-the-money region, 
and small negative value 
in the out-of-money region. 
The graph shows only the average absolute value of the optimal martingale, 
but the martingale value sign 
can be inferred 
by whether the stock path centroid 
is in-the-money 
or out-of-the-money. 

\begin{figure}[ht] 

  \begin{center} 
      \textbf{NNM (1, 1) [0.00] \{2.88\}} 
  \end{center} 

  \includegraphics[width=\textwidth]{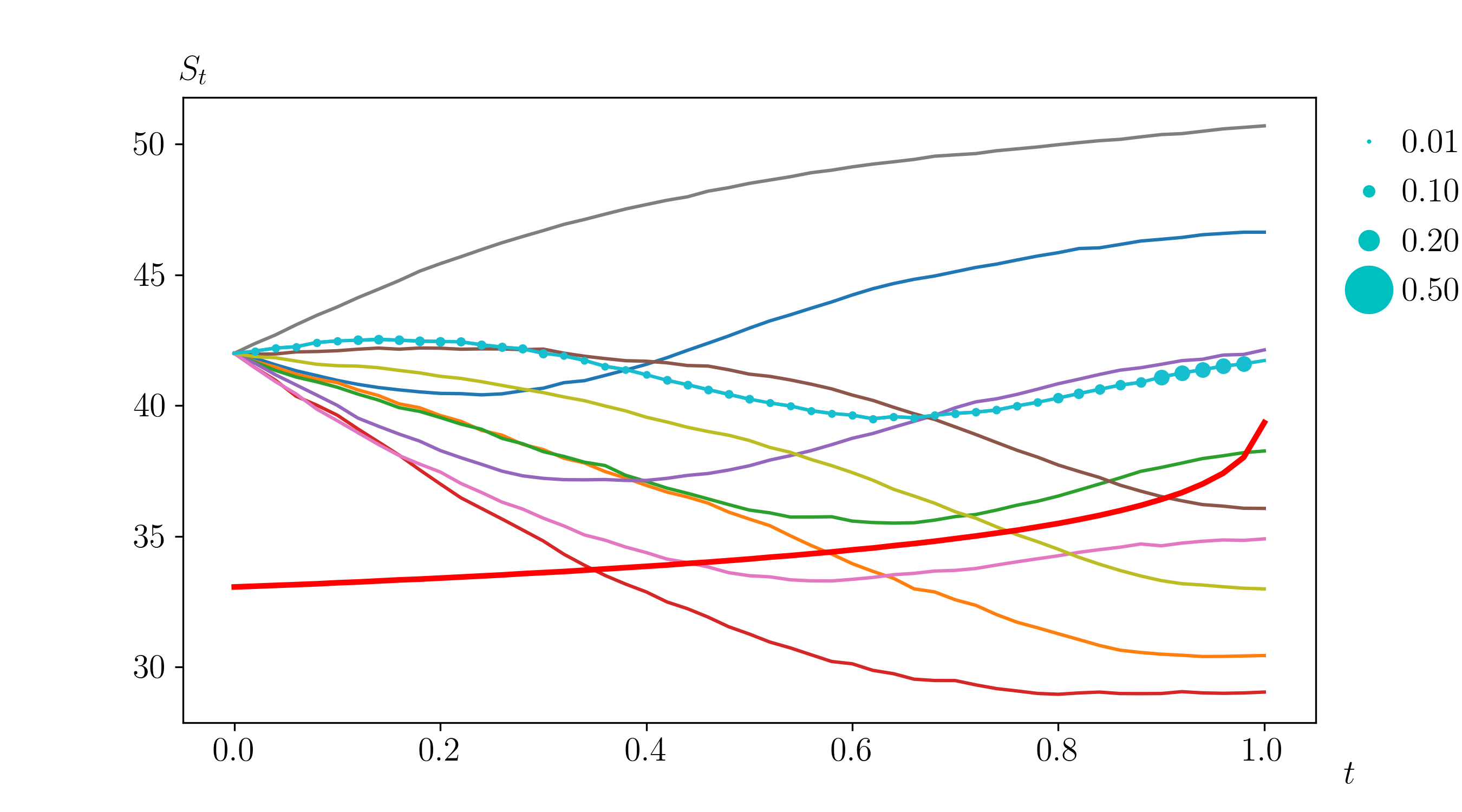} 
  \includegraphics[width=\textwidth]{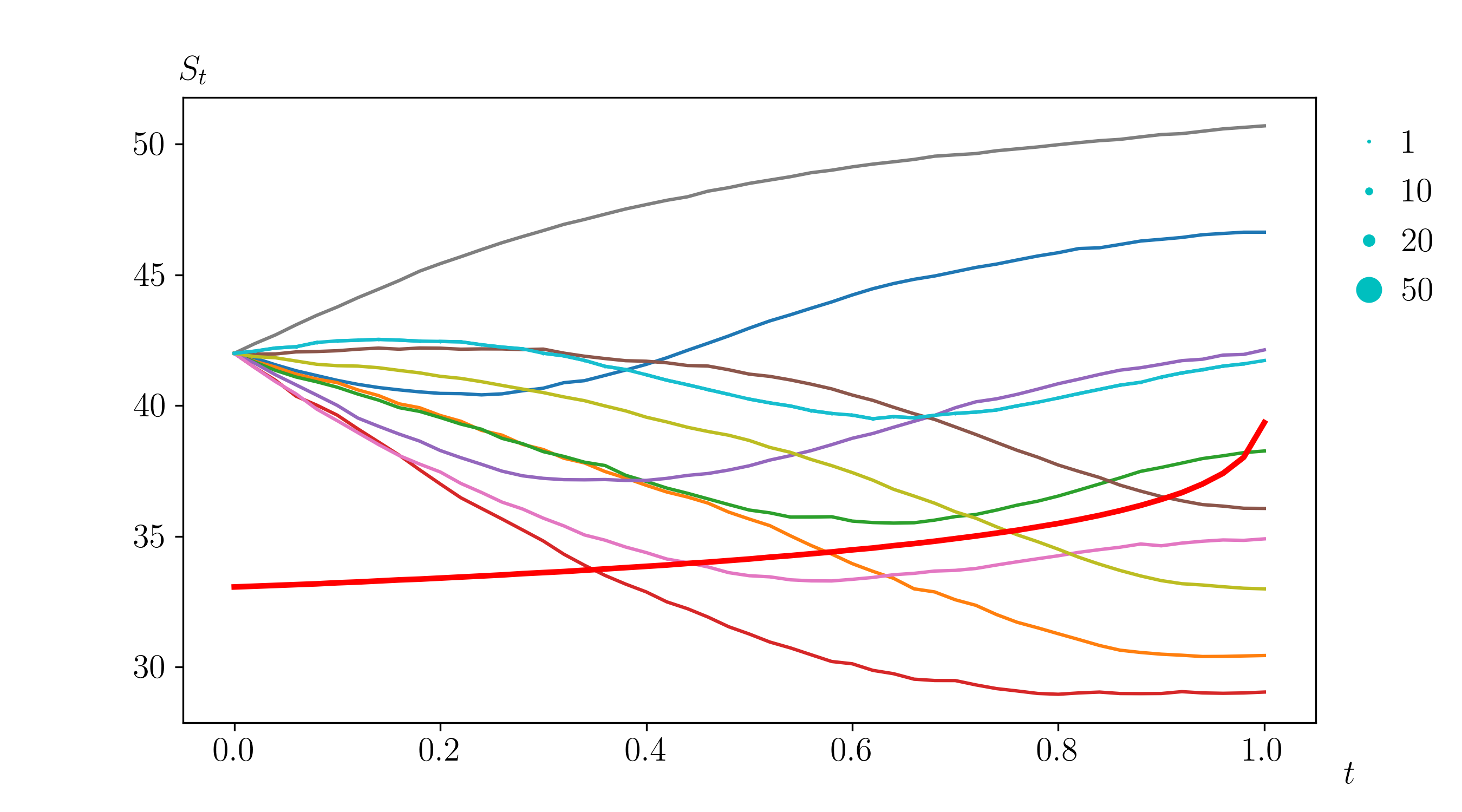} 

  \captionsetup{font={footnotesize}} 
  \caption{ 
    \label{aao_p1_q1_martingale_part} 
    The dual exercise distribution 
    and the $\ell_1$-average 
    of the NNM martingale $(p,q)=(1,1)$ 
    over a quantization of the optimal martingale. 
    See Figure \ref{aao_optimal_martingale_part} 
    for the quantization and notation details. 
    In the figure title, 
    the number in bracket is the filtration energy, 
    and the number in brace is the relaxed dual value 
    of the NNM martingale. 
    NNM uses 
    $N=1,000$ simulations, 
    and 
    an energy validation sample of size $N'=10,000$. 
  } 

\end{figure}

To measure how close 
a candidate martingale is to the optimal martingale 
we can look at the dual metrics. 
For the comparison to be meaningful, 
the reference Voronoi tesselation should be fixed. 
We use the optimal martingale tesselation 
as the reference tesselation. 
See \hyperref[meth]{Methods} for more details. 
Figure \ref{aao_p1_q1_martingale_part} 
shows 
the dual metrics for the NNM martingale with 
nearest-neighbor basis size $(p,q)=(1,1)$. 
By comparing the dual exercise distribution 
in Figure \ref{aao_p1_q1_martingale_part} and Figure \ref{aao_optimal_martingale_part}, 
we see 
that the $(1,1)$-martingale 
does not prevent dual exercise in the in-the-money region. 
The dual exercise distribution 
for the martingale is uniform over the timeline. 
This difference in dual exercise distribution 
is explained by the 
$\ell_1$-average of the $(1,1)$-martingale: 
the martingale is zero on every stock centroid. 
Dual exercise is hence possible everywhere. 
Note that Figure \ref{aao_p1_q1_martingale_part} 
shows that the option is dually exercised 
when the stock path centroids is out-of-the-money. 
This is because 
the Voronoi cell is an aggregation of several stock paths 
and many stock path in the cell can be in-the-money. 
In fact, for the $(1,1)$-martingale, 
every martingale path is identically zero, 
so that all stock paths fall in the same Voronoi cell. 

\begin{figure}[ht] 

  \begin{center} 
      \textbf{NNM (10, 350) [0.20] \{1.65\}} 
  \end{center} 

  \includegraphics[width=\textwidth]{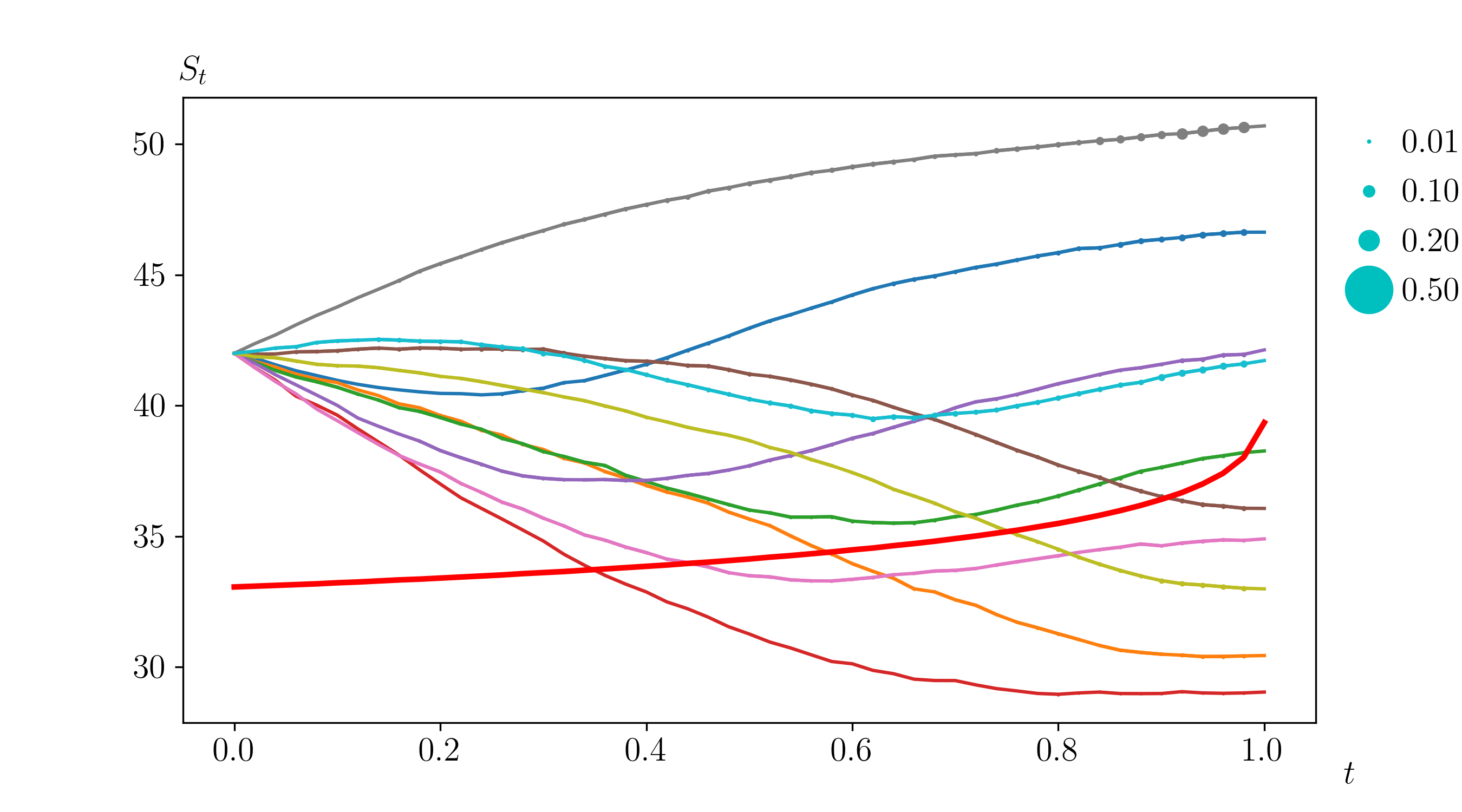} 
  \includegraphics[width=\textwidth]{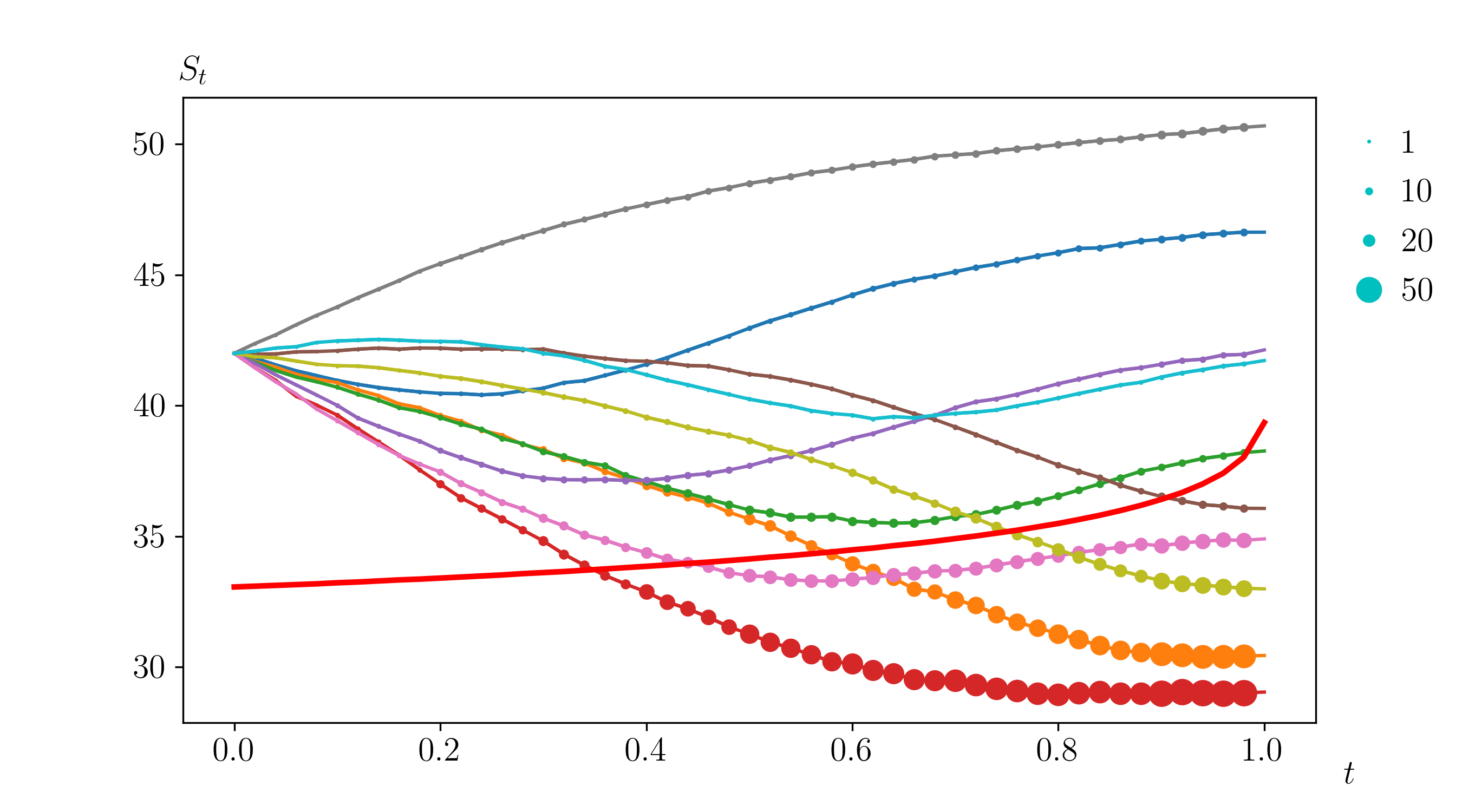} 

  \captionsetup{font={footnotesize}} 
  \caption{ 
    \label{aao_p10_q350_martingale_part} 
    The dual exercise distribution 
    and the $\ell_1$-average 
    of the NNM martingale $(p,q)=(10,350)$ 
    over a quantization of the optimal martingale. 
    See Figure \ref{aao_optimal_martingale_part} 
    for the quantization and notation details. 
    In the figure title, 
    the number in bracket is the filtration energy 
    of the NNM martingale, 
    and the number in brace is the relaxed dual value. 
    NNM uses 
    $N=1,000$ simulations, 
    and 
    an energy validation sample of size $N'=10,000$. 
  } 

\end{figure}

Figure \ref{aao_p10_q350_martingale_part} 
shows 
the dual metric for a high precision NNM martingale. 
A first observation from this figure 
is that 
both the dual exercise time 
and the $\ell_1$-average metrics 
of the martingale 
are very close to the optimal martingale. 
They are some small difference 
as to where the NNM pushes the dual exercise time, 
and in the $\ell_1$-magnitude of the martingale, 
but the relaxed dual value accuracy is preserved. 
This is expected from the convergence theory of the previous section. 
NNM is an algorithm with 
probabilistic guarantees, 
and such guarantees tolerate small difference in distribution. 

\begin{figure}[ht] 

  \begin{center} 
      \textbf{NNM (10, 100) [0.03] \{2.25\}} 
  \end{center} 

  \includegraphics[width=\textwidth]{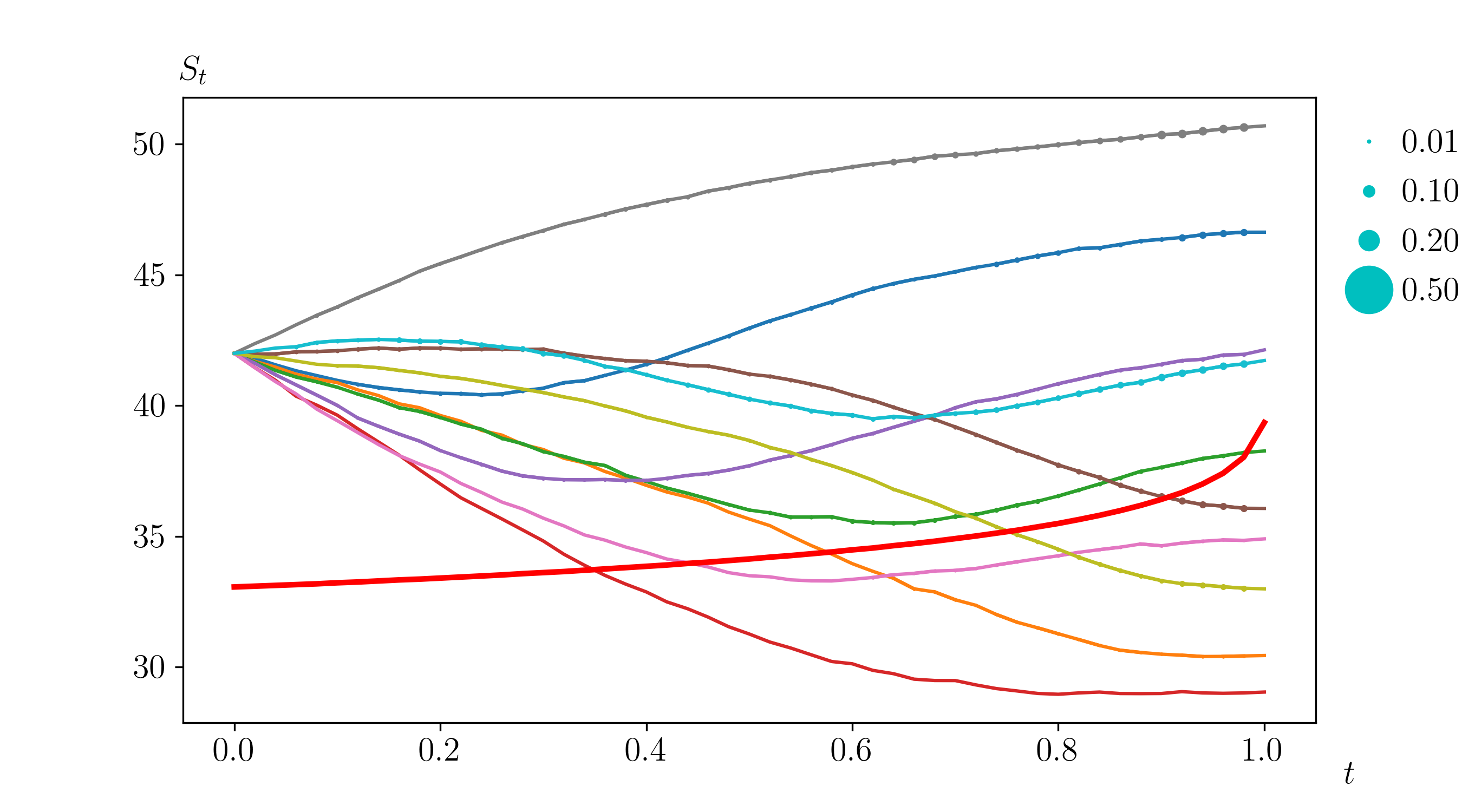} 
  \includegraphics[width=\textwidth]{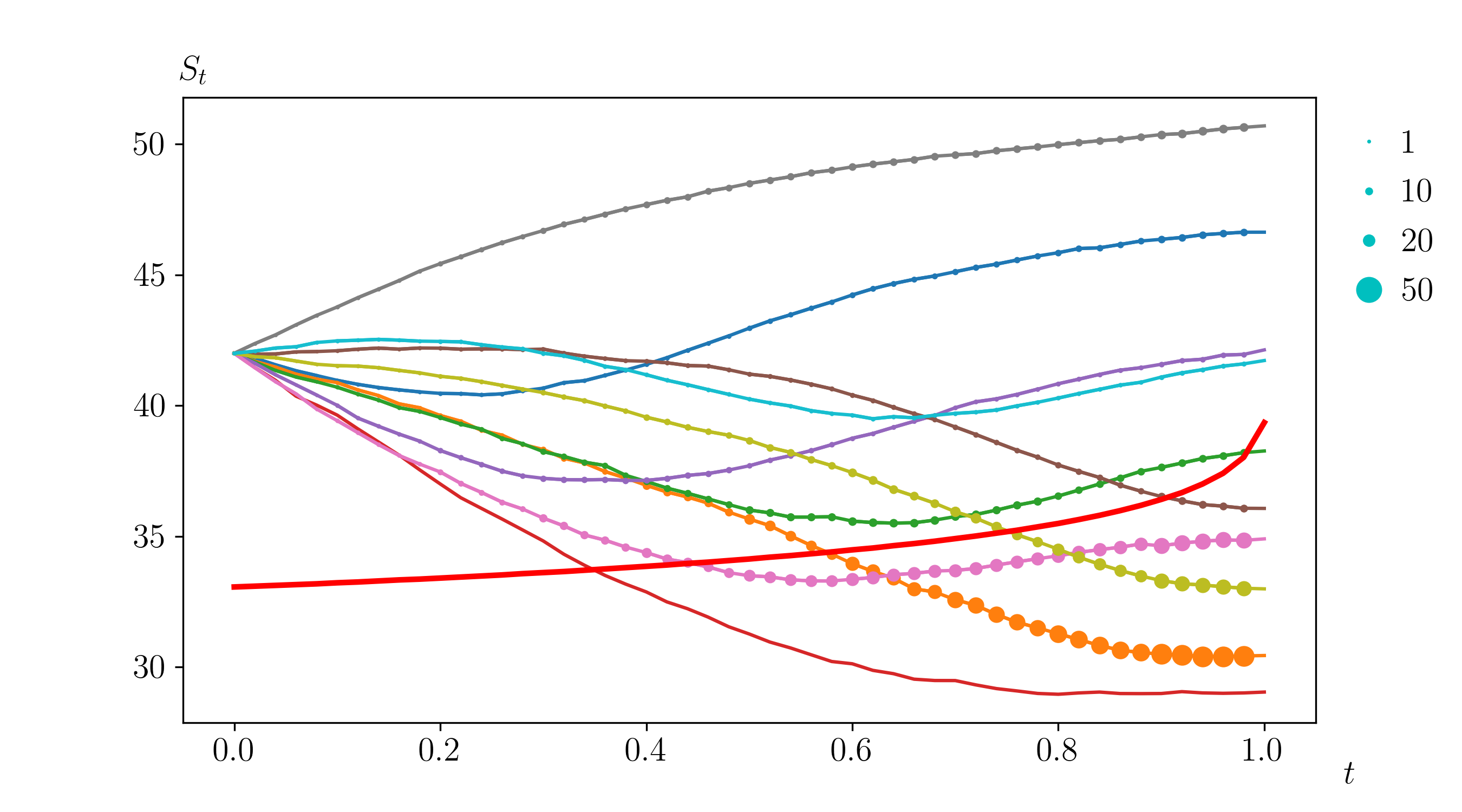} 

  \captionsetup{font={footnotesize}} 
  \caption{ 
    \label{aao_p10_q100_martingale_part} 
    The dual exercise distribution 
    and the $\ell_1$-average 
    of the NNM martingale $(p,q)=(10,100)$ 
    over a quantization of the optimal martingale. 
    See Figure \ref{aao_optimal_martingale_part} 
    for the quantization and notation details. 
    In the figure title, 
    the number in bracket is the filtration energy 
    of the NNM martingale, 
    and the number in brace is the relaxed dual value. 
    NNM uses 
    $N=1,000$ simulations, 
    and 
    an energy validation sample of size $N'=10,000$. 
  } 

\end{figure}

Comparing the $\ell_1$-average of an NNM martingale 
to the optimal martingale 
gives an idea of the NNM projected martingale quality. 
Indeed, if the two $\ell_1$-averages are close, 
we can infer that 
the nearest-neighbor basis of the projected martingale is 
sufficiently fine 
for the projected martingale to match in average the optimal martingale. 
This proxy for the projected martingale quality 
allows to draw some conclusions 
on the relation between the filtration energy 
and the projected martingale quality. 
To draw these conclusions, 
Figure \ref{aao_p10_q100_martingale_part} 
shows the dual metrics for a low precision NNM martingale. 
By comparing this figure 
to Figure \ref{aao_p10_q350_martingale_part}, 
we see that 
the projected martingale quality 
is more important 
then the filtration energy 
for the relaxed dual value accuracy. 
Indeed, the filtration energy is a metric 
for the accuracy of the projected Rogers operator, 
and the operator can only be as much accurate 
as the projected martingale. 
In particular, 
the $(10,100)$-martingale 
nearest-neighbor basis 
does not capture 
one of the optimal martingale centroid, 
and the dual exercise time for the stock path 
in this centroid is uncontrolled. 
This can be seen by 
the distribution of the dual exercise time 
being uniform on the middle stock centroid. 

  \clearpage 

\subsubsection{Estimates Volatility} 

\begin{figure}[ht] 

  \includegraphics[width=\textwidth]{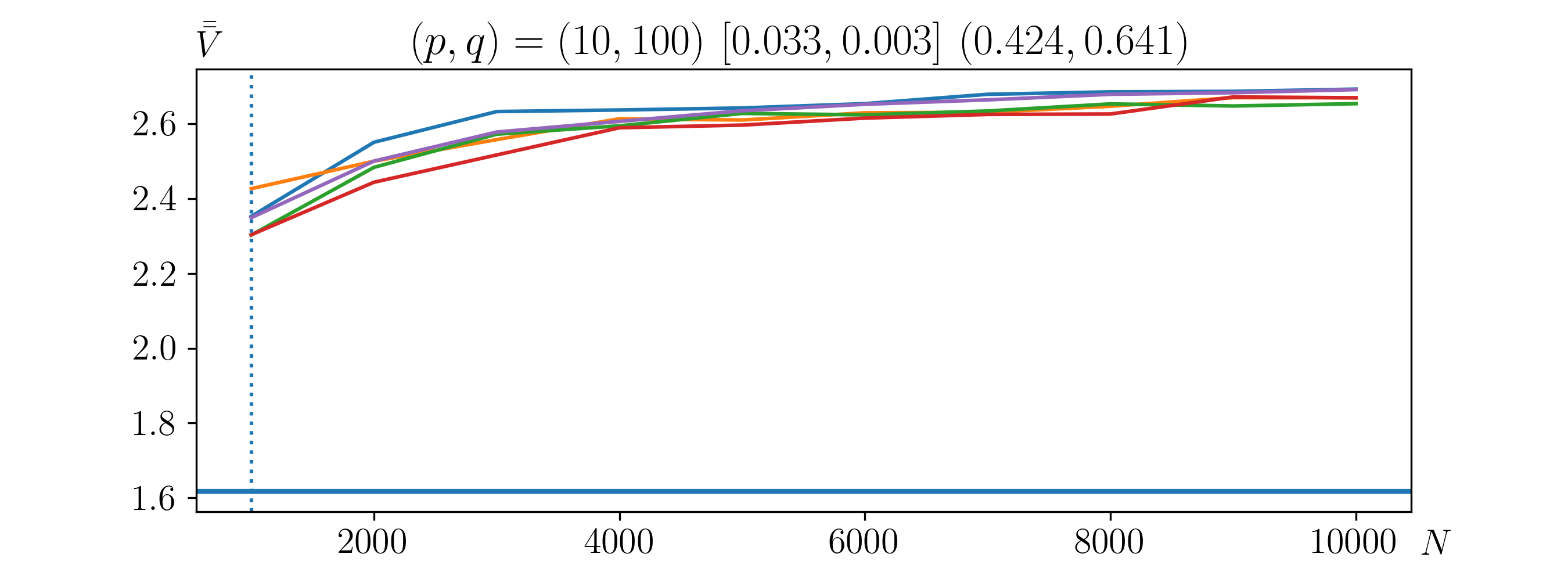} 
  \includegraphics[width=\textwidth]{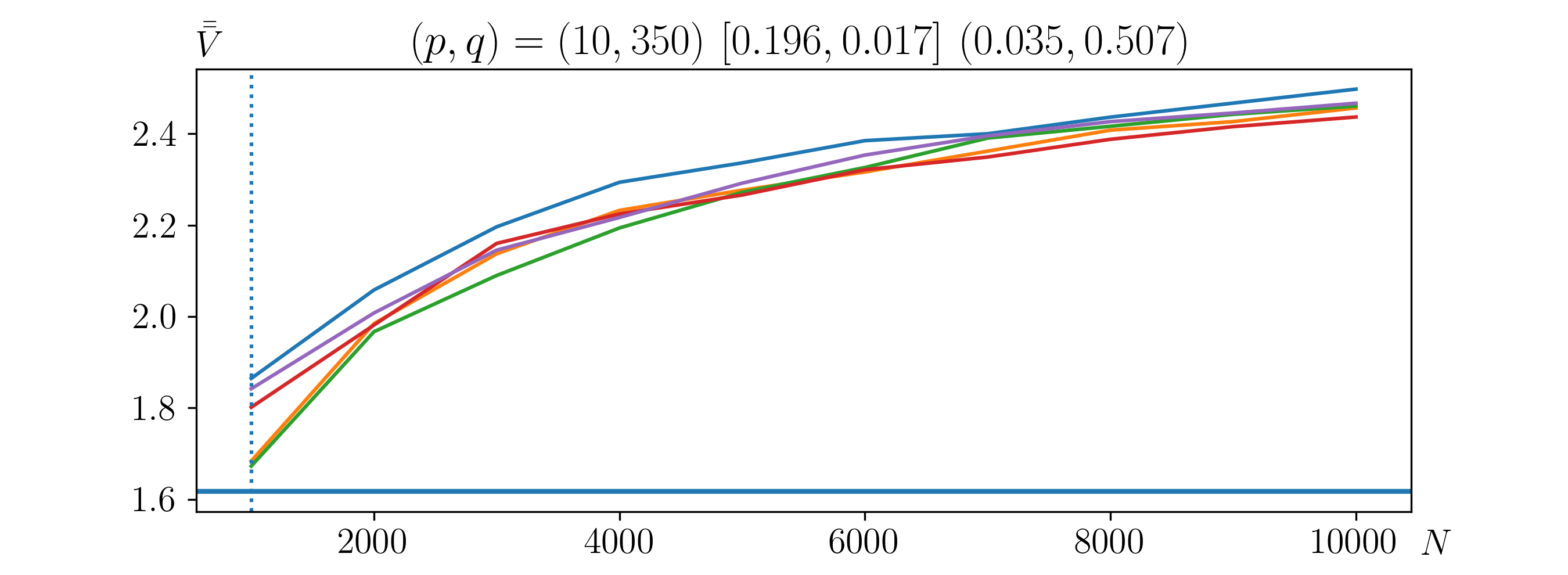} 
  \includegraphics[width=\textwidth]{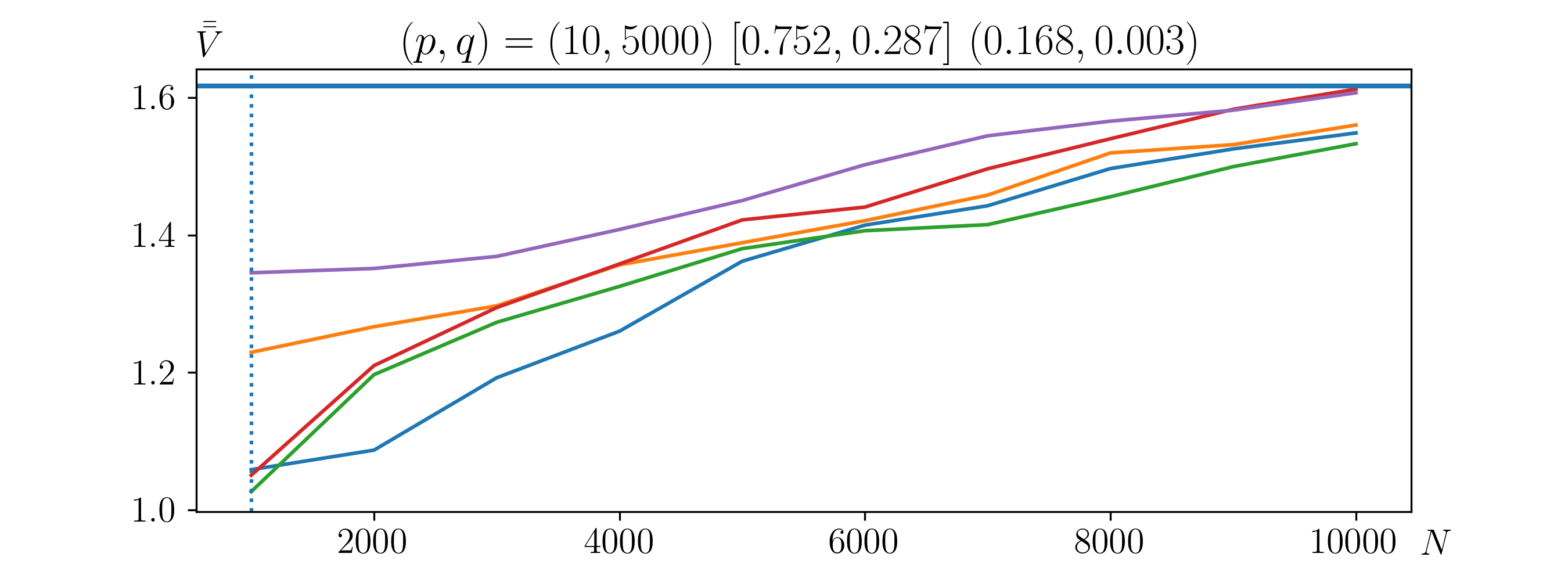} 

  \captionsetup{font={footnotesize}} 
  \caption{ 
    \label{aao_estimates_volatility_nnm} 
    Convergence of the relaxed dual value 
    for different random seed. 
    The sample size $N$ in the projected Rogers operator, 
    and 
    the nearest-neighbor basis size $(p,q)$ 
    are as indicated. 
    The energy of the hyperparameters 
    is indicated in square bracket. 
    The first number in bracket is 
    the energy with a sample size of $N=1,000$, 
    and an energy validation sample of $N'=10,000$. 
    The second number in bracket is 
    the energy with a sample size of $N=10,000$, 
    and an energy validation sample of $N'=100,000$. 
    The finite difference price is the horizontal line. 
    The best relative difference 
    between the relaxed dual value 
    and the finite difference price 
    is indicated in parenthesis. 
    The first number in parenthesis is 
    the best difference 
    for a sample size of $N=1,000$, 
    and the second number is for a sample size of $N=10,000$. 
    The vertical line indicates the $N=1,000$ sample size level. 
    The option can be exercised 50 times per year, 
    the strike price is $40$, 
    the risk-free rate is $0.06$, 
    the volatility is $0.20$, 
    the maturity is one year, 
    and the initial stock price is $42$. 
  } 

\end{figure}

Figure \ref{aao_estimates_volatility_nnm} 
shows the convergence of NNM 
for various random seeds 
and for various hyperparameters. 
A first observation from this figure 
is that when the filtration energy is too low, 
the relaxed dual value is far above the dual value. 
From the previous section, 
we can infer that this is due to the projected martingale 
quality 
that is not sufficient 
to capture the optimal martingale behavior. 
A second observation is 
that when the nearest-neighbor size is large 
such as $(p,q)=(10, 5000)$, 
an equally large sample size $N$ 
is needed for the convergence of the relaxed dual value. 
This observation can be contrasted 
with the observation that 
a good accuracy can be obtained with a small sample size $N=1,000$ 
when the filtration energy is adequate. 

\begin{figure}[ht] 

  \includegraphics[width=\textwidth]{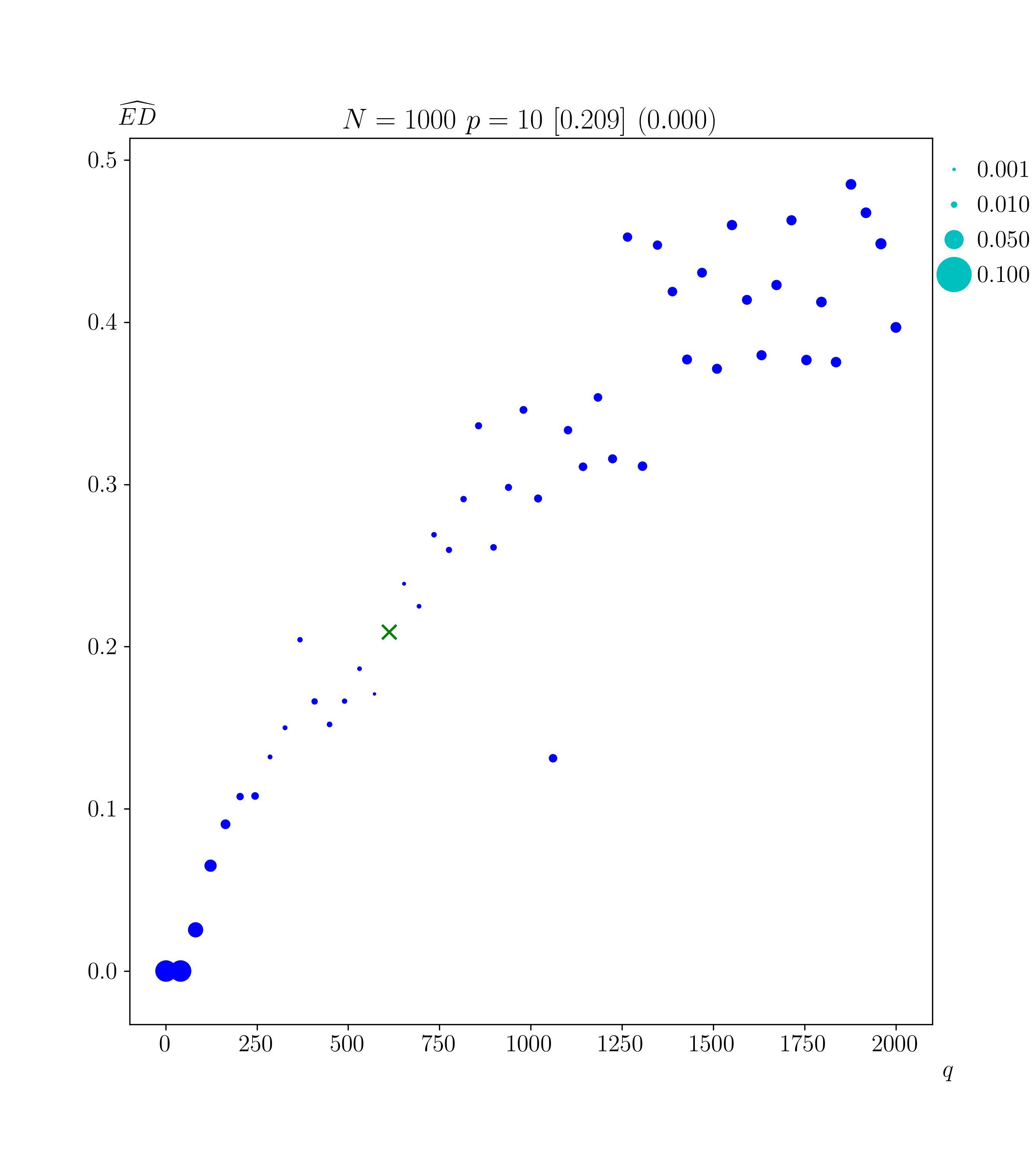} 

  \captionsetup{font={footnotesize}} 
  \caption{ 
    \label{aao_correct_convergence_plane_nnm} 
    Convergence of NNM 
    in the nearest-neighbor size 
    and energy plane. 
    The sample size $N$ in the projected Rogers operator, 
    and 
    the conditioning nearest-neighbor size $p$ 
    are as indicated. 
    The x-axis is the next nearest-neighbor size $q$, 
    and 
    the y-axis is the filtration energy $\widehat{ED}$. 
    The energy validation sample is of size $N'=10,000$. 
    The point size gradient 
    indicates the relative difference 
    between the relaxed dual value 
    and the finite difference price. 
    The best relative difference 
    is indicated by a cross, 
    and, in the title, 
    this best relative difference is indicated in parenthesis, 
    along with the corresponding filtration energy in bracket. 
    The NNM relaxed dual values 
    are obtained 
    by taking the best relative difference of 
    five different random seeds. 
    The option can be exercised 50 times per year, 
    the strike price is $40$, 
    the risk-free rate is $0.06$, 
    the volatility is $0.20$, 
    the maturity is one year, 
    and the initial stock price is $42$. 
  } 

\end{figure}

Figure \ref{aao_estimates_volatility_nnm} is informative for hyperparameters tuning, 
but the figure misses the key convergence behavior of NNM. 
Indeed, the figure approaches NNM as a probabilistic algorithm 
and the figure may give the impression 
that NNM does not converge with the sample size $N$. 
However, unlike SPLS, 
NNM is not a probabilistic algorithm, 
even if the convergence proof of NNM rests on probabilistic arguments. 
The Rogers operator is deterministic 
and the adequate perspective 
to study the convergence of NNM 
is by fixing the sample size $N$. 
From this perspective, 
the correct plane to analyse the convergence 
of NNM is not the sample size and dual value plane $(N,V)$. 
Instead, fix the sample size $N$ 
and look at the nearest-neighbor size, energy and dual plane 
$((p,q), \widehat{ED}, V)$. 
A pattern then emerges 
and shows that the convergence of NNM rests mainly on 
the filtration energy. 
Figure \ref{aao_correct_convergence_plane_nnm} 
displays such a plane. 

  \clearpage 

\section{A Swing Option} \label{aso} 

This section applies SPLS and NNM to 
a simple multiple exercise American option. 
The option is an American call 
on the price of electricity, 
with a constraint on the total number of exercises. 
This example is taken from \textcite{Meinshausen04}. 

Consider a multiple exercise American call option 
on the electricity price $S$ with a strike price of zero. 
The option can be exercised at most on $n$ days 
till the maturity of 50 days. 
With $X$ the exercise decision, 
the stochastic program for the option 
can be written with 
\begin{maxi!}[0] 
{X}
{ \Ebb{ \sum_{t=0}^{50} S_t X_t }  \tag{\theparentequation}\label{aso_aso}}
{\renewcommand{\theequation}{\theparentequation.\arabic{equation}} \label{}}
{  }
\addConstraint{ \sum_{t=0}^{50} X_t }{ \le  n \quad \label{aso_aso_cst}}{}
\end{maxi!} 
Assume further that the risk-neutral dynamics of the electricity price 
is given by the following AR(1) model 
\begin{align*} 
	\log S_t = 0.1 \log S_{t-1} + \frac{1}{2} W_t, \stepcounter{equation}\tag{\theequation}\label{aso_aso_ar}
\end{align*} 
with $S_{0} = 1$ and $W$ a standard Brownian process. 

In \textcite{Meinshausen04}, 
the lower bound 
is obtained with 
the least-square Monte Carlo method of \textcite{longstaff01}, 
and 
the upper bound is obtained 
by specializing the 
dual approach of \textcite{rogers02} and \textcite{haugh04} 
to the option. 
The approach of \textcite{Meinshausen04} 
enables the pricing 
of this example through a theoretical extension 
of single exercise methods to the multiple exercise case. 
In contrast, SPLS and NNM are fully general algorithms 
and can be applied as is to this example. 
To further show the versatility of our algorithms, 
we will use a different hyperparameters tuning method 
than previously presented. 

For SPLS, we tune the hyperparameters as follows. 
Fix a sample size $N$ for the lookahead operator. 
Then, increase the nearest-neighbor size $m$ 
until the projected dual value doesn't increase. 
This tuning approach 
aims at finding a better lower bound by increasing solely the projected strategy quality. 
For NNM, we proceed as follows. 
Fix a conditioning nearest-neighbor size $p$, 
and increase the next nearest-neighbor size $q$ 
until the first drop in relaxed dual value is recorded. 
When this drop is recorded, 
if the relaxed dual value is lower then the SPLS lower estimate, 
increase the sample size $N$, 
and repeat the procedure. 
This approach aims at finding the typing point 
in the accuracy of NMM, as suggested by 
Figure \ref{aao_correct_convergence_plane_nnm}. 
We run this tuning procedure once, 
for the case when only a single exercise is allowed. 
Then, we keep the same hyperparameters for 
cases with multiple exercises. 
Table \ref{aso} shows the results of this approach. 

\sisetup{zero-decimal-to-integer} 

\begin{table}[h] 

\singlespacing 
\tabcolsep=0.11cm 
\centering 

\begin{tabular}{ 
  >{\centering}p{0.1\textwidth} 
  >{\centering}p{0.25\textwidth} 
  >{\centering}p{0.25\textwidth} 
  >{\centering \arraybackslash}p{0.25\textwidth} 
  } 

  \toprule 

  $n$ & Least-Square Monte Carlo & SPLS & NNM \\ 

  \midrule 

  1 & 
  2.750 & 
  2.748 (-0.00) & 
  2.818 (0.02) 
  \\ 

  2 & 
  5.156 & 
  5.157 (0.00) & 
  5.176 (0.00) 
  \\ 

  3 & 
  7.306 & 
  7.305 (-0.00) & 
  7.301 (-0.00) 
  \\ 

  4 & 
  9.336 & 
  9.348 (0.00) & 
  9.277 (-0.01) 
  \\ 

  5 & 
  11.230 & 
  11.255 (0.00) & 
  11.121 (-0.01) 
  \\ 

  10 & 
  19.556 & 
  19.704 (0.01) & 
  19.183 (-0.02) 
  \\ 

  15 & 
  26.488 & 
  26.860 (0.01) & 
  25.892 (-0.02) 
  \\ 

  20 & 
  32.326 & 
  33.044 (0.02) & 
  31.763 (-0.02) 
  \\ 

  25 & 
  37.697 & 
  38.458 (0.02) & 
  37.007 (-0.02) 
  \\ 

  30 & 
  42.420 & 
  43.289 (0.02) & 
  41.829 (-0.01) 
  \\ 

  35 & 
  46.337 & 
  47.627 (0.03) & 
  46.118 (-0.00) 
  \\ 

  40 & 
  50.299 & 
  51.435 (0.02) & 
  50.051 (-0.00) 
  \\ 

  45 & 
  53.335 & 
  54.755 (0.03) & 
  53.439 (0.00) 
  \\ 

  50 & 
  56.765 & 
  57.439 (0.01) & 
  55.898 (-0.02) 
  \\ 

  \bottomrule 

\end{tabular} 

\captionsetup{font={footnotesize}} 
\caption{ 
  \label{aso_table_aso} 
	Comparison of SPLS and NNM 
	with a least-square Monte Carlo method 
	for a swing option 
	with a constraint on the total number of exercises. 
	The option is a zero-strike American call 
	on the electricity price 
	that can be exercised at most $n$ times till the maturity of 50 days. 
	The least-square Monte Carlo price is obtained 
	with the method of \textcite{Meinshausen04} 
  and $1,000$ simulations. 
  The relative difference to the least-square Monte Carlo price 
  is reported in parenthesis. 
  SPLS uses 
  $\widetilde{N}=1,000$ simulations, 
  a sample size of $N=100$, 
  a nearest-neighbor size of $m=250$, 
  and an energy validation sample of size $N'=1,000$. 
  NNM uses 
  $N=5,000$ simulations, 
  a nearest-neighbor size of $(p,q)=(5, 60)$, 
  and an energy validation sample of size $N'=10,000$. 
  The energy for all SPLS estimates is on average $0.666$, 
  and the energy for all NNM estimates is on average $0.007$ 
} 

\end{table} 

Table \ref{aso} 
shows that the SPLS and NNM estimates 
are very accurate, 
despite the simple hyperparameters tuning procedure used. 
This shows 
that adapting SPLS and NNM 
to a particular option 
rests solely on finding appropriate hyperparameters. 
Further, hyperparameters tuning can be done in many ways 
allowing to meet any computational, accuracy or speed constraints. 

\clearpage 

\section{An Asian Option} \label{aso_an_asian_option} 

This section applies SPLS and NNM to 
a multiple exercise constrained window Asian option 
driven by 
a Levy diffusion. 
The multiple exercise right 
is subject to a local limit, a global limit, 
and a refraction period. 
An option of this complexity has never been priced 
in the literature. 

Consider a multiple exercise Asian call option 
on the energy price $S$ with a strike price of $K$. 
The option can be exercised $T$ times per year, 
up to the maturity of one year. 
The average of the stock price is calculated over the 
last five periods. 
The option can be exercised at most $n$ times, 
and between each exercise a minimum waiting time of $R$ period is required. 
At each exercise, the option delivers 
at most $q$ of the payoff, 
and the total amount delivered cannot exceed $Q$. 
With $X$ the exercise decision, 
and 
$Y$ the ordered quantity of energy, 
the stochastic program for the option 
can be written with 
\begin{maxi!}[0] 
{X,Y}
{ \Ebb{ \sum_{t=4}^{T} e^{-r t} \pp{A_t - K} Y_t X_t }  \tag{\theparentequation}\label{ano_ano}}
{\renewcommand{\theequation}{\theparentequation.\arabic{equation}} \label{}}
{  }
\addConstraint{ Y_t }{\in [0, q]  \quad \label{ano_ano_0}}{ t=0,1, \ldots ,50}
\addConstraint{ \sum_{t=0}^T Y_t }{< [0, Q] \quad \label{ano_ano_1}}{}
\addConstraint{ \sum_{s=0}^R X_{t+s} }{ \le  1  \quad \label{ano_ano_2}}{ t=0,1, \ldots ,50}
\addConstraint{ \sum_{t=0}^T X_t }{ \le  n \quad \label{ano_ano_3}}{}
\addConstraint{ X }{\in \pe{0,1} \quad \label{ano_ano_4}}{}
\end{maxi!} 
where $A_t = \frac15 \sum_{s=t-4}^t S_s$, 
and $r$ is the interest rate. 
The stochastic dynamic for the energy price 
follows the diffusion equation 
\begin{align*} 
	dS = \theta_{} (\mu - S) dt + \sigma dW + dJ, \stepcounter{equation}\tag{\theequation}\label{ano_sde}
\end{align*} 
where 
$\theta_{}$ is the reversion speed, 
$\mu$ the long-term mean, 
$\sigma$ the price volatility, 
$W$ is a standard Brownian motion, 
and 
$J$ is an independent Poisson process with 
exponential jump. 
The L\'{e}vy measure of the compound Poisson process is 
$\nu(dj) = \lambda \alpha e^{-\alpha j}$. 

The stochastic program of this option 
is non-linear 
as the objective contains a product of decision variables. 
However, the program can be linearized 
by defining a third decision variable, 
and adding constraints to force the third variable 
to be equal to the product of the variables. 
The procedure is as follows. 
Substitute every occurrence of the product $y x$ 
by the variable $z$, and add the following three constraints 
\begin{align*} 
  z & \le  y, \\
  z & \le  a x, \\
  z & \ge  y + a x - a,
\end{align*} 
where $a$ is an upper bound on $y$, 
$x$ is binary and $y$ is nonnegative. 
This and other \textit{integer programming tricks} 
can be found in \textcite{AimsBook}. 
Such tricks allow 
SPLS and NNM to handle complex option 
with a linear solver. 

Table \ref{ano_option_price} 
displays the SPLS and NMM 
estimates for various pricing parameters. 
The estimates are obtained with a bootstrap procedure only: 
the hyperparameters of the algorithms are fixed, 
only multiple random starts are used, 
and no grid search is made. 
If the option price is taken as the midpoint between SPLS and NNM, 
the table shows that the estimates narrow the option price within 5\% in average. 
As we used a simple hyperparameters tuning method, 
some of the estimates are less precise. 
The convergence theory of the algorithms 
and the previous examples 
guarantee that any less precise estimate 
can be made sharper with more computational effort. 

\sisetup{zero-decimal-to-integer} 

\begin{table}[h] 

\singlespacing 
\tabcolsep=0.11cm 
\centering 

\begin{tabular}{ 
	c 
	c 
  c@{\hskip 2em} 
  c 
  c 
  c@{\hskip 2em} 
  c 
  c 
  c 
  c 
} 

	\toprule 

	&&& \multicolumn{3}{c}{\hspace{-2em}$\lambda=0$} & \multicolumn{3}{c}{$\lambda=0.1$} \tabularnewline 

	$n$ & $R$ & $\sigma$ & SPLS && NNM & SPLS && NNM \tabularnewline 

	\midrule 

	1 & 
	0 & 
	5 & 

	1.09 & 
	(0.11) & 
	1.21 & 

	1.14 & 
	(0.30) & 
	1.47 

	\tabularnewline 

	 & 
	 & 
	10 & 

	2.20 & 
	(0.21) & 
	2.66 & 

	2.25 & 
	(0.24) & 
	2.78 

	\tabularnewline 

	 & 
	2 & 
	5 & 

	1.10 & 
	(0.62) & 
	1.78 & 

	1.13 & 
	(0.04) & 
	1.17 

	\tabularnewline 

	 & 
	 & 
	10 & 

	2.22 & 
	(0.18) & 
	2.62 & 

	2.22 & 
	(0.55) & 
	3.43 

	\tabularnewline 

	\hline 

	5 & 
	0 & 
	5 & 

	9.74 & 
	(0.06) & 
	10.30 & 

	10.11 & 
	(0.11) & 
	11.27 

	\tabularnewline 

	 & 
	 & 
	10 & 

	19.48 & 
	(0.09) & 
	21.17 & 

	19.72 & 
	(0.07) & 
	21.07 

	\tabularnewline 

	 & 
	2 & 
	5 & 

	9.13 & 
	(0.14) & 
	10.40 & 

	9.43 & 
	(0.03) & 
	9.68 

	\tabularnewline 

	 & 
	 & 
	10 & 

	18.21 & 
	(0.09) & 
	19.85 & 

	18.59 & 
	(0.08) & 
	20.14 

	\tabularnewline 

	\hline 

	10 & 
	0 & 
	5 & 

	19.60 & 
	(0.10) & 
	21.65 & 

	20.39 & 
	(0.09) & 
	22.19 

	\tabularnewline 

	 & 
	 & 
	10 & 

	39.29 & 
	(0.06) & 
	41.77 & 

	40.03 & 
	(0.02) & 
	41.01 

	\tabularnewline 

	 & 
	2 & 
	5 & 

	16.67 & 
	(0.02) & 
	17.04 & 

	17.24 & 
	(0.07) & 
	18.48 

	\tabularnewline 

	 & 
	 & 
	10 & 

	33.38 & 
	(0.04) & 
	34.66 & 

	33.85 & 
	(0.03) & 
	35.00 

	\tabularnewline 

	\hline 

	20 & 
	0 & 
	5 & 

	36.51 & 
	(0.05) & 
	38.18 & 

	37.80 & 
	(0.28) & 
	48.48 

	\tabularnewline 

	 & 
	 & 
	10 & 

	72.99 & 
	(0.05) & 
	76.89 & 

	74.23 & 
	(0.06) & 
	78.49 

	\tabularnewline 

	 & 
	2 & 
	5 & 

	21.95 & 
	(0.07) & 
	23.51 & 

	22.73 & 
	(0.11) & 
	25.16 

	\tabularnewline 

	 & 
	 & 
	10 & 

	43.89 & 
	(0.04) & 
	45.64 & 

	44.54 & 
	(0.12) & 
	50.01 

	\tabularnewline 

	\hline 

	25 & 
	0 & 
	5 & 

	43.63 & 
	(0.08) & 
	47.00 & 

	45.23 & 
	(0.03) & 
	46.41 

	\tabularnewline 

	 & 
	 & 
	10 & 

	87.28 & 
	(0.10) & 
	95.78 & 

	88.54 & 
	(0.03) & 
	91.64 

	\tabularnewline 

	 & 
	2 & 
	5 & 

	21.95 & 
	(0.13) & 
	24.90 & 

	22.73 & 
	(0.11) & 
	25.13 

	\tabularnewline 

	 & 
	 & 
	10 & 

	43.90 & 
	(0.05) & 
	46.05 & 

	44.54 & 
	(0.10) & 
	48.80 

	\tabularnewline 

	\bottomrule 

\end{tabular} 

\captionsetup{font={footnotesize}} 
\caption{ 
	\label{ano_option_price} 
	SPLS and NNM 
	estimates 
	for 
	a multiple exercise constrained window Asian option 
	driven by 
	a Levy diffusion. 
	The option can be exercised 50 times per year, 
	with a maturity of one year. 
	The maximum number of exercise $n$, 
	the refraction period $R$, 
	the volatility $\sigma$, 
	and the jump intensity $\lambda$ is as indicated. 
	The risk-free rate is 0.06, 
	the reversion speed $\theta_{}$ is 0.02, 
	and 
	the jump severity is exponentially distributed 
	with rate $\alpha=0.5$, 
	The strike price $K$, 
	the initial stock price $S_{0}$, 
	and 
	the long-term mean $\mu$ have the same value of 36. 
	The global volume limit 
	is $Q=n-0.5$. 
	The relative difference 
	between the NNM and SPLS estimates is in parenthesis. 
	SPLS uses 
	$\widetilde{N}=1,000$ simulations, 
	a sample size in the projected lookhead operator of $N=100$, 
	a nearest-neighbor size of $m=50$, 
	and 
	an energy validation sample of size $N'=1,000$. 
	NNM uses 
	$N=1,500$ simulations, 
	a nearest-neighbor size of $(p,q)=(10, 60)$, 
	an energy validation sample of size $N'=10,000$, 
	and 
	a lower bound barrier of $1.02$. 
	The SPLS estimate is the best estimate 
	from 10 different random seeds, 
	while the NNM estimate 
	is the best estimate from 30 different random seeds. 
	The filtration energy is $0$ for SPLS, 
	and $0.02$ for NNM. 
} 

\end{table} 

\clearpage 

\section{A Passport Option} \label{apo} 

This section applies SPLS and NNM 
to a passport option on three assets. 
The passport option is stylized 
so that 
it includes 
all the exotic rights presented in 
\textcite{penaud99exo}, 
for a total of 8 rights. 
In \textcite{penaud99exo}, 
these features were priced one at a time. 
The complexity of this example 
is beyond the available algorithm in the literature. 

Consider a passport option 
on three stocks $S^i,i=1,2,3$. 
The option can be exercised 12 times per year, 
up to the maturity of one year. 
At each time $t$, 
the investor can trade an amount $Y^i_t \in [-Q,Q]$ 
in only one of the stock $i$. 
The number of switch 
from a position in one stock 
to a position in another stock is limited to $n$. 
The investor is allowed once 
to change the position limit to $2 Q$ 
for a duration of $L$ periods. 
The change in position value is limited to $|Y^i_t-Y^i_{t+1}|<q$, 
so that change in position are smooth. 
The investor is allowed 3 exceptions 
to this smooth change constraint. 
Denote the trading account value by $A_t$. 
After 6 months, 
the investor needs to choose 
between a payoff at maturity of 
$\max(A_T,0)$ 
or 
$\max(-A_T,0)$. 
If the trading account ever reaches the barrier value of $B$, 
the investor receives $B$ at maturity 
and the option expires. 
The investor is allowed 
to reset the account value to zero $3$ times. 
Finally, 
the investor is allowed once 
to mark the account value, 
and then to restore the account to the marked value at a later time. 

To write the stochastic program for the option, 
let $X$ be the exercise decision, 
$Y$ the position amount, 
$PL$ the position limit extension right, 
$SL$ the smooth limit extension right, 
$P_{long}$ the long or short decision, 
$R$ the reset right, 
$MP$ the magic potion mark right, 
and 
$MPX$ the magic potion restore right. 
Also, for a binary decision $Z$, let $\overline{Z}=1-Z$ be the complement. 
The stochastic program for the option can then be written with 

{ 
\singlespacing 
\begin{maxi!}[2] 
{X,Y, PL, SL, P_{long}}
{ \Ebb{ e^{-r T} \bigg( \overline{P_B} P_{long} \max(A_T,0) + \overline{P_B} \, \overline{P_{long}} \max(-A_T,0) + P_B B \bigg)}  \tag{\theparentequation}\label{apo_apo}}
{\renewcommand{\theequation}{\theparentequation.\arabic{equation}} \label{}}
{  }
\addConstraint{\text{ Trading volume I } \nonumber}{  }{}
\addConstraint{ Y^i_t  \nonumber}{\in [-Q - Q  U_t, Q + Q U_t ]   }{ i=1,2,3 \,\, t=0,1, \ldots ,T-1}
\addConstraint{ U_t  \nonumber}{= \sum_{s=t-L}^t P\!L_s   }{ t=0,1, \ldots ,T-1}
\addConstraint{ \sum_{t=0}^{T-1} P\!L_t  \nonumber}{ \le  1  }{}
\addConstraint{\text{ Trading volume II } \nonumber}{  }{}
\addConstraint{ \pa{Y^i_t - Y^i_{t-1}}  \nonumber}{ \le  q + 4 Q S\!L_t   }{ i=1,2,3 \,\, t=1,2, \ldots ,T-1}
\addConstraint{ \sum_{t=1}^{T-1} S\!L_t  \nonumber}{ \le  3  }{}
\addConstraint{\text{ Trading timing } \nonumber}{  }{}
\addConstraint{ \sum_{i=1,2,3} X^i_t  \nonumber}{ \le  1   }{ t=0,1, \ldots ,T-1}
\addConstraint{ \sum_{t=1}^{T-1} \sum_{i=1,2,3} \pa{X^i_t - X^i_{t-1}}  \nonumber}{ \le  n  }{}
\addConstraint{\text{ Barrier } \nonumber}{  }{}
\addConstraint{ P_B  \nonumber}{= \bigvee_{t=1}^{T} \ib{|A_t|  \ge  B}  }{}
\addConstraint{\text{ Trading account } \nonumber}{  }{}
\addConstraint{ A_t  \nonumber}{= \overline{R_t} \, \overline{M\!P\!X_t} \, A^{a}_t + \overline{R_t} \, M\!P\!X_t \, A^{b}_t    }{ t=1,2, \ldots ,T}
\addConstraint{ A^{a}_t  \nonumber}{= A_{t-1} +\sum_{i=1,2,3} \p{S^i_t-S^i_{t-1}} Y^i_{t-1} X^i_{t-1}    }{ t=1,2, \ldots ,T}
\addConstraint{ A^{b}_t  \nonumber}{=  \sum_{s=0}^{t-1} M\!P_s A_s     }{ t=1,2, \ldots ,T}
\addConstraint{ A_{0}  \nonumber}{= 0  }{}
\addConstraint{ \sum_{t=0}^{T-1} R_t  \le  3\enskip ; \enskip \sum_{t=0}^{T-1} M\!P_t  \le  1\enskip ; \enskip \sum_{t=1}^{T} M\!P\!X_t  \le  1  \nonumber}{  }{}
\addConstraint{\text{ Binary Decisions } \nonumber}{  }{}
\addConstraint{ X^i_t  \nonumber}{\in \pe{0,1}   }{ i=1,2,3 \,\, t=0,1, \ldots ,T-1}
\addConstraint{ P\!L_t, S\!L_t, M\!P_t, M\!P\!X_t  \nonumber}{\in \pe{0,1}   }{ t=0,1, \ldots ,T-1}
\addConstraint{ P_{long}  \nonumber}{\in \pe{0,1}   }{ t=T/2 }
\end{maxi!} 
} 

The risk-neutral dynamic for the stock price 
follows the stochastic differential equation 
\begin{align*} 
	dS^i = r S^i dt + \sigma_i S^i dW^i, \stepcounter{equation}\tag{\theequation}\label{apo_sde}
\end{align*} 
where 
$r$ is the risk-free-rate, 
$\sigma_i$ is the volatility, 
and 
$W^i$ is an independent standard Brownian motion. 

Table \ref{apo_option_price} 
displays the SPLS and NMM 
estimates for various pricing parameters. 
The weighting stopping time used 
is the logical OR of every binary decision 
\eqref{cmeao_form_logical_or}. 
This weighting stopping time at time $t$ can be written with 
\begin{align*} 
    \p{\bigvee_{i=1}^3 X^i_t} \bigvee P\!L_t \bigvee S\!L_t \bigvee M\!P_t \bigvee M\!P\!X_t \bigvee \ib{Plong, t=T/2}.
\end{align*} 
The runtime of one SPLS estimate with a single processor 
is 10 hours, we use 100 processors to bring the runtime down to 5 minutes. 
For NNM, the runtime of one estimate with a single processor 
is 5 hours, we use 100 processors to bring the runtime down to 3 minutes. 
With 100 processors, the runtime of the entire table is 20 hours. 
To maintain the total runtime small, we opt for coarse estimates of the option price. 
The previous sections guarantee that 
with a greater computational effort very precise estimates of the option price can be obtained. 
For example, with $1,000$ processors, the runtime of the entire table would be 2 hours, 
and fine-grain estimates can be obtained. 

\sisetup{zero-decimal-to-integer} 

\begin{table}[h] 

\singlespacing 
\tabcolsep=0.11cm 
\centering 

\begin{tabular}{ 
	c 
	c 
  c@{\hskip 2em} 
  c@{\hskip 2em} 
  c@{\hskip 2em} 
  c@{\hskip 2em} 
	c 
} 

	\toprule 

	&&& 
	\multicolumn{2}{c}{\hspace{-2em}$B=\infty$} & 
	\multicolumn{2}{c}{$B=20$} 
	\tabularnewline 

	$n$ & $Q$ & $q$ & 
	$L=0$ & 
	$L=2$ & 
	$L=0$ & 
	$L=2$ 
	\tabularnewline 

	\midrule 

	1 & 
	1 & 
	1 & 

	4.05 (0.04) 4.21 & 

	4.70 (0.37) 6.44 & 

  3.44 (0.12) 3.85 & 

  3.23 (0.45) 4.70 

	\tabularnewline 

	 & 
	 & 
	2 & 

	5.28 (0.21) 6.37 & 

	5.68 (0.86) 10.56 & 

  3.90 (0.08) 4.21 & 

  4.15 (0.07) 4.46 

	\tabularnewline 

	 & 
	10 & 
	1 & 

	41.60 (0.03) 42.94 & 

	44.48 (0.06) 47.34 & 

  11.46 (0.22) 13.93 & 

  12.15 (0.56) 19.01 

	\tabularnewline 

	 & 
	 & 
	2 & 

	44.60 (0.08) 48.34 & 

	44.76 (0.06) 47.63 & 

  11.70 (0.61) 18.84 & 

  10.97 (0.08) 11.89 

	\tabularnewline 

	\hline 

	5 & 
	1 & 
	1 & 

	4.46 (0.06) 4.72 & 

	4.11 (0.35) 5.56 & 

  3.78 (1.97) 11.22 & 

  3.77 (1.89) 10.88 

	\tabularnewline 

	 & 
	 & 
	2 & 

	5.35 (0.78) 9.50 & 

	5.68 (0.02) 5.80 & 

  4.30 (0.25) 5.36 & 

  4.28 (0.42) 6.07 

	\tabularnewline 

	 & 
	10 & 
	1 & 

	31.59 (0.43) 45.06 & 

	32.62 (0.61) 52.67 & 

  17.03 (1.34) 39.89 & 

  18.09 (0.53) 27.72 

	\tabularnewline 

	 & 
	 & 
	2 & 

	36.55 (0.24) 45.27 & 

	36.56 (0.39) 50.91 & 

  16.51 (0.33) 21.88 & 

  17.67 (0.03) 18.19 

	\tabularnewline 

	\bottomrule 

\end{tabular} 

\captionsetup{font={footnotesize}} 
\caption{ 
	\label{apo_option_price} 
	SPLS and NNM estimates for 
	an exotic passport option. 
	The option can be exercised 12 times per year, 
	up to the maturity of one year. 
	The number of switches 
	from a position in one stock 
	to a position in another stock is limited to $n$. 
	The position limit is $Q$, 
	and 
	the maximal change in position is $q$. 
	The position limit can be changed once to $2 Q$ 
	for a duration of $L$ periods. 
	If the option reaches the barrier value of $B$, 
	the investor receives $B$ and the option expires. 
	The SPLS estimate is on the left of the parenthesis, 
	the NNM estimate is on the right of the parenthesis, 
	and 
	the relative difference 
	between the NNM and SPLS estimates is in parenthesis. 
	The volatility is 
	$(\sigma_1, \sigma_2, \sigma_3)=(0.05, 0.25, 0.5)$, 
	the initial stock price is 
	$S^1_0=S^2_0=S^3_0=36$, 
	and the risk-free-rate is $0.06$. 
	SPLS uses 
	$\widetilde{N}=1,000$ simulations, 
	a sample size in the projected lookhead operator of $N=1$, 
	a nearest-neighbor size of $m=1$, 
	and 
	an energy validation sample of size $N'=1,000$. 
	NNM uses 
	$N=1,000$ simulations, 
	a nearest-neighbor size of $(p,q)=(5, 10)$, 
	an energy validation sample of size $N'=10,000$, 
	and 
	a lower bound barrier of $1.02$. 
	Both the SPLS and NNM estimates 
	are the best estimates 
	from 5 different random seeds. 
	The filtration energy is $0$ for both SPLS and NNM. 
} 

\end{table} 

\clearpage 

\section{Conclusion} \label{conclu} 

This article presents an alternative 
to dynamic programming 
for the pricing 
of multiple exercise American option 
with constrained rights. 
The algorithms provide lower and upper estimates 
of the option price 
with convergence guarantees provided 
through a Vapnik-Chernovenkis dimension analysis. 
The algorithms are fast, fully general, 
and applicable with no adjustment 
to a large class of options. 
We illustrate the algorithm 
with two realistic examples 
including 
a swing option with four constraints, 
and a passport option with 16 constraints. 

The ability to value constrained multiple exercise American derivatives 
has many important advantages. 
In particular, the exercise constraints 
are at the heart of the negotiations of such derivative contract. 
A general valuation approach greatly facilitates those negotiations. 
Further, such a valuation approach simplifies risk management 
and can broaden the hedging instruments offering for those derivatives. 
The presented algorithms 
provide such a general valuation approach. 

\section{Methods} \label{meth} 

This section collects 
the technical details behind the various results presented. 
The name of each subsection contains in parenthesis 
the equation number being proved. 
The last subsection 
details our implementation of SPLS and NNM. 
In this section, 
the Vapnik-Chernovenkis dimension is denoted by \textit{VC-dimension}, 
and 
the shattering coefficient of a class $\mathcal{A}$ 
with a sample of size $n$ 
is denoted by $s(\mathcal{A},n)$. 
The $\ell_1$-covering 
$\mathcal{N}(\epsilon,A)$ 
of a set $A$ 
is the cardinality 
of the smallest $\epsilon$-net necessary to cover the set 
with the norm being the $\ell_1$-norm 
divided by the dimension of the element in the set. 
In covering number estimates, 
we use a universal constant $c$ to collect 
every constants, so that $cx$ is equivalent to the big-O notation 
$\mathcal{O}(x)$. 

\subsection{Dual Problem Derivation (\ref{cmeao_dual})} \label{meth_dual_problem_derivation_(cmeao_dual)} 

Define the value function 
\begin{align*} 
  V_t = \max_{X_t, Y_t} \, e^{-r t} f_t(Y_{0:t}, S_{0:t}) X_t + \Ep{V_{t+1}|F_t},
\end{align*} 
where the maximum is taken over the admissible decisions $X_t$ and $Y_t$, 
and 
$V_{T+1}$ is assumed to be zero. 
Since the decision $X_t$ is in $\pe{0,1}$, the value function is a supermartingale 
and admits the following Doob decomposition 
\begin{align*} 
  V_t = V_0 + M_t - A_t,
\end{align*} 
where $M_t$ is a martingale vanishing at time zero, 
and $A_t$ is a previsible increasing process vanishing at time zero. 
Now, we have 
\begin{align*} 
V_0
&= \max_{X, Y} \Ebb{ \sum_{t=0}^{T} e^{-r t} f_t(Y_{0:t}, S_{0:t}) X_t } \\
& = \max_{X, Y} \Ebb{ \sum_{t=0}^{T} e^{-r t} f_t X_t - M_t X_t } \stepcounter{equation}\tag{\theequation}\label{meth_mgal_diff} \\
&  \le  \Ebb{ \max_{x, y} \sum_{t=0}^{T} e^{-r t} f_t x_t - M_t x_t } \stepcounter{equation}\tag{\theequation}\label{meth_max_in} \\
&  \le  \Ebb{ \max_{x, y} \sum_{t=0}^{T} V_t x_t - M_t x_t } \stepcounter{equation}\tag{\theequation}\label{meth_v_in} \\
& = \Ebb{ \max_{x, y} \sum_{t=0}^{T} (V_t - M_t) x_t } \stepcounter{equation}\tag{\theequation}\label{meth_rewrite} \\
& = \Ebb{ \max_{x, y} \sum_{t=0}^{T} (V_0 - A_t) x_t} \stepcounter{equation}\tag{\theequation}\label{meth_compensator} \\
& = V_0  \stepcounter{equation}\tag{\theequation}\label{meth_zero}
\end{align*} 
\eqref{meth_mgal_diff} holds by the optional sampling theorem. 
\eqref{meth_max_in} inserts the maximum into the expectation. 
\eqref{meth_v_in} holds by definition of the value function 
and by noting that $a<b$ implies $a y <b y$, 
whenever $y$ is in $\pe{0,1}$. 
\eqref{meth_rewrite} is a simple rewrite. 
\eqref{meth_compensator} uses the Doob decomposition of the value function. 
\eqref{meth_zero} 
holds by monotonicity of the compensator $A$, 
and 
assumes that the decision $x_0$ at time zero is always admissible. 

The dual formulation \eqref{cmeao_dual} follows by 
using \eqref{meth_max_in} and 
observing that 
\begin{align*} 
V_0
 \le  \min_M \Ebb{ \max_{x, y} \sum_{t=0}^{T} e^{-r t} f_t x_t - M_t x_t}
 \le  \Ebb{ \max_{x, y} \sum_{t=0}^{T} e^{-r t} f_t x_t - M_t x_t}
= V_0,
\end{align*} 
where the last equality holds 
when $M$ is taken as the martingale part of the value function. 

\subsection{SPLS Projected Strategy Convergence (\ref{spls_projected_strategy_rate})} \label{meth_spls_projected_strategy_convergence_(spls_projected_strategy_rate)} 

We have 
\begin{align*} 
& P\bigg( \sup_A \frac{1}{m_{s-t}} \sum_{i=1}^{m_{s-t}} \ib{\bar{B}_i \in A} > \epsilon \bigg)  \\
& =
\Ebb{ \Pc{ \sup_A \frac{1}{m_{s-t}} \sum_{i=1}^{m_{s-t}} \ib{\bar{B}_i \in A} > \epsilon }{  \{\bar{B}_i\}_{i=1}^{m_{s-t}} } } \stepcounter{equation}\tag{\theequation}\label{meth_conditioning_spsc} \\
&  \le
s(\mathcal{A}, m_{s-t})
\Ebb{ \sup_A \Pc{ \frac{1}{m_{s-t}} \sum_{i=1}^{m_{s-t}} \ib{\bar{B}_i \in A} > \epsilon }{ \{\bar{B}_i\}_{i=1}^{m_{s-t}} } } \stepcounter{equation}\tag{\theequation}\label{meth_shattering_spsc} \\
&  \le
2 s(\mathcal{A}, m_{s-t}) e^{- 2 \epsilon^2 m_{s-t}}. \stepcounter{equation}\tag{\theequation}\label{meth_hoeffding_spsc}
\end{align*} 
\eqref{meth_conditioning_spsc} is a simple conditioning on the Voronoi cell $\{\bar{B}_i\}_{i=1}^{m_{s-t}}$, 
\eqref{meth_shattering_spsc} holds by definition of the shattering coefficient $s(\mathcal{A},m_{s-t})$, 
and 
\eqref{meth_hoeffding_spsc} holds by Hoeffding's inequality. 
To obtain the shatter coefficient, 
note that 
\begin{align*} 
s(\mathcal{A}, m_{s-t})
& \le  \prod_{i=1}^{\alpha_{s-t}} s(2^{B_i}, m_{s-t}) \\
& \le  \prod_{i=1}^{\alpha_{s-t}} {m_{s-t}}^{ (d(s-t)+1) (\alpha_{s-t}-1)} \\
& \le  {m_{s-t}}^{3 d(s-t) \alpha^2_{s-t} }
\end{align*} 
as each set $2^{B_i}$ 
is the intersection of at most $\alpha_{s-t}-1$ hyperplanes, 
and the VC-dimension 
of each hyperplane is $d (s-t) + 1$. 
The factor of $3$ is used to bound the VC-dimension 
of each hyperplane so that the bound is greater than 2. 
See Theorem 
13.5, 13.9, 13.8 and 13.3 
in \textcite{devroye96}. 
The result follows. 

\subsection{SPLS Projected Lookahead Operator Convergence (\ref{spls_projected_lookahead_operator_rate})} \label{meth_spls_projected_lookahead_operator_convergence_(spls_projected_lookahead_operator_rate)} 

The shatter coefficient of the first component in $A$ can be bounded with 
\begin{align*} 
s(\prod_{s=t}^{T} \{\bar{B}^{s,i}\}_{i=1}^{m_{s-t}}, n)
&  \le  \prod_{s=t}^T \prod_{i=1}^{m_{s-t}} s(\bar{B}^{s,i}, n) \\
&  \le  \prod_{s=t}^T \prod_{i=1}^{m_{s-t}} n^{ (d(s-t)+1) (m_{s-t}-1) } \\
&  \le  n^{ d(T-t) \sum_{s=t}^T m^2_{s-t} }
\end{align*} 
as each set $\bar{B}^{s,i}$ 
is the intersection of at most $m_{s-t}-1$ hyperplanes, 
and the VC-dimension 
of each hyperplane is $d (s-t) + 1$. 
See Theorem 
13.5, 13.9 and 13.8 
in \textcite{devroye96}. 
By Theorem 13.8 in \textcite{devroye96}, 
the shatter coefficient for the second component in $A$ 
is $N^{2 d (T-t)}$, 
which we bound by $N^{3 d (T-t)}$ 
to make the VC-dimension greater then 2. 
The result then follows by Theorem 13.5, 13.3 and 12.5 in \textcite{devroye96}. 

\subsection{SPLS Projected Option Value (\ref{spls_projected_option_value_rate})} \label{meth_spls_projected_option_value_(spls_projected_option_value_rate)} 

In this proof, 
the projected lookahead operator 
$\bar{L}(\bar{X}_{t:T},F_t)$ 
at time $t$ 
is assumed to be a uniform Lipschitz function 
of the filtration, 
so that there exists a constant $C$ 
for which the operator 
is Lipschitz in the filtration 
for any strategy $\bar{X}_{t:T}$. 
This assumption implies 
that the operator 
can be viewed as strategy, 
with the internal structure of the operator abstracted. 
The proof hence proceed by 
considering the covering number 
induced by any strategy 
that is a Lipschitz function of the filtration. 

Denote by $J(X,S^{(\widetilde{N})})$ 
the vector $(J(X,S^{\widetilde{n}}),\widetilde{n}=1,2, \ldots ,\widetilde{N})$, 
we bound the $\ell_1$-covering 
$\mathcal{N}(\epsilon, J(X,S^{(\widetilde{N})}))$ 
with an $\ell_1$-covering of the strategy 
$\mathcal{N}(\epsilon, X)$. 
Let $X$ and $Y$ be two strategies, by the Lipschitz assumption, 
\begin{align*} 
\pn{J(X,S^{(\widetilde{N})}) - J(Y,S^{(\widetilde{N})})}
  &= \frac{1}{\widetilde{N}} \sum_{\widetilde{n}=1}^{\widetilde{N}} \pn{J(X,S^{\widetilde{n}}) - J(Y,S^{\widetilde{n}})} \\
  & \le  \frac{1}{\widetilde{N}} \sum_{\widetilde{n}=1}^{\widetilde{N}} K \pn{X - Y} \\
  &= K \pn{X - Y}.
\end{align*} 
Proving that $\mathcal{N}(\epsilon, J(X,S^{(\widetilde{N})}))  \le  \mathcal{N}(\frac{\epsilon}{K},X)$. 

With $F_t$ and $F^\prime_t$ two different filtrations, by the Lipschitz assumption, 
\begin{align*} 
  \pn{X(F) - X(F')}
    &= \frac{1}{T} \sum_{t=0}^T \pn{X_t(F_t) - X_t(F^\prime_t)} \\
    & \le  \frac{C}{T} \sum_{t=0}^T \pn{F_t - F^\prime_t},
\end{align*} 
This proves that 
\begin{align*} 
  \mathcal{N}(\frac{\epsilon}{K},X) & \le  \prod_{t=0}^T \mathcal{N}(\frac{\epsilon}{C K},F_t),
\end{align*} 

The minimum information necessary in the filtration 
are the past decision $X_{0:t-1}$ 
and the stock path history $S_{0:t}$. 
As each decision is a vector in $\R^a$, 
and each stock is a vector in $\R^d$, 
the filtration is at most a vector in $\R^{(a+d) t}$, giving that 
\begin{align*} 
  \mathcal{N}(\frac{\epsilon}{K},X)
    & \le  \prod_{t=0}^T \p{\frac{C K}{\epsilon}}^{(a+d) t} \\
    & \le  \p{\frac{C K}{\epsilon}}^{(a+d) T^2}.
\end{align*} 
The result then follows 
by Theorem 29.1 in \textcite{devroye96}. 
\subsection{SPLS Consistency (\ref{spls_consistency})} \label{meth_spls_consistency_(spls_consistency)} 

Consider the lookahead operator $L(X_{t:T}, F_t)$ at time $t$, 
and define the estimation error 
for a projected strategy with 
\begin{align*} 
  Z_t = \pa{ \frac{1}{N} \sum_{n=1}^N J(\bar{L}(\bar{X}_{t:T},S^n) -  \E{J(L(X_{t:T},F_t)} },
\end{align*} 
where $\bar{X}$ is the projection of $X$. 
Denote by $X \sim \bar{X}$, 
the event that the projected strategy 
has unambiguous cells. 
By using \eqref{spls_projected_strategy_rate}, 
this event can be written with 
\begin{align*} 
\p{X \sim \bar{X}} = \p{\sup_A \frac{1}{m_{s-t}} \sum_{i=1}^{m_{s-t}} \ib{\bar{B}_i \in A}  < \epsilon, t=0,1, \ldots ,T}.
\end{align*} 
By conditioning, the probability of error becomes 
\begin{align*} 
  P(Z_t>\epsilon)
    &= \Pc{Z_t>\epsilon}{X \sim \bar{X}} P(X \sim \bar{X})
      + \Pc{Z_t>\epsilon}{X \not \sim \bar{X} } P(X \not \sim \bar{X}).
\end{align*} 
By \eqref{spls_projected_strategy_rate}, 
the second term tends to zero when $m$ is large. 
For the first term, 
the conditioning event is assumed to imply that 
any strategy $X$ can be accurately projected 
to a strategy $\bar{X}$. 
The magnitude of the estimation error $Z_t$ is then due solely to 
a discrepancy in distribution. 
By using \eqref{spls_projected_lookahead_operator_rate}, 
this discrepancy can be controlled with 
the expected estimation error 
\begin{align*} 
  \E{Z_t}  \le  B  \E{\sup_A |\nu_N(A) - \nu(A)|},
\end{align*} 
so that the estimation error $Z_t$ 
converges in mean to zero. 
Markov inequality then implies that 
the first term tends to zero with $N$ 
and with $m_{\max}^2 < o(N)$. 

To complete the proof, 
let $Z<\epsilon$ be the event that all the estimation error $Z_t$ 
are less than $\epsilon$. 
The probability of an error in option value can then be written with 
\begin{align*} 
  P(|\bar{V} - V| > \epsilon)
    &= \Pc{|\bar{V} - V|>\epsilon}{Z<\epsilon}P(Z<\epsilon)
        + \Pc{|\bar{V} - V|>\epsilon}{Z>\epsilon}P(Z>\epsilon).
\end{align*} 
By the previous, the second term tends to zero. 
For the first term, 
the conditioning event is assumed to imply that 
the lookahead operator $L(X_{t:T},F_t)$ 
can be approximated arbitrarily well by a 
projected lookahead operator $\bar{L}(\bar{X}_{t:T},F_t)$. 
When this is the case, 
the option value $V$ 
can be approximated arbitrarily well by 
the expected projected residual payoff $\E{J(\bar{L},S)}$, 
so that $V=\E{J(\bar{L},S)}$ in probability, 
for some projected lookahead operator $\bar{L}$. 
By \eqref{spls_projected_option_value_rate}, 
the first term then tends to zero with $\widetilde{N}$. 

\subsection{SPLS Convergence in Bayes-Value (\ref{spls_bayes_value_rate})} \label{meth_spls_convergence_in_bayes-value_(spls_bayes_value_rate)} 

Denote by $J(\bar{L}_{0:T},S^{(\widetilde{N})})$ 
the vector $(J(\bar{L}_{0:T},S^{\widetilde{n}}),\widetilde{n}=1,2, \ldots ,\widetilde{N})$. 
The following four steps 
provide 
an estimate for the expected covering number 
$\Epb{\mathcal{N}(\epsilon, J(\bar{L}_{0:T},S^{(\widetilde{N})}))}$. 
The estimate is crude and can be used 
to bound 
$\Epb{\mathcal{N}(\frac{\epsilon}{8}, J(\bar{L}_{0:T},S^{(\widetilde{N})}))}$. 
Using this estimate, 
the result follows by Theorem 29.1 in \textcite{devroye96}. 

\textbf{Step 1}. 
We bound the covering number 
of the residual payoff 
$\mathcal{N}(\epsilon, J(\bar{L}_{0:T},S^{(\widetilde{N})}))$ 
with the covering number of the projected lookahead process 
$\mathcal{N}(\epsilon, \bar{L}_{0:T})$. 
Let $\bar{L}$ and $\bar{L}^\prime$ be two projected lookahead processes, by the Lipschitz assumption, 
\begin{align*} 
\pn{J(\bar{L}_{0:T},S^{(\widetilde{N})}) - J(\bar{L}^\prime_{0:T},S^{(\widetilde{N})})}
  &= \frac{1}{\widetilde{N}} \sum_{\widetilde{n}=1}^{\widetilde{N}} \pn{J(\bar{L}_{0:T},S^{\widetilde{n}}) - J(\bar{L}^\prime_{0:T},S^{\widetilde{n}})} \\
  & \le  \frac{1}{\widetilde{N}} \sum_{\widetilde{n}=1}^{\widetilde{N}} K \pn{\bar{L}_{0:T} - \bar{L}^\prime_{0:T}} \\
  &= K \pn{\bar{L}_{0:T} - \bar{L}^\prime_{0:T}}.
\end{align*} 
Proving that 
\begin{align*} 
\mathcal{N}(\epsilon, J(X,S^{(\widetilde{N})}))  \le  \mathcal{N}(\frac{\epsilon}{K},\bar{L}_{0:T}).
\end{align*} 

\textbf{Step 2}. 
We bound 
the covering number of 
the projected lookahead process 
with 
the covering number of 
the projected lookahead operator. 
With $\bar{L}$ and $\bar{L}^\prime$ two projected lookahead processes, we have 
\begin{align*} 
\pn{\bar{L}_{0:t} - \bar{L}^\prime_{0:t}}
  & \le  \sum_{t=0}^T \pn{\bar{L}_t - \bar{L}^\prime_t}.
\end{align*} 
Proving that 
\begin{align*} 
\mathcal{N}(\epsilon, \bar{L}_{0:T})  \le  \prod_{t=0}^T \mathcal{N}(\epsilon,\bar{L}_t).
\end{align*} 

\textbf{Step 3}. 
From \eqref{spls_lookahead_operator_bar}, 
the projected lookahead operator $\bar{L}_t$ can be written with 
\begin{align*} 
  \bar{L}(\bar{X}_{t:T}, F_t) = \p{\argmax_{\bar{X}_{t:T}} \frac{1}{N} \sum_{n=0}^N J(\bar{X}_{t:T}, S^n_{t:T})}_t .
\end{align*} 
The minimum information necessary in the filtration 
are the past decision $X_{0:t-1}$ and the stock path history $S_{0:t}$. 
The projected lookahead operator 
is hence a function of the form 
\begin{align*} 
 (\bar{X}_{t:T}, X_{0:t-1}, S_{0:t}) \to \p{\argmax_{\bar{X}_{t:T}} \frac{1}{N} \sum_{n=0}^N J(\bar{X}_{t:T}, S^n_{t:T})}_t.
\end{align*} 
By viewing a function as a subset of the product of 
the function domain and codomain, we have 
\begin{align*} 
  \mathcal{N}(\epsilon,\bar{L}_t)  \le
    \mathcal{N}(\frac{\epsilon}{2}, \bar{X}_{t:T})
    \mathcal{N}(\frac{\epsilon}{2}, X_{0:t-1})
    \mathcal{N}(\frac{\epsilon}{2}, S_{0:t})
    \mathcal{N}(\frac{\epsilon}{2}, Z_t),
\end{align*} 
where $Z_t$ is the codomain of the projected lookahead operator. 
The next three sub-steps 
derive the covering numbers for each element in the right-hand side above. 

\textbf{Step 3.a}, $\bar{X}_{t:T}$. 
The covering number for a decision $X_s$ in $X_{t:T}$ 
is the covering number for 
a step-function constructed 
with a nearest-neighbor basis for the stock path \eqref{spls_step_fct}. 
Such a function is of the form 
$\cup_{i=1}^{m_{s-t}} B_i \times x^i$, 
where $B_i$ 
is a Voronoi cell in $\R^{d(s-t)}$, 
and $x^i$ is in $\R^a$. 
The covering number of the decision 
can hence be bounded by the product 
of the covering numbers, giving 
\begin{align*} 
  \mathcal{N}(\epsilon, X_s)  \le  \p{\frac{X_{\max}}{\epsilon}}^a \mathcal{N}(\frac{\epsilon}{2}, \cup_{i=1}^{m_{s-t}} B_i),
\end{align*} 
where $X_{\max}$ is an upper bound on a decision. 

We bound the covering number of the Voronoi tesselation 
$\cup_{i=1}^{m_{s-t}} B_i$ 
in term of its VC-dimension. 
Each cell $B_i$ in the tesselation 
is the intersection of at most $m_{s-t}-1$ hyperplanes, 
and the VC-dimension of each hyperplane 
is $d(s-t)+1$. 
Using Theorem 1.1 in \textcite{vaart09}, 
the VC-dimension of a cell is bounded by 
 $3 d (s-t) m_{s-t} \log(4 m_{s-t})$. 
Another application of this theorem 
gives that the VC-dimension of the tesselation is bounded 
by $9 d (s-t) m^2_{s-t} \log(4 m_{s-t}) \log(4 m_{s-t})$, 
or simply $c d (s-t) m^3_{s-t}$. 
Using Theorem 2.6.4 in \textcite{vaart96}, 
the covering number for the Voronoi tesselation is then given by 
\begin{align*} 
  \mathcal{N}(\epsilon, \cup_{i=1}^{m_{s-t}} B_i)
    & \le  c d (s-t) m^3_{s-t} \p{\frac{4e}{\epsilon}}^{2 c d (s-t) m^3_{s-t} } \\
    & \le  c d T m^3_{s-t} \p{\frac{c}{\epsilon}}^{c d T m^3_{s-t} }.
    \stepcounter{equation}\tag{\theequation}\label{meth_voronoi_tesselation_covering_number}
\end{align*} 

We then have 
\begin{align*} 
  \mathcal{N}(\epsilon,\bar{X}_{t:T})
    & \le  \prod_{s=t}^T \mathcal{N}(\epsilon, X_s ) \\
    & \le  \prod_{s=t}^T \p{\frac{X_{\max}}{\epsilon}}^a \mathcal{N}(\frac{\epsilon}{2}, \cup_{i=1}^{m_{s-t}} B_i) \\
    & \le  \p{\frac{X_{\max}}{\epsilon}}^{a T} \prod_{s=t}^T  c d T m^3_{s-t} \p{\frac{c}{\epsilon}}^{c d T m^3_{s-t} } \\
    & \le  (c d T m_{\max}^3)^T \p{\frac{X_{\max}}{\epsilon}}^{c (a+d) T^2 m_{\max}^3},
\end{align*} 
where $m_{\max}=\pn{m}_\infty$ is the largest nearest-neighbor size. 

\textbf{Step 3.b}, $X_{0:t-1}$ and $S_{0:t}$. 
The covering number for the past decision $X_{0:t-1}$ 
is the covering number for a vector in $\R^{a(t-1)}$ 
and can be written with 
\begin{align*} 
  \mathcal{N}(\epsilon,X_{0:t-1})
    = \p{\frac{X_{\max}}{\epsilon}}^{a (t-1)}
     \le  \p{\frac{X_{\max}}{\epsilon}}^{a T}.
\end{align*} 
Similarly, the observed stock path $S_{0:t}$. 
is a vector in $\R^{d t}$ 
with covering number 
\begin{align*} 
  \mathcal{N}(\epsilon,S_{0:t})
    = \p{\frac{S_{\max}}{\epsilon}}^{a t}
     \le  \p{\frac{S_{\max}}{\epsilon}}^{a T},
\end{align*} 
where $S_{\max}$ is an upper bound on the stock price. 

\textbf{Step 3.c}, $Z_t$. 
Since the selection operator $(\cdot)_t$ is an injection, 
and $\argmax$ is an injection with an appropriate tie-breaking rule, 
we can bound the covering number of the codomain $Z_t$ 
with a covering number 
for the best empirical average residual payoff. 
This best empirical average can be written with 
\begin{align*} 
  A^*_N = \max_{\bar{X}_{t:T}} \frac{1}{N} \sum_{n=0}^N J(\bar{X}_{t:T}, S^n_{t:T}) .
\end{align*} 
Denote by $A_{N,t}$ the empirical average 
\begin{align*} 
  A_{N,t} = \sum_{n=0}^N J(\bar{X}_{t:T}, S^n_{t:T}),
\end{align*} 
and denote by $A_t$ 
the expected average 
\begin{align*} 
  A_t = \Ec{J(\bar{X}_{t:T}, S^n_{t:T})}{F_t}.
\end{align*} 

We bound the covering number of the best empirical average 
with a uniform deviation from the expected average 
(\citeauthor{devroye96} \cite*{devroye96}, Lemmma 8.2). 
Let $A_{N,t}$ and ${A^\prime}^*_{N,t}$ be two best averages 
computed from different samples, then 
\begin{align*} 
  |A^*_{N,t} - {A^\prime}^*_{N,t}|
    &= |
        \max_{\bar{X}_{t:T}} \frac{1}{N} \sum_{n=0}^N J(\bar{X}_{t:T}, S^n_{t:T})
          - \max_{\bar{X}_{t:T}} \frac{1}{N} \sum_{n=0}^N J(\bar{X}_{t:T}, {S^\prime}^n_{t:T})
      | \\
    &= | A^*_{N,t} - A^*_t + A^*_t - {A^\prime}^*_{N,t} | \\
    & \le  2 \max_{\bar{X}_{t:T},X_{0:t-1}} |A_{N,t} - A_t|.
\end{align*} 

Using the uniform law of large numbers 
(\citeauthor{devroye96} \cite*{devroye96}, Theorem 29.1), 
we can now estimate the probability 
that two best averages are closed by $\epsilon$ 
\begin{align*} 
  P( |A^*_{N,t} - {A^\prime}^*_{N,t}|  \le  \epsilon)
  & \le  P( \max_{\bar{X}_{t:T},X_{0:t-1}} |A_{N,t} - A_t|  \le  \frac{\epsilon}{2}) \\
  &= 1 - P( \max_{\bar{X}_{t:T},X_{0:t-1}} |A_{N,t} - A_t| > \frac{\epsilon}{2}) \\
  & \ge  1 - 8 \Epb{\mathcal{N}(\frac{\epsilon}{16},(\bar{X}_{t:T},X_{0:t-1}))} e^{ - \epsilon^2 N / (512 B^2)},
\end{align*} 
where $B$ is an upper bound on the residual payoff. 
By viewing the expected covering number 
of the best empirical average 
as a geometric distribution, 
we have 
\begin{align*} 
  \Epb{\mathcal{N}(\epsilon,A^*_{N,t})}
    &= \p{P( |A^*_{N,t} - {A^\prime}^*_{N,t}|  \le  \epsilon)}^{-1} \\
    & \le  \p{1 - 8 \Epb{\mathcal{N}(\frac{\epsilon}{16},(\bar{X}_{t:T},X_{0:t-1}))} e^{ - \epsilon^2 N / (512 B^2)}}^{-1} \\
    &= \exp\p{ 8 \Epb{\mathcal{N}(\frac{\epsilon}{16},(\bar{X}_{t:T},X_{0:t-1}))} e^{ - \epsilon^2 N / (512 B^2)} },
\end{align*} 
where the identity $1-x  \le  e^{-x}$ 
was used. 

By using Step 3.a and 3.b, the product bound gives 
\begin{align*} 
  \mathcal{N}(\frac{\epsilon}{16},(\bar{X}_{t:T},X_{0:t-1}))
  & \le  \mathcal{N}(\frac{\epsilon}{16},\bar{X}_{t:T}) \mathcal{N}(\frac{\epsilon}{16},X_{0:t-1})) \\
  & \le
     \p{\frac{16 X_{\max}}{\epsilon}}^{a T}
    (c d T m_{\max}^3)^T \p{\frac{X_{\max}}{\epsilon}}^{c (a+d) T^2 m_{\max}^3} \\
  & \le  (c d T m_{\max}^3)^T \p{\frac{X_{\max}}{\epsilon}}^{c (a+d) T^2 m_{\max}^3}
\end{align*} 
proving that 
\begin{align*} 
  \Epb{\mathcal{N}(\epsilon,A^*_{N,t})}
  & \le  \exp\p{
    8
    (c d T m_{\max}^3)^T
    \p{\frac{X_{\max}}{\epsilon}}^{c (a+d) T^2 m_{\max}^3}
    e^{ - \epsilon^2 N / (512 B^2)}
  }.
\end{align*} 

\textbf{Step 4}. 
By combining all the previous results, we have 
\begin{align*} 
& \Epb{\mathcal{N}(\epsilon, J(X,S^{(\widetilde{N})}))}  \\
  & \le  \Epb{\mathcal{N}(\frac{\epsilon}{K},\bar{L}_{0:T})} \\
  & \le  \Epb{\prod_{t=0}^T \mathcal{N}(\frac{\epsilon}{K},\bar{L}_t)} \\
  & \le  \Epb{\prod_{t=0}^T
        \mathcal{N}(\frac{\epsilon}{2 K}, \bar{X}_{t:T})
        \mathcal{N}(\frac{\epsilon}{2 K}, X_{0:t-1})
        \mathcal{N}(\frac{\epsilon}{2 K}, S_{0:t})
        \mathcal{N}(\frac{\epsilon}{2 K}, Z_t)
      } \\
  & \le
    (c d T m_{\max}^3)^{T^2} \p{\frac{K X_{\max}}{\epsilon}}^{c (a+d) T^3 m_{\max}^3}
    \p{\frac{K X_{\max}}{\epsilon}}^{a T^2 }
    \\
    &
    \p{\frac{K S_{\max}}{\epsilon}}^{a T^2}
     \exp\p{
      8 T
      (c d T m_{\max}^3)^T
      \p{\frac{K X_{\max}}{\epsilon}}^{c (a+d) T^2 m_{\max}^3}
      e^{ - \epsilon^2 N / (512 B^2 K^2)}
    }
      \\
  & \le
   \exp\p{
    8 T
    (c d T m_{\max}^3)^{T}
    \p{\frac{K X_{\max} S_{\max}}{\epsilon}}^{c (a+d) T^2 m_{\max}^3}
    e^{ - \epsilon^2 N / (512 B^2 K^2)}
  }.
\end{align*} 

\subsection{SPLS Universal Bayes-Value Consistency and Consistency (\ref{spls_bayes_value})} \label{meth_spls_universal_bayes-value_consistency_and_consistency_(spls_bayes_value)} 

To prove the universal Bayes-value efficiency, 
assume that 
the lookahead operator 
is restricted to 
the class of strategy implied by 
the projected lookahead operator. 
When this is the case, we can write 
\begin{align*} 
  V = \sup_{\bar{L}} \Epb{J(\bar{L}_{0:T},S)}.
\end{align*} 
Use Lemma 8.2 in \textcite{devroye96} to write 
\begin{align*} 
  \bar{V} - V
  & = \p{\sup_{\bar{L}} \frac{1}{\widetilde{N}} \sum_{\widetilde{n}=1}^{\widetilde{N}} J(\bar{L}_{0:T}, S^{\widetilde{n}})} - \p{\sup_{\bar{L}} \Epb{J(\bar{L}_{0:T},S)}} \\
  & \le   2 \sup_{\bar{L}} | \frac{1}{\widetilde{N}} \sum_{\widetilde{n}=1}^{\widetilde{N}} J(\bar{L}_{0:T}, S^{\widetilde{n}}) - \Epb{J(\bar{L}_{0:T},S)} |.
\end{align*} 
By using \eqref{spls_bayes_value_rate}, 
and Problem 12.1 in \textcite{devroye96}, 
the expectation of the above right-hand side is bounded by 
\begin{align*} 
  \sqrt{\frac{
    c T
    (c d T m_{\max}^3)^{T}
    \p{\frac{K X_{\max} S_{\max}}{\epsilon}}^{c (a+d) T^2 m_{\max}^3}
    e^{ - \epsilon^2 N / (512 B^2 K^2)}
    }{
      \widetilde{N} / (128 B^2)
    }}.
\end{align*} 
This bound implies that 
$\Epb{|\bar{V}-V|}$ 
converges to zero 
with 
$\widetilde{N}$, $N$ and $m$ 
going to infinity, 
and $m_{\max}^3 < o(N)$, 
so that the projected option value 
converges in mean to the option value. 
As convergence in mean implies 
convergence in $\ell_1$-norm, 
the universal consistency 
result 
\eqref{spls_bayes_value} 
follows. 

To prove consistency of SPLS, 
note that 
convergence in mean 
implies convergence in probability, 
and the consistency result follows. 

\subsection{NNM Projected Martingale in a Binomial World (Section \ref{nnm_uniform_convergence_in_measure})} \label{meth_nnm_projected_martingale_in_a_binomial_world_(*nnm_uniform_convergence_in_measure)} 

For a vanilla American put 
in a binomial world, 
the optimal martingale has size 
$(\alpha, \beta) = ((2^{t-1},2), t=1, 2,  \ldots , T)$, 
and 
a good projected martingale can be obtained with 
$(p,q) = ( (\mathcal{O}(t), 2), t=1, 2,  \ldots , T)$. 
The stated size for the optimal martingale 
is by construction of the binomial world. 
At time $t$, 
the maximum number of different stock paths 
leading to time $t$ is $2^{t-1}$, 
and, given the path history up to time $t-1$, 
the maximum number of different stock prices 
at time $t$ is 2. 
The claimed size for the conditioning part 
of a good projected martingale 
can be verified numerically. 
To do so quantize with Lloyd's method 
all the $2^{t-1}$ path leading to time $t$. 
Repeat this quantization over time, 
and note that a quantization with a small 
mean square error 
is obtained when 
the number of centroids grow linearly over time. 

\subsection{NNM Projected Martingale Convergence (\ref{nnm_conditioning_part_rate}, \ref{nnm_next_part_rate})} \label{meth_nnm_projected_martingale_convergence_(nnm_conditioning_part_rate,nnm_next_part_rate)} 

For the conditioning part, 
\begin{align*} 
& P\bigg( \sup_A \frac{1}{p_t} \sum_{i=1}^{p_t} \ib{\bar{B}_i \in A} > \epsilon \bigg)  \\
& =
\Ebb{ \Pc{ \sup_A \frac{1}{p_t} \sum_{i=1}^{p_t} \ib{\bar{B}_i \in A} > \epsilon }{  \{\bar{B}_i\}_{i=1}^{p_t} } } \stepcounter{equation}\tag{\theequation}\label{meth_conditioning_npmc} \\
&  \le
s(\mathcal{A}, p_t)
\Ebb{ \sup_A \Pc{ \frac{1}{p_t} \sum_{i=1}^{p_t} \ib{\bar{B}_i \in A} > \epsilon }{ \{\bar{B}_i\}_{i=1}^{p_t} } } \stepcounter{equation}\tag{\theequation}\label{meth_shattering_npmc} \\
&  \le
2 s(\mathcal{A}, p_t) e^{- 2 \epsilon^2 p_t} \stepcounter{equation}\tag{\theequation}\label{meth_hoeffding_npmc}
\end{align*} 
\eqref{meth_conditioning_npmc} is a simple conditioning on the Voronoi cell $\{\bar{B}_i\}_{i=1}^{p_t}$, 
\eqref{meth_shattering_npmc} holds by definition of the shattering coefficient $s(\mathcal{A},p_t)$, 
and 
\eqref{meth_hoeffding_npmc} holds by Hoeffding's inequality. 
To obtain the shatter coefficient, 
note that 
\begin{align*} 
s(\mathcal{A}, p_t)
& \le  \prod_{i=1}^{\alpha_t} s(2^{B_i}, p_t) \\
& \le  \prod_{i=1}^{\alpha_t} {p_t}^{ (d t + 1) (\alpha_t-1)} \\
& \le  {p_t}^{3 d t \alpha_t^2 }
\end{align*} 
as each set $B_i$ 
is the intersection of at most $\alpha_t-1$ hyperplanes, 
and the VC-dimension 
of each hyperplane is $d t + 1$. 
The factor of $3$ is used to bound the VC-dimension 
of each hyperplane so that the bound is greater than 2. 
See Theorem 
13.5, 13.9, 13.8 and 13.3 
in \textcite{devroye96}. 

Similarly, for the current part, 
\begin{align*} 
& P\bigg( \sup_A \frac{1}{q_t} \sum_{i=1}^{q_t} \ib{\bar{B}_i \in A} > \epsilon \bigg)  \\
& =
\Ebb{ \Pc{ \sup_A \frac{1}{q_t} \sum_{i=1}^{q_t} \ib{\bar{B}_i \in A} > \epsilon }{  \{\bar{B}_i\}_{i=1}^{q_t} } } \stepcounter{equation}\tag{\theequation}\label{meth_conditioning} \\
&  \le
s(\mathcal{A}, q_t)
\Ebb{ \sup_A \Pc{ \frac{1}{q_t} \sum_{i=1}^{q_t} \ib{\bar{B}_i \in A} > \epsilon }{ \{\bar{B}_i\}_{i=1}^{q_t} } } \stepcounter{equation}\tag{\theequation}\label{meth_shattering} \\
&  \le
2 s(\mathcal{A}, q_t) e^{- 2 \epsilon^2 q_t}. \stepcounter{equation}\tag{\theequation}\label{meth_hoeffding}
\end{align*} 
To obtain the shatter coefficient, 
note that 
\begin{align*} 
s(\mathcal{A}, q_t)
& \le  \prod_{i=1}^{\beta_t} s(2^{B_i}, q_t) \\
& \le  \prod_{i=1}^{\beta_t} {q_t}^{ (d+1) (\beta_t-1)} \\
& \le  {q_t}^{3 d \beta_t^2 }
\end{align*} 
as each set $B_i$ 
is the intersection of at most $\beta_t-1$ hyperplanes, 
and the VC-dimension 
of each hyperplane is $d+1$. 
The factor of $3$ is used to bound the VC-dimension 
of each hyperplane so that the bound is greater than 2. 
See Theorem 
13.5, 13.9, 13.8 and 13.3 
in \textcite{devroye96}. 

\subsection{NNM Relaxed Rogers Operator (\ref{nnm_relaxed_rogers_operator_rate_1}, \ref{nnm_relaxed_rogers_operator_rate_2})} \label{meth_nnm_relaxed_rogers_operator_(nnm_relaxed_rogers_operator_rate_1,nnm_relaxed_rogers_operator_rate_2)} 

For the stochastic process estimate, 
the shatter coefficient of the class can be bounded with 
\begin{align*} 
s( \{B \times [a,b]\},  N)
&  \le  \p{\prod_{t=1}^T \prod_{i=1}^{p_t q_t} s(\bar{B}^t_i, N)} s([a,b], N) \\
&  \le  N^{2T} \prod_{t=1}^T \prod_{i=1}^{p_t q_t} N^{(d (t-1) + 1) (p_t-1) + (d+1) (q_t-1)} \\
&  \le  N^{ 3 d T \sum_{t=1}^T t p_t^2 q_t^2 }
\end{align*} 
Indeed, 
the VC-dimension of an interval in $\R^T$ 
is $2T$. 
Each set $\bar{B}^t_i$ 
can be written as $U \times W$. 
The conditioning part $U$ 
is the intersection of at most $p_t-1$ hyperplanes, 
and the VC-dimension 
of each hyperplane is $d (t-1) + 1$. 
The current part $W$ 
is the intersection of at most $q_t-1$ hyperplanes, 
and the VC-dimension 
of each hyperplane is $d+1$. 
The factor of $3$ is used to obtain a simple final bound 
that is greater than 2. 
See Theorem 
13.5, 13.9, 13.8 and 13.3 
in \textcite{devroye96}. 

For the point estimates, 
the same steps as above 
give the following shatter coefficient 
\begin{align*} 
s( \{B \times [a,b]\},  N)
&  \le  N^{ 3 d \sum_{t=1}^T t p_t^2 q_t^2 }
\end{align*} 
as the interval $[a,b]$ 
is now an interval in $\R^d$ 
with a VC-dimension of $2d$. 
\subsection{NNM Projected Dual Value (\ref{nnm_relaxed_dual_value_rate})} \label{meth_nnm_projected_dual_value_(nnm_relaxed_dual_value_rate)} 

Denote by $D(\bar{M},S^{(\widetilde{N})})$ 
the vector $(D(\bar{M},S^{\widetilde{n}},\widetilde{n}=1,2, \ldots ,\widetilde{N})$, 
and 
let $\bar{M}$ and $\bar{M}^\prime$ be two projected martingale processes. 
By the Lipschitz assumption 
\begin{align*} 
\pn{D(\bar{M},S^{(\widetilde{N})}) - D(\bar{M}^\prime,S^{(\widetilde{N})})}
  &= \frac{1}{\widetilde{N}} \sum_{\widetilde{n}=1}^{\widetilde{N}} \pn{D(\bar{M},S^{\widetilde{n}}) - D(\bar{M}^\prime,S^{\widetilde{n}})} \\
  & \le  \frac{1}{\widetilde{N}} \sum_{\widetilde{n}=1}^{\widetilde{N}} K \pn{\bar{M} - \bar{M}^\prime} \\
  &= K \pn{\bar{M} - \bar{M}^\prime}.
\end{align*} 
Proving that $\mathcal{N}(\epsilon, D(\bar{M},S^{(\widetilde{N})}))  \le  \mathcal{N}(\frac{\epsilon}{K},\bar{M})$. 

We bound the covering number of 
the projected martingale process 
with 
the covering number 
of the projected martingale. 
Indeed, 
with $\bar{M}$ and $\bar{M}^\prime$ two projected martingales 
\begin{align*} 
\pn{\bar{M}_{0:t} - \bar{M}^\prime_{0:t}}
  &= \frac{1}{T} \sum_{t=0}^T \pn{\bar{M}_t - \bar{M}^\prime_t}.
\end{align*} 
Proving that $\mathcal{N}(\epsilon, \bar{M}_{0:T})  \le  \prod_{t=0}^T \mathcal{N}(\epsilon,\bar{M}_t)$. 

Each projected martingale $M_t$ 
is a step-function 
on a nearest-neighbor basis for the stock path 
and can be written 
$\cup_{i=1}^{p_t q_t} B_i \times m^i$ 
where 
$m^i$ is bounded by $U$, 
and 
$B_i$ 
is the intersection of two Voronoi cells, 
one in $\R^{d(t-1)}$ 
and one in $\R^d$. 
The covering number of the decision 
can hence be bounded by the product 
of the covering numbers, giving 
\begin{align*} 
  \mathcal{N}(\epsilon, \bar{M}_t)  \le  \p{\frac{2 U}{\epsilon}} \mathcal{N}(\frac{\epsilon}{2}, \cup_{i=1}^{p_t q_t} B_i).
\end{align*} 
We bound the covering number of the Voronoi tesselation 
$\cup_{i=1}^{p_t q_t} B_i$ 
in term of its VC-dimension. 
Each cell $B_i$ in the tesselation 
is the intersection of at most $p_t q_t-1$ hyperplanes, 
and the VC-dimension of each hyperplane 
is $d t + 1$. 
Using Theorem 1.1 in \textcite{vaart09}, 
the VC-dimension of each cell 
is bounded by $3 d t p_t q_t \log(4 p_t q_t)$. 
Another application of this theorem 
gives that the VC-dimension of the tesselation 
is bounded by $9 d t p_t^2 q_t^2 \log(4 p_t q_t) \log(4 p_t q_t)$, 
or simply $c d t p_t^3 q_t^3$. 
By 
Theorem 2.6.4 
in 
\textcite{vaart96}, 
the covering number for the Voronoi tesselation is then given by 
\begin{align*} 
  \mathcal{N}(\epsilon, \cup_{i=1}^{p_t q_t} B_i)
    & \le  c d t p_t^3 q_t^3 \p{\frac{4e}{\epsilon}}^{c d t p_t^3 q_t^3 } \\
    & \le  c d T p_t^3 q_t^3 \p{\frac{c}{\epsilon}}^{c d T p_t^3 q_t^3 }.
\end{align*} 
Proving that 
\begin{align*} 
  \mathcal{N}(\epsilon,\bar{M}_t)
    & \le  \p{\frac{2 U}{\epsilon}} \mathcal{N}(\frac{\epsilon}{2}, \cup_{i=1}^{p_t q_t} B_i) \\
    & \le  \p{\frac{2 U}{\epsilon}} c d T p_t^3 q_t^3 \p{\frac{c}{\epsilon}}^{c d T p_t^3 q_t^3 } \\
    & \le  c d T p_t^3 q_t^3 \p{\frac{c U}{\epsilon}}^{c d T p_t^3 q_t^3 }.
\end{align*} 

By combining all the previous results, we have 
\begin{align*} 
\mathcal{N}(\epsilon, D(\bar{M},S^{(N)}))
  & \le  \mathcal{N}(\frac{\epsilon}{K},\bar{M}) \\
  & \le  \prod_{t=0}^T \mathcal{N}(\frac{\epsilon}{K},\bar{M}_t) \\
  & \le  \prod_{t=0}^T c d T p_t^3 q_t^3 \p{\frac{c U K}{\epsilon}}^{c d T p_t^3 q_t^3 } \\
  & \le  (c d T p_{\max}^3 q_{\max}^3)^T \p{\frac{U K}{\epsilon}}^{c d T^2 p_{\max}^3 q_{\max}^3 },
  \stepcounter{equation}\tag{\theequation}\label{meth_projected_martingale_covering_number}
\end{align*} 
where $(p_{\max}, q_{\max})=(\pn{p}_\infty, \pn{q}_\infty)$ 
is the largest nearest-neighbor size. 
Our estimate for 
the covering number 
$\mathcal{N}(\epsilon, D(\bar{M},S^{(N)}))$ 
is crude 
and can be used 
for 
$\mathcal{N}(\frac{\epsilon}{8}, D(\bar{M},S^{(N)}))$. 
The result now follows 
by Theorem 29.1 in \textcite{devroye96}. 

\subsection{NNM Consistency (\ref{nnm_consistency_1}, \ref{nnm_consistency_2})} \label{meth_nnm_consistency_(nnm_consistency_1,nnm_consistency_2)} 

We start with the consistency of the relaxed dual value 
\eqref{nnm_consistency_2}. 
Let $M$ be the optimal martingale 
and 
consider the estimation error 
in the relaxed dual value 
\begin{align*} 
  |\bar{\bar{V}} - V| = \pa{ \frac{1}{N} \sum_{n=1}^N D(\bar{R}^m(\bar{Z}),S^n) -  \E{D(M,S)}  }.
\end{align*} 
Denote by $U$ 
the event that the projected martingale 
has unambiguous cells. 
By using 
\eqref{nnm_conditioning_part_rate} 
and \eqref{nnm_next_part_rate}, 
this event can be written with 
\begin{align*} 
U = \p{\sup_A \frac{1}{p_t} \sum_{i=1}^{p_t } \ib{\bar{B}_i \in A}  > \epsilon, \sup_A \frac{1}{q_t} \sum_{i=1}^{q_t } \ib{\bar{B}_i \in A}  > \epsilon, t=1, \ldots ,T}.
\end{align*} 
Denote by $W$ 
the event that the distribution of the relaxed Rogers operator is accurate. 
By using \eqref{nnm_relaxed_rogers_operator_rate_2}, 
this event can be written with 
\begin{align*} 
  W = \p{ \sup_A | \nu_n(A) - \nu(A) |> \epsilon }.
\end{align*} 
By conditioning, the probability of error becomes 
\begin{align*} 
  P(|\bar{\bar{V}} - V|>\epsilon)
    &= \Pc{|\bar{\bar{V}} - V|>\epsilon}{U,W} P(U,W)
      + \Pc{|\bar{\bar{V}} - V|>\epsilon}{\overline{(U,W)}} P(\overline{(U,W)} ),
\end{align*} 
where the notation $(U,W)$ 
means that both the event $U$ and $W$ hold, 
and the notation $\overline{(U,W)}$ 
is the complementary event. 
By 
\eqref{nnm_conditioning_part_rate}, 
\eqref{nnm_next_part_rate}, 
\eqref{nnm_relaxed_rogers_operator_rate_2}, 
and 
the union bound, 
the second term 
tends to zero 
when $(p,q)$ and $N$ are large, 
with 
$p_{\max}^2 q_{\max}^2 < o(N)$. 
For the first term, 
the event $U$ is assumed to imply that 
the optimal martingale can be approximated accurately 
with a projected martingale. 
The estimation error can hence be written with 
\begin{align*} 
  |\bar{\bar{V}} - V| = \pa{ \frac{1}{N} \sum_{n=1}^N D(\bar{R}^m(\bar{Z}),S^n) -  \E{D(\bar{M},S)}  }.
\end{align*} 
where $\bar{M}$ is the projection of the optimal martingale, 
and the equality is in probability. 
The event $W$ is assumed to imply 
that the distribution underlying the relaxed Rogers operator 
is accurate, 
so that the 
following equality in distribution holds 
\begin{align*} 
  \frac{1}{N} \sum_{n=1}^N D(\bar{R}^m(\bar{Z}),S^n) = \E{D(\bar{M},S)}.
\end{align*} 
As the last equality is 
an equality in distribution for scalars, 
the equality also holds in probability, 
and $\Pc{|\bar{\bar{V}} - V|>\epsilon}{U,W}$ vanishes 
when $(p,q)$ and $N$ are large, 
with 
$p_{\max}^2 q_{\max}^2 < o(N)$. 
The consistency result \eqref{nnm_consistency_2} now follows. 

For the consistency of the projected dual value 
\eqref{nnm_consistency_1}, 
the steps are the same as above, 
except that an additional step is needed. 
Indeed, the estimation error is now evaluated 
with an out-of-sample test, 
and the estimation error can be written with 
\begin{align*} 
  |\bar{V} - V| = \pa{ \frac{1}{\widetilde{N}} \sum_{\widetilde{n}=1}^{\widetilde{N}} D(\bar{R}^m(\bar{Z}),S^{\widetilde{n}}) -  \E{D(M,S)}  }.
\end{align*} 
Apply the above steps, 
with the modification 
that the event $W$ 
now implies the following equality in distribution 
\begin{align*} 
  \bar{R}^m(\bar{Z}) = \bar{M}.
\end{align*} 
It hence remains to show 
that 
$\Pc{|\bar{V} - V|>\epsilon}{U,W}$ 
vanishes. 
To this end, 
use \eqref{nnm_relaxed_dual_value_rate} 
to show 
that the quantity vanishes 
when $(p,q)$ and $N$ are large, 
with 
$p_{\max}^3 q_{\max}^3 < o(N)$. 
The consistency result \eqref{nnm_consistency_1} now follows. 

\subsection{NNM Convergence in Bayes-Value (\ref{nnm_projected_dual_value_bayes_rate})} \label{meth_nnm_convergence_in_bayes-value_(nnm_projected_dual_value_bayes_rate)} 

Denote by $D(\bar{R}(S^{(N)}),S^{(\widetilde{N})})$ 
the vector $(D(\bar{R}(S^{(N)}),S^n),n=1,2, \ldots ,\widetilde{N})$, 
we bound the covering number 
of the dual payoff 
$\mathcal{N}(\epsilon, D(\bar{R}(S^{(N)}),S^{(\widetilde{N})}))$ 
with the covering number of the general relaxed Rogers operator. 
Let $S^{(N)}$ and ${S^\prime}^{(N)}$ be two random sample, by the Lipschitz assumption, 
\begin{align*} 
\pn{D(\bar{R}(S^{(N)}),S^{(\widetilde{N})}) - D(\bar{R}({S^\prime}^{(N)}),S^{(\widetilde{N})})}
  &= \frac{1}{\widetilde{N}} \sum_{\widetilde{n}=1}^{\widetilde{N}} \pn{D(\bar{R}(S^{(N)}),S^{\widetilde{n}}) - S(\bar{R}({S^\prime}^{(N)}),S^{\widetilde{n}})} \\
  & \le  \frac{1}{\widetilde{N}} \sum_{\widetilde{n}=1}^{\widetilde{N}} K \pn{\bar{R}(S^{(N)}) - \bar{R}({S^\prime}^{(N)})} \\
  &= K \pn{\bar{R}(S^{(N)}) - \bar{R}({S^\prime}^{(N)})}.
\end{align*} 
Proving that 
\begin{align*} 
  \mathcal{N}(\epsilon, D(\bar{R}(S^{(N)}),S^{(\widetilde{N})}))  \le  \mathcal{N}(\frac{\epsilon}{K},\bar{R}(S^{(N)})).
\end{align*} 

The general relaxed Rogers operator is a mapping 
from a random sample to a projected martingale. 
The operator is hence a function of the form 
$S^{(N)} \times \bar{M}$. 
The covering number of a function can be bound by the product 
of the covering number of the domain and the codomain, giving 
\begin{align*} 
  \mathcal{N}(\epsilon, \bar{R}(S^{(N)}))  \le  \mathcal{N}(\frac{\epsilon}{2},S^{(N)}) \mathcal{N}(\frac{\epsilon}{2},\bar{M}).
\end{align*} 
The covering number for a random sample $S^{(N)}$ 
is equivalent to the covering number for a random Voronoi tesselation. 
By \eqref{meth_voronoi_tesselation_covering_number}, we have 
\begin{align*} 
  \mathcal{N}(\epsilon,S^{(N)})
   \le
  c d T N^3 \p{\frac{c}{\epsilon}}^{c d T N^3 }.
\end{align*} 
The covering number for a projected martingale 
is given by \eqref{meth_projected_martingale_covering_number} and is 
\begin{align*} 
\mathcal{N}(\epsilon, \bar{M})  \le
  (c d T p_{\max}^3 q_{\max}^3)^T \p{\frac{U}{\epsilon}}^{c d T^2 p_{\max}^3 q_{\max}^3 }.
\end{align*} 
By combining all the previous results, we have 
\begin{align*} 
  \mathcal{N}(\epsilon, D(\bar{R}(S^{(N)}),S^{(\widetilde{N})}))
    & \le  \mathcal{N}(\frac{\epsilon}{K},\bar{R}(S^{(N)})) \\
    & \le  \mathcal{N}(\frac{\epsilon}{2 K},S^{(N)}) \mathcal{N}(\frac{\epsilon}{2 K},\bar{M}) \\
    & \le  c d T N^3 \p{\frac{c K}{\epsilon}}^{c d T N^3 }
       (c d T p_{\max}^3 q_{\max}^3)^T \p{\frac{U K}{\epsilon}}^{c d T^2 p_{\max}^3 q_{\max}^3 } \\
    & \le  (c d T p_{\max}^3 q_{\max}^3 N^3)^T \p{\frac{U K}{\epsilon}}^{c d T^2 p_{\max}^3 q_{\max}^3 N^3}.
\end{align*} 
The result now follows by Theorem 29.1 in \textcite{devroye96}. 

\subsection{NNM Universal Bayes-Value Consistency (\ref{nnm_bayes_value_1}, \ref{nnm_bayes_value_2})} \label{meth_nnm_universal_bayes-value_consistency_(nnm_bayes_value_1,nnm_bayes_value_2)} 

To prove the universal Bayes-value efficiency, 
assume that 
the Rogers operator 
is restricted to 
the class of martingale implied by 
the relaxed Rogers operator. 
For the projected dual value 
this assumption allows to write 
\begin{align*} 
  V = \sup_{\bar{R}(S^{(N)})} \Epb{D(\bar{R}(S^{(N)}),S)}.
\end{align*} 
For the relaxed dual value, 
this assumption allows to write 
\begin{align*} 
  V = \sup_{\bar{M}} \Epb{D(\bar{M},S)}.
\end{align*} 

For the projected dual value, 
use 
Lemma 8.2 
in \textcite{devroye96} to write 
\begin{align*} 
  & \p{\sup_{\bar{R}(S^{(N)})} \frac{1}{\widetilde{N}} \sum_{n=1}^{\widetilde{N}} D(\bar{R}(S^{(N)}), S^n)} - \p{\sup_{\bar{L}} \Epb{D(\bar{R}(S^{(N)}),S)}} \\
  &  \le   2 \sup_{\bar{R}(S^{(N)})} | \frac{1}{\widetilde{N}} \sum_{n=1}^{\widetilde{N}} D(\bar{R}(S^{(N)}), S^n) - \Epb{D(\bar{R}(S^{(N)}),S)} |.
\end{align*} 
By using \eqref{nnm_projected_dual_value_bayes_rate}, 
and Problem 12.1 in \textcite{devroye96}, 
the expectation of the right-hand side is bounded by 
\begin{align*} 
  \sqrt{\frac{
    \log \p{
    8 e
    (c d T p_{\max}^3 q_{\max}^3 N^3)^T \p{\frac{U K}{\epsilon}}^{c d T^2 p_{\max}^3 q_{\max}^3 N^3}
  }
    }{\widetilde{N} / (128 B^2)}}.
\end{align*} 
This bound implies that 
$\Epb{|\bar{V}-V|}$ 
converges to zero. 
The universal consistency 
\eqref{nnm_bayes_value_1} 
follows. 

Similarly, for the relaxed dual value, write 
\begin{align*} 
  \p{\sup_{\bar{M}} \frac{1}{N} \sum_{n=1}^{N} D(\bar{M}, S^n)} - \p{\sup_{\bar{M}} \Epb{D(\bar{M},S)}}
  & \le   2 \sup_{\bar{M}} | \frac{1}{N} \sum_{n=1}^{N} D(\bar{M}, S^n) - \Epb{D(\bar{M},S)} |.
\end{align*} 
Then, using \eqref{nnm_relaxed_dual_value_rate}, 
the expectation of the right-hand side is bounded by 
\begin{align*} 
  \sqrt{\frac{
    \log \p{
      8 e
      (c d T p_{\max}^3 q_{\max}^3)^T \p{\frac{U K}{\epsilon}}^{c d T^2 p_{\max}^3 q_{\max}^3 }
  }
    }{N / (128 B^2)}}.
\end{align*} 
The universal consistency 
\eqref{nnm_bayes_value_2} 
follows. 

\subsection{An American Option: Value Function Martingale Part (Section \ref{aao_nnm_martingale})} \label{meth_an_american_option:_value_function_martingale_part_(*aao_nnm_martingale)} 

For an American put, 
the value function $v_t$ is a function of the last observed stock price $S_t$. 
This value function can be written with 
\begin{align*} 
  v_t(S_t) = \max_{X_{t:T}} \Ec{\sum_{s=t}^T e^{-r t} \pp{K - S_s} X_s }{S_t},
\end{align*} 
and can be found with an implicit finite difference scheme. 
To simulate a sample martingale path $M$ 
of the value function martingale part, the procedure is as follows. 
First, sample a stock path $S$ and obtain 
a sample value function path with 
\begin{align*} 
  V = (v_0(S_{0}), v_1(S_{1}),  \ldots , v_T(S_T)),
\end{align*} 
Second, set $M_0=0$, and for each time $t=0,1, \ldots ,T-1$ 
obtain a sample of the next period stock price 
conditional on the current stock price. 
This sample can be written with 
\begin{align*} 
  \{S^i_{t+1}|S_t\}_{i=1}^n.
\end{align*} 
For each next period price find the value function realization 
\begin{align*} 
  \{V^i_{t+1}|S_t\}_{i=1}^n = \{v_t(S^i_{t+1})|S_t\}.
\end{align*} 
Set the martingale realization in the next period as 
the current martingale plus 
the centered value function 
\begin{align*} 
  M_{t+1} =  M_t + V_{t+1} - \frac{1}{n} \sum_{i=1}^n V^i_{t+1}.
\end{align*} 

\subsection{An American Option: Dual Metrics (\ref{aao_dual_exercise_time}, \ref{aao_dual_abs_average})} \label{meth_an_american_option:_dual_metrics_(aao_dual_exercise_time,aao_dual_abs_average)} 

To quantize a dual martingale 
use two samples: 
A sample of the stock path 
$\{S^i\}_{i=1}^n$ 
and a corresponding sample of the martingale 
$\{M^i\}_{i=1}^n$ 
generated by the stock path. 
Run Lloyd's algorithm (\citeauthor{lloyd82} \cite*{lloyd82}, \citeauthor{arthur07} \cite*{arthur07}) 
to split the martingale sample into 10 Voronoi cells. 
Denote these cells by $\{B_{1}, B_{2},  \ldots , B_{10}\}$. 
As the martingale sample is generated by the stock sample, 
these Voronoi cells also define a tesselation on 
the stock path. 
This implicit tesselation can be written with 
\begin{align*} 
  B^\prime_i = \pe{S^i : M^i \in B_i},
\end{align*} 
where $B^\prime_i$ is the $i$-th cell of the implicit tesselation, 
for $i=1,2, \ldots ,10$. 
This implicit tesselation allows to map 
any metrics to the stock domain, 
and in turn, the exercise boundary domain. 
To visualize a metric 
in the exercise boundary domain, 
find the stock centroid in the implicit tesselation. 
This centroid can be written with 
\begin{align*} 
  \hat{S}^i = \frac{1}{m} \sum_{j=1}^m S^j,
\end{align*} 
where the addition is vectorial, 
$m$ is the size of the implicit cell $B^\prime_i$, 
and $S^j \in B^\prime_i$ is a stock path in the implicit cell $B^\prime_i$. 
Now, any metric on a martingale cell $B_i$ 
can be visualized in the exercise boundary domain. 
It suffices to map the metric to the corresponding 
implicit cell $B^\prime_i$, 
and to represent the metric 
along the stock centroid $\hat{S}^i$. 

To compare the metrics of multiple dual martingales, 
fix the martingale tesselation 
and compute the metric conditional on the fixed tesselation. 
For example, let $\{B_i\}$ 
be a tesselation of a sample $\{M_i\}$ of optimal martingale path. 
Such a sample can be obtained 
with the method presented in the previous section. 
Consider the dual exercise distribution metric. 
This metric $\mu_1(\{M^j\}, \{S^j\})$ 
is a function 
of a martingale path sample $\{M^j\}$, 
and its generating stock path sample $\{S^j\}$. 
This metric can be written with 
\begin{align*} 
  \mu_1(\{M^j\}, \{S^i\}) = \mathcal{H}\pe{
    (S^j_{t^j}, t^j) :
      t^j \text{ is a dual exercise time}
  },
\end{align*} 
where $\mathcal{H}$ denotes the histogram operator, 
and the dual exercise time $t^j$ of a stock path $S^j$ 
is the time $t$ that maximizes the dual payoff 
\begin{align*} 
  \max_x \sum_{t=0}^T e^{-r t} \pp{K - S^j_t} x_t   - x_t M^j_t.
\end{align*} 
The histogram operator 
gives a set $\{(S_a, S_b, t_a, t_b, p)\}$ 
such that the probability 
that the dual exercise fall in the interval $[S_a,S_b) \times [t_a, t_b)$ 
is $p$. 
The dual exercise metric is a set-valued estimate, 
and can be represented along 
the implicit stock centroid $\hat{S}^i$ 
with a linewidth gradient propotional to the probability $p$. 
Similarly, the $\ell_1$-average metric 
$\mu_2(t, \{M^j\}, B_i)$ 
can be written with 
\begin{align*} 
  \mu_2(t, \{M^j\}, B_i) = \frac{1}{n} \sum_{j=1}^n \pa{M^j},
\end{align*} 
where 
the sum is over the martingale path $M^j$ 
that falls in the reference cell $B_i$, 
and 
$n$ is the number of such path. 
The $\ell_1$-average metric is a point estimate 
and can be represented with a linewidth gradient 
along 
the reference implicit stock centroid $\hat{S}^i$. 

\subsection{Implementation} \label{meth_implementation} 

To run SPLS and NNM, 
we use a 100 processors machine with 100 GB RAM in the \textcite{googlecloud20}, 
and 
we use the linear programming solver \textcite{gurobi20}. 
To find the nearest-neighbor of a path among a set of centroids 
we use the multiple random projection technique 
of \textcite{hyvonen16}. 

\printbibliography[heading=bibintoc,title={References}]

@article{
	longstaff01,
	title={Valuing American options by simulation: a simple least-squares approach},
	author={Longstaff, Francis A and Schwartz, Eduardo S},
	journal={Review of Financial studies},
	volume={14},
	number={1},
	pages={113-147},
	year={2001},
	publisher={Soc Financial Studies}
}

@article{
	BlackScholes73,
	title={The pricing of options and corporate liabilities},
	author={Black, Fischer and Scholes, Myron},
	journal={The journal of political economy},
	pages={637-654},
	year={1973},
	publisher={JSTOR}
}

@article{
	merton73,
	title={Theory of rational option pricing},
	author={Merton, Robert C},
	journal={The Bell Journal of economics and management science},
	pages={141-183},
	year={1973},
	publisher={JSTOR}
}

@article{
	HarrisonKreps79,
	title={Martingales and arbitrage in multiperiod securities markets},
	author={Harrison, J Michael and Kreps, David M},
	journal={Journal of Economic theory},
	volume={20},
	number={3},
	pages={381-408},
	year={1979},
	publisher={Elsevier}
}

@article{
	crr79,
	title={Option pricing: A simplified approach},
	author={Cox, John C and Ross, Stephen A and Rubinstein, Mark},
	journal={Journal of financial Economics},
	volume={7},
	number={3},
	pages={229--263},
	year={1979},
	publisher={Elsevier}
}

@article{
	rogers02,
	title={Monte Carlo valuation of American options},
	author={Rogers, L.C.G.},
	journal={Mathematical Finance},
	volume={12},
	number={3},
	pages={271--286},
	year={2002},
	publisher={Wiley Online Library}
}

@article{
haugh04,
title={Pricing American options: a duality approach},
author={Haugh, Martin B and Kogan, Leonid},
journal={Operations Research},
volume={52},
number={2},
pages={258--270},
year={2004},
publisher={Informs}
}

@article{
	andersen04,
	title={Primal-dual simulation algorithm for pricing multidimensional American options},
	author={Andersen, Leif and Broadie, Mark},
	journal={Management Science},
	volume={50},
	number={9},
	pages={1222--1234},
	year={2004},
	publisher={Informs}
}

@article{
	Meinshausen04,
	title={Monte Carlo methods for the valuation of multiple-exercise options},
	author={Meinshausen, Nicolai and Hambly, B.M.},
	journal={Mathematical Finance},
	volume={14},
	number={4},
	pages={557--583},
	year={2004},
	publisher={Wiley Online Library}
}

@article{
	hambly10,
	title={A dual approach to multiple exercise option problems under constraints},
	author={Aleksandrov, Nikolay and Hambly, B.M.},
	journal={Mathematical methods of operations research},
	volume={71},
	number={3},
	pages={503--533},
	year={2010},
	publisher={Springer}
}

@article{
	schoenmakers12,
	title={A pure martingale dual for multiple stopping},
	author={Schoenmakers, John},
	journal={Finance and Stochastics},
	volume={16},
	number={2},
	pages={319--334},
	year={2012},
	publisher={Springer}
}

@article{
bender11a,
  title={Dual pricing of multi-exercise options under volume constraints},
  author={Bender, Christian},
  journal={Finance and Stochastics},
  volume={15},
  number={1},
  pages={1--26},
  year={2011},
  publisher={Springer}
}

@article{
	bender11b,
	title={Primal and dual pricing of multiple exercise options in continuous time},
	author={Bender, Christian},
	journal={SIAM Journal on Financial Mathematics},
	volume={2},
	number={1},
	pages={562--586},
	year={2011},
	publisher={SIAM}
}

@article{
	bender15,
	title={Dual representations for general multiple stopping problems},
	author={Bender, Christian and Schoenmakers, John and Zhang, Jianing},
	journal={Mathematical Finance},
	volume={25},
	number={2},
	pages={339--370},
	year={2015},
	publisher={Wiley Online Library}
}

@article{
	carmona08a,
	title={Optimal multiple stopping and valuation of swing options},
	author={Carmona, Ren{\'e} and Touzi, Nizar},
	journal={Mathematical Finance},
	volume={18},
	number={2},
	pages={239--268},
	year={2008},
	publisher={Wiley Online Library}
}

@article{
	marshall11,
	title={Forest of stochastic meshes: A new method for valuing high-dimensional swing options},
	author={Marshall, T James and Reesor, R Mark},
	journal={Operations Research Letters},
	volume={39},
	number={1},
	pages={17--21},
	year={2011},
	publisher={Elsevier}
}

@article{
	pages09,
	title={Optimal quantization for the pricing of swing options},
	author={Bardou, Olivier and Bouthemy, Sandrine and Pages, Gilles},
	journal={Applied Mathematical Finance},
	volume={16},
	number={2},
	pages={183--217},
	year={2009},
	publisher={Taylor \& Francis}
}

@article{
	osterlee12,
	title={Pricing high-dimensional Bermudan options using the stochastic grid method},
	author={Jain, Shashi and Oosterlee, Cornelis W},
	journal={International Journal of Computer Mathematics},
	volume={89},
	number={9},
	pages={1186--1211},
	year={2012},
	publisher={Taylor \& Francis}
}

@article{
	bender06,
	title={An iterative method for multiple stopping: convergence and stability},
	author={Bender, Christian and Schoenmakers, John},
	journal={Advances in applied probability},
	pages={729--749},
	year={2006},
	publisher={JSTOR}
}

@article{
  carmona08b,
  title={Optimal multiple stopping of linear diffusions},
  author={Carmona, Ren{\'e} and Dayanik, Savas},
  journal={Mathematics of Operations Research},
  volume={33},
  number={2},
  pages={446--460},
  year={2008},
  publisher={INFORMS}
}

@book{
  devroye96,
  title={A probabilistic theory of pattern recognition},
  author={Devroye, Luc and Gy{\"o}rfi, L{\'a}szl{\'o} and Lugosi, G{\'a}bor},
  year={1996},
  publisher={Springer Science}
}

@article{
  vapnik71,
  title={On the Uniform Convergence of Relative Frequencies of Events to Their Probabilities},
  author={Vapnik, VN and Chervonenkis, A Ya},
  journal={Theory of Probability and its Applications},
  volume={16},
  number={2},
  pages={264},
  year={1971},
  publisher={Society for Industrial and Applied Mathematics}
}

@article{
  Frank56,
  title={An algorithm for quadratic programming},
  author={Frank, Marguerite and Wolfe, Philip},
  journal={Naval research logistics quarterly},
  volume={3},
  number={1-2},
  pages={95--110},
  year={1956},
  publisher={Wiley Online Library}
}

@techreport{
  selvaprabu13,
  title={Improved least squares Monte Carlo for term structure option valuation with energy applications},
  author={Nadarajah, Selvaprabu and Margot, Fran{\c{c}}ois and Secomandi, Nicola},
  year={2013},
  institution={Technical report 2012-E54, Tepper School of Business, Carnegie Mellon University, Pittsburgh, PA}
}

@article{
  jaillet04,
  title={Valuation of commodity-based swing options},
  author={Jaillet, Patrick and Ronn, Ehud I and Tompaidis, Stathis},
  journal={Management science},
  volume={50},
  number={7},
  pages={909--921},
  year={2004},
  publisher={INFORMS}
}

@article{
  lavassani01,
  title={A discrete valuation of swing options},
  author={Lari-Lavassani, Ali and Simchi, Mohamadreza and Ware, Antony},
  journal={Canadian applied mathematics quarterly},
  volume={9},
  number={1},
  pages={35--74},
  year={2001}
}

@article{
  barrera06,
  title={Numerical methods for the pricing of swing options: a stochastic control approach},
  author={Barrera-Esteve, Christophe and Bergeret, Florent and Dossal, Charles and Gobet, Emmanuel and Meziou, Asma and Munos, R{\'e}mi and Reboul-Salze, Damien},
  journal={Methodology and computing in applied probability},
  volume={8},
  number={4},
  pages={517--540},
  year={2006},
  publisher={Springer}
}

@article{
  dahlgren05b,
  title={A continuous time model to price commodity-based swing options},
  author={Dahlgren, M},
  journal={Review of derivatives research},
  volume={8},
  number={1},
  pages={27--47},
  year={2005},
  publisher={Springer}
}

@article{
  dahlgren05a,
  title={The swing option on the stock market},
  author={Dahlgren, Martin and Korn, Ralf},
  journal={International Journal of Theoretical and Applied Finance},
  volume={8},
  number={01},
  pages={123--139},
  year={2005},
  publisher={World Scientific}
}

@article{
  wilhelm08,
  title={Finite element valuation of swing options},
  author={Wilhelm, Martina and Winter, Christoph},
  journal={Journal of Computational Finance},
  volume={11},
  number={3},
  pages={107},
  year={2008}
}

@article{
	penaud99exo,
  title={Exotic passport options},
  author={Penaud, Antony and Wilmott, Paul and Ahn, Hyungsok},
  journal={Asia-Pacific Financial Markets},
  volume={6},
  number={2},
  pages={171--182},
  year={1999},
  publisher={Springer}
}

@book{
	AimsBook,
	author={Bisschop, J.},
	title={Aimms Optimization Modeling},
	year={2016},
	url = {https://aimms.com/english/developers/resources/manuals/optimization-modeling/}
}

@article{
	silver17,
  title={Mastering the game of go without human knowledge},
  author={Silver, David and Schrittwieser, Julian and Simonyan, Karen and Antonoglou, Ioannis and Huang, Aja and Guez, Arthur and Hubert, Thomas and Baker, Lucas and Lai, Matthew and Bolton, Adrian and others},
  journal={Nature},
  volume={550},
  number={7676},
  pages={354},
  year={2017},
  publisher={Nature Publishing Group}
}

@article{
	browne12,
  title={A survey of monte carlo tree search methods},
  author={Browne, Cameron B and Powley, Edward and Whitehouse, Daniel and Lucas, Simon M and Cowling, Peter I and Rohlfshagen, Philipp and Tavener, Stephen and Perez, Diego and Samothrakis, Spyridon and Colton, Simon},
  journal={IEEE Transactions on Computational Intelligence and AI in games},
  volume={4},
  number={1},
  pages={1--43},
  year={2012},
  publisher={IEEE}
}

@article{
	thompson95,
  title={Valuation of path-dependent contingent claims with multiple exercise decisions over time: The case of take-or-pay},
  author={Thompson, Andrew C},
  journal={Journal of Financial and Quantitative Analysis},
  volume={30},
  number={2},
  pages={271--293},
  year={1995},
  publisher={Cambridge University Press}
}

@article{
	becker19,
  title={Deep Optimal Stopping.},
  author={Becker, Sebastian and Cheridito, Patrick and Jentzen, Arnulf},
  journal={Journal of Machine Learning Research},
  volume={20},
  number={74},
  pages={1--25},
  year={2019}
}

@article{
	lelong16,
  title={Pricing American options using martingale bases},
  author={Lelong, J{\'e}r{\^o}me},
  journal={arXiv preprint arXiv:1604.03317},
  year={2016}
}

@article{
	rogers07,
  title={Pathwise stochastic optimal control},
  author={Rogers, L.C.G.},
  journal={SIAM Journal on Control and Optimization},
  volume={46},
  number={3},
  pages={1116--1132},
  year={2007},
  publisher={SIAM}
}

@article{
	pichler12,
  title={A distance for multistage stochastic optimization models},
  author={Pflug, Georg Ch and Pichler, Alois},
  journal={SIAM Journal on Optimization},
  volume={22},
  number={1},
  pages={1--23},
  year={2012},
  publisher={SIAM}
}

@article{
	heitsch06,
  title={Stability of multistage stochastic programs},
  author={Heitsch, Holger and R{\"o}misch, Werner and Strugarek, Cyrille},
  journal={SIAM Journal on Optimization},
  volume={17},
  number={2},
  pages={511--525},
  year={2006},
  publisher={SIAM}
}

@inproceedings{
	hyvonen16,
  title={Fast nearest neighbor search through sparse random projections and voting},
  author={Hyv{\"o}nen, Ville and Pitk{\"a}nen, Teemu and Tasoulis, Sotiris and J{\"a}{\"a}saari, Elias and Tuomainen, Risto and Wang, Liang and Corander, Jukka and Roos, Teemu},
  booktitle={Big Data (Big Data), 2016 IEEE International Conference on},
  pages={881--888},
  year={2016},
  organization={IEEE}
}

@misc{
	gurobi20,
  author = "Gurobi",
  title = "Gurobi Optimizer Reference Manual",
  year = 2020,
  url = "http://www.gurobi.com"
}

@misc{
	googlecloud20,
  author = {{Google Cloud}},
  title = "Google Cloud Platform",
  year = 2020,
  url = "http://cloud.google.com"
}

@article{
	ramdas17,
  title={On wasserstein two-sample testing and related families of nonparametric tests},
  author={Ramdas, Aaditya and Trillos, Nicol{\'a}s and Cuturi, Marco},
  journal={Entropy},
  volume={19},
  number={2},
  pages={47},
  year={2017},
  publisher={Multidisciplinary Digital Publishing Institute}
}

@article{
	baringhaus04,
  title={On a new multivariate two-sample test},
  author={Baringhaus, Ludwig and Franz, Carsten},
  journal={Journal of multivariate analysis},
  volume={88},
  number={1},
  pages={190--206},
  year={2004},
  publisher={Elsevier}
}

@article{
	szekely04,
  title={Testing for equal distributions in high dimension},
  author={Sz{\'e}kely, G{\'a}bor J and Rizzo, Maria L and others},
  journal={InterStat},
  volume={5},
  number={16.10},
  pages={1249--1272},
  year={2004}
}

@article{
	vapnik81,
  title={Necessary and sufficient conditions for the uniform convergence of means to their expectations},
  author={Vapnik, Vladimir N and Chervonenkis, A Ya},
  journal={Theory of Probability \& Its Applications},
  volume={26},
  number={3},
  pages={532--553},
  year={1982},
  publisher={SIAM}
}

@book{
	vaart96,
  title={Weak convergence and empirical processes: with applications to statistics},
  author={Van Der Vaart, Aad and Wellner, Jon A},
  year={1996},
  publisher={Springer}
}

@article{
	vaart09,
  title={A note on bounds for VC dimensions},
  author={Van Der Vaart, Aad and Wellner, Jon A},
  journal={Institute of Mathematical Statistics collections},
  volume={5},
  pages={103},
  year={2009},
  publisher={NIH Public Access}
}

@article{
	carr92,
  title={Alternative characterizations of American put options},
  author={Carr, Peter and Jarrow, Robert and Myneni, Ravi},
  journal={Mathematical Finance},
  volume={2},
  number={2},
  pages={87--106},
  year={1992},
  publisher={Wiley Online Library}
}

@inproceedings{
	arthur07,
  title={k-means++: The advantages of careful seeding},
  author={Arthur, David and Vassilvitskii, Sergei},
  booktitle={Proceedings of the eighteenth annual ACM-SIAM symposium on Discrete algorithms},
  pages={1027--1035},
  year={2007},
  organization={Society for Industrial and Applied Mathematics}
}

@article{
	lloyd82,
  title={Least squares quantization in PCM},
  author={Lloyd, Stuart},
  journal={IEEE transactions on information theory},
  volume={28},
  number={2},
  pages={129--137},
  year={1982},
  publisher={IEEE}
}

\end{document}